\documentclass[12pt,a4paper]{article}
\usepackage[utf8]{inputenc}
\usepackage{amsmath}
\usepackage{amsfonts}
\usepackage{mathtools}
\usepackage{graphicx}
\usepackage{color}
\usepackage[hidelinks]{hyperref}
\usepackage{tabularx}
\usepackage{float}
\usepackage{array}
\usepackage{multirow}
\usepackage{makecell}
\usepackage{url}
\usepackage{dsfont}
\usepackage{stmaryrd}
\usepackage{natbib}
\usepackage{caption}
\usepackage{titlesec}
\usepackage{lscape}
\usepackage{multicol}
\usepackage{bm}
\usepackage{diagbox}
\usepackage{subcaption}
\usepackage{tabularx}

\renewcommand{\baselinestretch}{1.3}

\makeatletter
\@addtoreset{equation}{section}

\makeatother

\newcounter{Fig}[figure]

\newcounter{Tab}[table]

  {\refstepcounter{Tab}
   \addtocounter{table}{1}
   \samepage\vspace{0.2cm}
   \centerline{#1} \nobreak
   \begin{verse}{Table \thesection.\arabic{Tab}:}
  }%
  {\end{verse} \vspace{0.2cm}}

\relax

\newcommand{\bqa}{\begin{eqnarray*}}
\newcommand{\eqa}{\end{eqnarray*}}
\newcommand{\bqan}{\begin{eqnarray}}
\newcommand{\eqan}{\end{eqnarray}}
\newcommand{\bqt}{\begin{quote}}
\newcommand{\eqt}{\end{quote}}
\newcommand{\bt}{\begin{tabbing}}
\newcommand{\et}{\end{tabbing}}
\newcommand{\bit}{\begin{itemize}}
\newcommand{\eit}{\end{itemize}}
\newcommand{\ben}{\begin{enumerate}}
\newcommand{\een}{\end{enumerate}}
\newcommand{\beq}{\begin{equation}}
\newcommand{\eeq}{\end{equation}}
\newcommand{\bdefi}{\begin{definition}}
\newcommand{\edefi}{\end{definition}}
\newcommand{\bpro}{\begin{proposition}}
\newcommand{\epro}{\end{proposition}}
\newcommand{\blem}{\begin{lemma}}
\newcommand{\elem}{\end{lemma}}
\newcommand{\bth}{\begin{theorem}}
\newcommand{\eth}{\end{theorem}}
\newcommand{\bco}{\begin{corollary}}
\newcommand{\eco}{\end{corollary}}
\newcommand{\bdes}{\begin{description}}
\newcommand{\edes}{\end{description}}
\newcommand{\bre}{\begin{remark}}
\newcommand{\ere}{\end{remark}}

\newtheorem{definition}{Definition}[section]
\newtheorem{proposition}[definition]{Proposition}
\newtheorem{lemma}[definition]{Lemma}
\newtheorem{theorem}[definition]{Theorem}
\newtheorem{corollary}[definition]{Corollary}
\newtheorem{remark}[definition]{Remark}

\usepackage[T1]{fontenc}

\begin{document}

\begin{titlepage}

\title{Cloud failure and cyber insurance: stress scenarios and diversification}

\author{Olivier Lopez$^{1}$, Daniel Nkameni$^{1,2}$}

\date{\today}
\maketitle

\renewcommand{\baselinestretch}{1.1}

\begin{abstract}
The expansion of the cyber insurance market remains exposed to the threat of accumulation events that could simultaneously affect a large number of policyholders. Although few such catastrophes have been observed so far, apart from worldwide cyberattacks such as WannaCry and NotPetya in 2017, the nature of cyber risk makes their occurrence plausible. Stress-testing tools are therefore needed to assess whether an insurance portfolio can withstand such crises. In this perspective, the European Insurance and Occupational Pensions Authority (EIOPA) has identified cloud outage as one of the key scenarios to consider in cyber insurance stress-testing frameworks. In this paper, we propose a framework to model and calibrate cloud-outage scenarios and to measure the diversification of a cyber insurance portfolio. We also show how this diversification can protect against accumulation risk and provide underwriting guidelines to reduce the vulnerability of a portfolio to cloud-outage scenarios.
\end{abstract}

\vspace*{1cm}


\vspace*{1cm}

\noindent{\bf Short title:} Cloud failure and cyber insurance.

\vspace*{.5cm}

{\small
\parindent 0cm
$^{1}$ CREST, CNRS, Ecole polytechnique, Groupe ENSAE-ENSAI, ENSAE Paris, Institut Polytechnique de Paris, Palaiseau, France\\
$^{2}$ Detralytics, Paris La Défense Cedex, France.

E-mails: olivier.lopez@ensae.fr, daniel.nkameni@ensae.fr
}

\end{titlepage}

\tiny
\normalsize
\addtocounter{page}{1}

\section{Introduction}

The adoption of cloud solutions has greatly increased over the last decades (see \cite{flexera2024} and \cite{naldi2016economic}). Many organizations nowadays strongly rely on cloud services to perform their everyday activities, and this goes far beyond the tech sector. Cloud services have multiple uses. The two most common uses are the processing of complex code on remote servers and the efficient storage and sharing of data within organizations (see \cite{hsu2022deeper}). This increasing dependence on cloud solutions exposes organizations to potential issues related to the servers and other tools on which these cloud solutions rely. Indeed, these tools could stop working, be physically destroyed, or be hacked (see \cite{bisong2011overview}). A well-known example among recent incidents is the fire at OVHcloud’s\footnote{OVHcloud, founded in 1999, is a leading European cloud computing company headquartered in Roubaix, France, offering a wide range of cloud services and solutions.} data center in 2021. Although this was an accident, this incident shows the potential weaknesses of the sector. In this particular case, the existence of backups did not protect some customers, since some of these backups were stored in the building that burned down (see \cite{website2021} and \cite{transatlantic2021}).

The risks associated with cloud infrastructures have also been documented in the computer science literature. The study by \cite{li2013cloud} provides one of the earliest surveys of public cloud service outages. Unsurprisingly, their study reveals that cloud interruption is a major concern for users, even though these users are aware that ``zero downtime'' is unrealistic for large-scale Internet services. Their work classifies outage root causes and shows that cloud failures can originate from diverse mechanisms, including software bugs, configuration errors, network failures, human errors, power failures, and cascading dependencies. The last cause also complicates the resolution of the cloud outage once an issue occurs. Indeed, \cite{wang2021fast} stress that modern cloud platforms are composed of many interdependent services, so that the diagnosis and mitigation of an outage may require identifying chains of service dependencies rather than a single isolated failure. This could be extremely time-consuming and costly for users.

From a cyber insurance perspective, these cloud outages, and particularly the scenario of a massive cloud failure, threaten the applicability and viability of cloud interruption insurance. The particular case of a massive interruption affecting a large proportion of an insurance portfolio (accumulation) could instantly generate a large number of claims. This massive interruption would also disrupt the independence between policyholders and hence risk mutualization, which is at the core of the insurance business. The projection of the total impact of such an episode on a whole economy made by \cite{elingscenar} also demonstrates the huge losses that could be generated by such an incident. This is one of the reasons why cloud outage is mentioned by EIOPA as one of the risks that require particular attention when performing cyber insurance stress testing (see \cite{EIOPA}).  

This concern is consistent with the broader literature on systemic cyber risk and cyber accumulation risk. \cite{bohme2006models} argue that cyber risks may exhibit correlation both within firms and across firms, which creates important challenges for insurance pooling. According to the authors, correlation across firms, which is a catalyst for systemic damages, influences insurers' decisions to set premiums and hence to offer cyber insurance products. More recently, \cite{awiszus2023modeling} distinguish idiosyncratic, systematic, and systemic cyber risks, and emphasize that systemic cyber events require modeling approaches that go beyond classical actuarial independence assumptions. The authors insist on the need to work with risk measures that are adapted to accumulation events. This is particularly relevant for cloud failures, where a single technological dependency may simultaneously affect many insured entities (see \cite{aaa2022cyber} and \cite{ga2023cyber}).

The present paper aims at providing a generic framework to model cloud interruption (or, more generally, critical infrastructure failure), and to measure the exposure of a portfolio to such a risk. The focus is not on pricing\footnote{See \cite{mastroeni2017insurance} for an example of pricing methodology applied to the specific insurance of cloud outages and \cite{mastroeni2022pricing} for the pricing of related CatBonds.}, but on defining a portfolio-level measure of diversification that guarantees an acceptable level of mutualization in a cloud insurance portfolio in the case of standard-regime losses but also, and more importantly, in the case of ``systemic''\footnote{The word ``systemic'' here and in the rest of the paper is not used in the sense of financial systemic risk, but is used to describe a situation in which a cloud interruption event affects all or a substantial part of the insurance portfolio.} losses. This work therefore addresses ``cyber underwriting risk'' as defined by the EIOPA from a collective portfolio point of view, and not at an individual policyholder level\footnote{The evaluation of the impact of the disruption of a cloud service on individual policyholders has been studied in numerous studies such as \cite{abualkishik2020disaster}.}.

With this portfolio vision in mind, one of the main contributions of this paper is to provide guidelines to underwriters to achieve a composition of the portfolio that is compatible with diversification requirements in standard-regime and accumulation situations. This is done by ensuring that the proportion of policyholders relying on the different available cloud solutions is sufficiently balanced. Similar approaches have been used in climate risk insurance, where insurers define underwriting guidelines in terms of geographical locations in order to avoid concentrating their risk in a specific geographical location, as this could lead to an accumulation scenario (see \cite{aerts2008dealing} and \cite{denaro2020insurance}).

Our methodological choice to tackle diversification by focusing on cloud service providers is not arbitrary. Indeed, several studies and regulatory analyses argue that the benefits of outsourcing to major cloud service providers must be weighed against the possibility that a disruption at a small number of dominant providers could become a single point of failure (see \cite{gozman2020cloud} and \cite{gulliver2023cloud}). In the financial sector, for example, \cite{asensio2022financial} study cloud outsourcing and financial stability risks and show, through a stylized model, that cloud service providers must be significantly more resilient than individual firms for outsourcing to improve system-wide resilience. Similarly, \cite{ting2023managing} note that cloud adoption raises questions about concentration risk, third-party dependencies, and the need for authorities to understand the systemic implications of critical cloud providers. Cloud service providers therefore have different characteristics that influence their probabilities of interruption, their speed of response after an interruption, the severity of the damages caused by interruptions, etc. To build an optimal portfolio in terms of diversification, these cloud-provider specificities need to be taken into account.

To achieve this, the present paper starts by modeling the consequences of a cloud failure, formalizes this notion of diversification, and accounts for exposure to systemic events. The idea is to distinguish two regimes:

\begin{itemize}
\item first, a situation in which isolated claims happen. This is the standard or classical insurance framework where mutualization operates due to the independence (or the weak form of dependence) between policyholders. This is the regime from which the insurance company expects to generate a significant amount of profitability;

\item a stressed regime, where a large part of the portfolio is simultaneously impacted by a cloud interruption. This stress-test framework measures the ability of the portfolio to withstand such systemic events.
\end{itemize}

Two risk measures are then defined and combined to define the resilience of the portfolio in both regimes, leading to a criterion to be optimized in order to reach the appropriate level of diversification. The underwriting guidelines mentioned above are derived from the solution of this optimization problem. The application of these guidelines is supposed to strengthen the insurance portfolio in standard-regime and stressed situations.

The rest of the paper is organized as follows. In section \ref{sec:cloud}, we introduce the modeling of the cyber insurance portfolio, distinguishing between the two regimes (the standard one and the systemic one). The risk measure used to quantify the exposure of the portfolio to cloud interruption in the standard regime is also presented in this section. In section \ref{sec:optim}, we define the risk measure used in the stressed regime and then define and discuss an optimization criterion to improve the diversification of a cloud insurance portfolio. A practical example of application of the method is given in section \ref{sec:empirical}. Section \ref{sec:conclusion} concludes the study.

\section{Cloud interruption model}
\label{sec:cloud}

Cloud interruptions have been approached through several complementary frameworks. A first stream focuses on the technical availability of cloud services and data centers, using reliability models, stochastic Petri nets, hierarchical models, or Bayesian networks to represent failures, redundancy, replication, and recovery mechanisms (see \cite{jammal2016availability}, \cite{bibartiu2024availability} and \cite{nguyen2019reliability}). For instance, \cite{jammal2016availability} propose a stochastic Petri net approach for analyzing the availability of cloud-deployed applications across geographically distributed data centers. \cite{bibartiu2024availability} use Bayesian networks to model redundant and replicated cloud services, including common-cause failures and communication failures. These approaches are valuable for understanding the engineering determinants of cloud availability, but they generally remain provider- or system-centric. In contrast, the present paper adopts an insurance portfolio perspective, where the key object is not only the availability of a technical system but also the aggregate insured loss generated by a common technological dependency. A well-known issue in cyber insurance is the scarcity of data on losses and policyholders' characteristics, which complicates the use of data-driven approaches to build insurance products (see \cite{malavasi2022cyber}). Model-based approaches, such as the one used in this paper, in which a mathematical model is used to describe the phenomenon being insured, are therefore a natural alternative (see \cite{biener2015insurability}).

In this section, we introduce the model used to describe the loss caused by cloud interruptions at a portfolio level. We start by explaining, in section \ref{sec:general}, the general framework used to describe the total loss at a portfolio level. From an analysis of the consequences of a cloud interruption for a policyholder, we explain how to model the financial cost of the interruption. The first step to achieve this is to model the timeline of the interruption (see section \ref{sec:timeline}). This timeline is then used in Section \ref{sec:loss} to model the losses suffered at all stages of the interruption. Section \ref{sec:profit} studies the characteristics of the portfolio in the case where these losses are standard and mutualization applies.

\subsection{Describing the loss: portfolio and individual level}
\label{sec:general}

Let us consider a market with $d$ cloud providers and a portfolio containing $n$ policyholders. We introduce a binary random variable $\delta_{i,j}\in \{0,1\}$ indicating whether policyholder $i \in \{1,\cdots,n\}$ suffered a financial loss due to an interruption of provider $j \in \{1,\cdots,d\}$. The damage at the portfolio level is therefore given by:

$$\frak{T}_s=\sum_{i=1}^n\sum_{j=1}^d \delta_{i,j}w_{i,j}\tau_i L^{(j)}_{i},$$

where $w_{i,j}\in [0,1]$ denotes the exposure of policyholder $i$ to cloud provider $j$ (that is, the proportion of its cloud-dependent business activities that relies on $j$). We assume throughout the rest of the study that $\sum_{j=1}^d w_{i,j}=1$. $\tau_i \in \mathds{R}^*_+$ denotes the turnover of policyholder $i$, and $L_{i}^{(j)} \in \mathds{R}_+$ denotes the unit financial loss associated with an interruption of service provider $j$ for policyholder $i$. The variables $L_i^{(j)}$ are supposed to have the same distribution as a variable $L^{(j)}$. This representation of the loss implies that the size of the company, materialized by its turnover $\tau_i$, is a scale factor for the loss.

We distinguish the following two cases for the cloud interruption indicator $\delta_{i,j}$:
\begin{itemize}
\item the standard regime, where $\delta_{i,j}$ is stochastic (see section \ref{sec:profit});
\item the stressed regime, corresponding to a systemic event where all users of $j$ are simultaneously affected. This corresponds to $\delta_{i,j}=1$ for all $i$.
\end{itemize}

To model the loss $L_{i}^{(j)}$ suffered during a cloud interruption, we need to understand the timeline of a business interruption.

\subsubsection{Timeline of a business interruption}
\label{sec:timeline}

We consider a situation where policyholder $i$ suffers a financial loss due to an interruption of provider $j$. We denote by $T_{i}^{(j)} \in \mathds{R}^*_+$ the total interruption duration. If the interruptions are non-isolated (stressed regime where $\delta_{i,j}=1$ for all $i$), then $T_{i}^{(j)}=T^{(j)}$ for all $i$. Otherwise, the interruption durations for policyholders $i$ and $i'$ with $i\neq i'$ can be considered independent. We characterize the timeline of an interruption using three variables:

\begin{enumerate}
\item the initial duration of unavailability of the cloud service $T_{i}^{(j)}$. In the case of a non-isolated interruption, this duration is common to every policyholder in the portfolio associated with cloud provider $j$;
\item the time $U_i\in \mathds{R}^*_+ \bigcup \{\infty\}$ after which the policyholder is able to react via the introduction of a backup plan. This backup plan does not fully replace the functions of the cloud servers, but helps resume a certain proportion of the policyholder's activity, thereby reducing the loss. Let us also note that this backup plan may not exist or may come too late if $U_i$ is larger than the total duration of the interruption. In these cases, we consider $U_i=\infty$;
\item the additional time $V_i\in \mathds{R}^*_+$ after which the activity resumes at full capacity after restoration of the service. Indeed, restoration of the cloud service does not always mean immediate recovery for the cloud user, who will usually need additional time before returning to normal operations. In cloud environments, recovery is generally not limited to the restoration of the provider's service: customers may also need to restore data, restart applications, verify consistency, redeploy workloads, and activate alternative operating procedures. Hence, for a given policyholder, the total duration of the event is $T_i^{(j)}+V_i$.
\end{enumerate}

It is important to note that, in this timeline, any backup or recovery plan is assumed to be activated only during the interruption period $T_i^{(j)}$, and not during the recovery period $V_i$. The role of the backup plan is to reduce damages while the cloud service remains unavailable. Once the interruption is resolved, the policyholder focuses on resuming normal activities, so activating a backup plan at this stage is no longer the priority and may even be counterproductive.

This decomposition of the timeline of the outage is aligned with the approach considered in \cite{lloyds}. The distinction between the initial interruption duration and the additional time needed to return to normal operations is consistent with the disaster recovery and business continuity literature. \cite{abualkishik2020disaster} review disaster recovery mechanisms in cloud computing and emphasize the role of backup, replication, and recovery time. This supports the need to distinguish the duration of the provider outage from the effective duration of business disruption experienced by the policyholder.

In practice, the distribution of the variables $U_i$ and $V_i$ may differ from one policyholder to another, and could also depend on the cloud provider affected by the outage. Covariates could be introduced to capture this heterogeneity. However, to simplify the notation and discussion, we assume that the policyholders are ``identical'', in the sense that $(U_i)_{1\leq i \leq n}$ and $(V_i)_{1\leq i \leq n}$ are identically distributed random variables. We also assume that the policyholders are independent, and that $T_i^{(j)}$ is independent of $U_i$ and $V_i$.

\subsubsection{Loss experienced by the policyholder}
\label{sec:loss}

The report by \cite{lloyds} introduces a rough way to translate a duration of service interruption into the value of the loss experienced by a policyholder affected by a cloud failure. The basic idea is to assume that one day of business interruption leads to a loss equal to $1/365$ of the annual turnover. This report also proposes restricting the turnover considered in this calculation to the online activities of the company. This restriction may not be legitimate: although a cloud interruption disrupting the information systems has direct consequences on online activities, it could also disrupt physical dimensions of the business. Indeed, a company could stop working because of a loss of access to some digital applications. We therefore retain our assumption that the loss is proportional to the turnover $\tau_i$ and to the duration of inactivity. We also retain the intuitive assumption that the magnitude of the loss for policyholder $i$ is proportional to the exposure of this policyholder to the affected cloud provider $j$ (that is, $w_{i,j}$, the proportion of cloud activities of $i$ depending on $j$).

The translation of cloud interruption into economic loss has also been studied from a customer-risk perspective. \cite{naldi2017evaluation} evaluates customer losses under cloud outages using loss exceedance probabilities and Value-at-Risk, thereby linking availability analysis to operational loss measurement. \cite{mastroeni2017insurance} then use this type of loss modeling to study insurance pricing and refund sustainability for cloud outages. \cite{mastroeni2022pricing}, on the other hand, propose a reinsurance scheme to compensate cloud providers, who will in turn compensate customers, in case of outage. These contributions are close to the present work in that they explicitly connect cloud service interruption, customer losses, and insurance mechanisms. The difference is that our model is designed for portfolio-level loss, accumulation, and underwriting analysis: instead of assessing the exposure of one cloud customer or one cloud provider, we aggregate losses across policyholders sharing common cloud dependencies in order to determine the best underwriting rules to ensure mutualization, including in stressed regimes.

With this in mind, we recall the definition of the financial loss suffered by policyholder $i$, which was given in Section \ref{sec:general} as $w_{i,j}\tau_{i}L^{(j)}_i$, where

\begin{equation}
L^{(j)}_i=\frak{a}_{i,j}(T_{i}^{(j)}-(T_{i}^{(j)}-U_i)_+(1-\frak{b}_{i}))+\frac{\frak{a}_{i,j}\left\{1-(1-\frak{b}_{i})\epsilon^{(j)}_i\right\}V_i}{2}
\label{loss}
\end{equation}

The variable $\frak{a}_{i,j} \in \mathds{R}^*_+$ is the average (over time) insured cost\footnote{Unlike raw financial losses, insured losses are usually subject to deductibles and contractual limits that prevent the insurer from bearing the full amount of extreme losses.} of one day of service interruption for policyholder $i$ in the absence of an active backup plan when cloud provider $j$ is unavailable. Making this cost depend on $j$ emphasizes the fact that cloud providers do not necessarily supply the same services: some may have a higher degree of criticality, which is reflected in a higher value of $\frak{a}_{i,j}$. The variable $1-\frak{b}_{i}\in [0,1]$ measures the efficiency of the backup plan of policyholder $i$. Equivalently, $\frak{b}_{i}$ represents the residual proportion of the loss after the backup plan is activated. We assume in the rest of the study that $(\frak{a}_{i,j},\frak{b}_{i})_{1\leq i \leq n}$ are independent, and that, for all $j$, $(\frak{a}_{i,j},\frak{b}_{i})_{1\leq i \leq n}$ have the same distribution.

$\epsilon_i^{(j)}=\mathbf{1}_{U_i\leq T_{i}^{(j)}}$ is an indicator function equal to 1 if the backup plan manages to activate before the end of the service interruption. As explained above, when a backup plan exists, it starts after a time $U_i < +\infty$. 

\begin{figure}[!h]
    \centering
    \includegraphics[width=\linewidth]{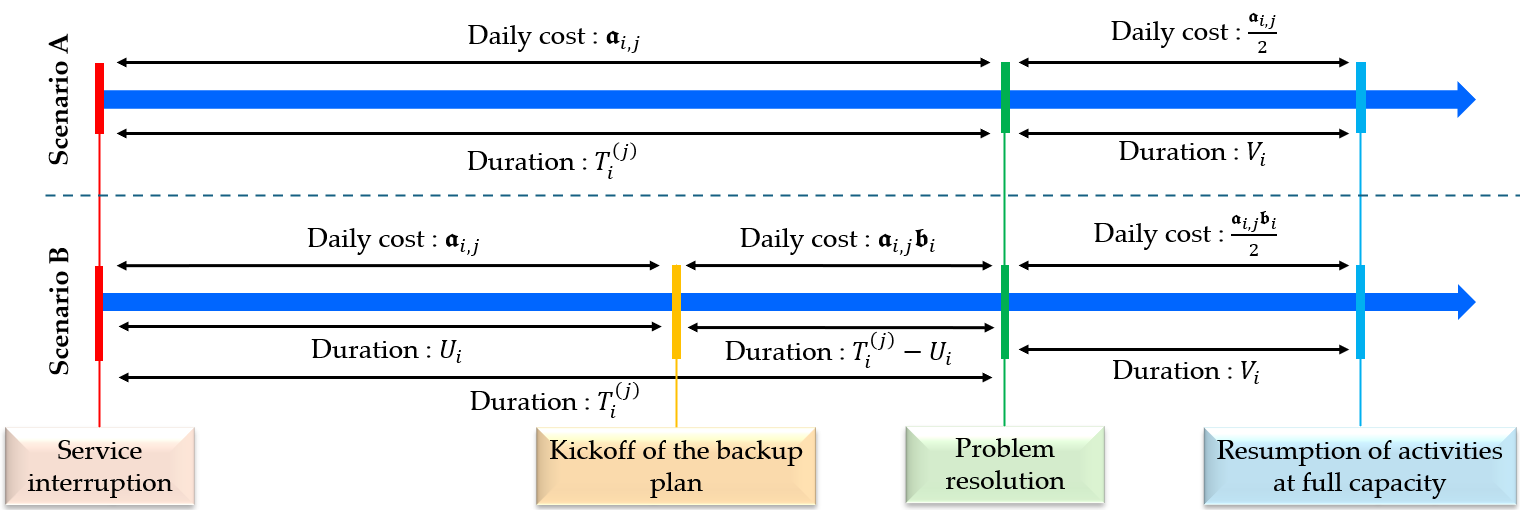}
    \\[20pt]
    \caption{Timeline of a cloud-related business interruption of provider $j$. In scenario A, policyholder $i$ does not have a backup plan, and in scenario B, policyholder $i$ has a backup plan with efficiency $1-\frak{b}_{i}$.}
    \label{fig:timeline_illust}
\end{figure}

The loss model illustrated by Figure \ref{fig:timeline_illust} is therefore a superposition of three periods:

\begin{itemize}
\item first, the policyholder suffers a loss equal to $\frak{a}_{i,j}$ per unit of time, for a duration $T_{i}^{(j)}-(T_{i}^{(j)}-U_i)_+;$
\item then, during $(T_{i}^{(j)}-U_i)_+$, the loss is reduced to $\frak{a}_{i,j}\frak{b}_{i};$
\item finally, the service is restored, and we assume a continuous and linear decrease in the loss, from $\frak{a}_{i,j}$ (resp. $\frak{a}_{i,j}\frak{b}_{i}$) to 0 over a period of time $V_i$ if the backup plan failed (resp. managed) to activate. This is represented by the last term of Equation \ref{loss}.
\end{itemize}

We assume that the random vectors $(U_i,V_i,\frak{a}_{i,j},\frak{b}_{i})$, $1\leq i \leq n$, are i.i.d. and independent of $T_{i}^{(j)}$, and that their four components are also mutually independent. It is of course possible to relax this last assumption of independence between $U_i$, $V_i$, $\frak{a}_{i,j}$ and $\frak{b}_{i}$ (for example, one can assume that if the backup plan starts soon enough, this has implications for the value of $V_i$, which should be shorter), but this would unnecessarily increase the complexity of the model, since the present version helps achieve the goals of our study.

\subsection{Profitability of the portfolio in the standard regime}
\label{sec:profit}

We first consider the standard case where we only have isolated claims. By using the term ``isolated'', we do not necessarily mean that there is no correlation between policyholders and/or incidents, but simply that we are working in the classical framework where mutualization is valid. In this framework, optimizing the portfolio composition is a matter of optimizing profitability.

To simplify the model, we consider that every policyholder pays a premium that is proportional to its expected loss under the standard regime, with the same loading factor. More precisely, this means that if a policyholder $i$ generates a loss $X_i$, the premium is $(1+\theta)E[X_i]$, with the same value of $\theta$ for all policyholders. In this framework, a higher expected loss corresponds to a higher expected premium, and therefore to a higher expected return, although the variance can of course give a more mixed picture. The approach that we consider can be extended to the case where the loading factors are not the same for all individuals, while also introducing more advanced pricing schemes that take correlation into account, as in \cite{mastroeni2023cyber}. In the present case, we keep the idea that imposing a constraint on the expected loss of the portfolio is equivalent to imposing a constraint on average profitability.

In the rest of the study, we assume that the probability of failure for provider $j$ is $\mathbb{P}(\delta_{i,j}=1)=p_j$, that is, that it only depends on cloud provider $j$ and not on the policyholder's use. In this framework, policyholder $i$ can experience at most one incident from provider $j$ over the time period considered. However, the model can be generalized by considering that $w_{i,j}\tau_iL_{i}^{(j)}$ is the aggregated amount of all losses generated by $j$ for $i$. 

Assuming that $\delta_{i,j}$ and $L_{i}^{(j)}$ are independent, the average loss of the portfolio becomes:

\begin{equation}\label{eq:moy}\sum_{j=1}^d   \sum_{i=1}^n p_j m_j w_{i,j}\tau_i=\sum_{j=1}^d \pi_j \bar{w}_j=\bm{\pi}'\bar{\mathbf{w}}=\frak{L}_s(\bar{\mathbf{w}}),
\end{equation}

where $m_j = \mathbb{E}(L^{(j)})$, $\pi_j=p_jm_j$, $\bm{\pi}=(\pi_j)_{j=1,\cdots,d}'$, $\bar{w}_j=\sum_{i=1}^nw_{i,j}\tau_i$, and $\bar{\mathbf{w}}=(\bar{w}_j)'_{1\leq j \leq d}$, with $\mathbf{z}'$ denoting the transpose of vector $\mathbf{z}$. Clearly, we have $\sum_{j=1}^d \bar{w}_j=\sum_{i=1}^n \tau_i$, which can be considered as the aggregated turnover of the portfolio, and $\bar{w}_j$ can be understood as the exposure of the insurance company, as an aggregator of individual risks, to cloud provider $j$.

It is clear that if the target is to maximize expected profitability under the premium structure described above, the insurer would be attracted to providers associated with higher expected losses, since they generate higher expected premiums. In such a situation, the insurance company may become too exposed to a systemic default of this provider. This could lead to a very high variance even in the standard regime. Following the classical quadratic view of modern portfolio theory as defined by \cite{markowitz1991foundations}, we therefore aim to control the variance.

At a portfolio level, the variance of the loss is

$$
\sum_{j,k=1}^d \sum_{i_1,i_2=1}^n w_{i_1,j}\tau_{i_1}Cov(\delta_{i_1,j}L_{i_1}^{(j)},\delta_{i_2,k}L_{i_2}^{(k)})w_{i_2,k}\tau_{i_2},
$$

where $Cov(X,Y)$ denotes the covariance between two random variables $X$ and $Y$.

We make the following assumption:

\begin{equation}
\label{hyp_independance}
Cov(\delta_{i_1,j}L_{i_1}^{(j)},\delta_{i_2,k}L_{i_2}^{(k)})=\sigma_j \sigma_k r_{j,k},
\end{equation}

where
$$\sigma^2_j=Var(\delta_{i,j}L_{i}^{(j)})= p_j (E[L^{(j)2}]-p_jE[L^{(j)}]^2).$$

Typically, Equation \ref{hyp_independance} means that the correlation between the costs of two incidents is driven by the cloud providers and that policyholders are otherwise uncorrelated except through these providers.

This assumption is relatively realistic if we consider that the size of the portfolio is large enough, and if we recall that we are in the standard regime, where we exclude the existence of systemic incidents that would simultaneously affect a significant part of the portfolio. This latter situation will be considered in a second step. The idea that, for a large cyber insurance portfolio in the absence of accumulation, the correlation between policyholders is mainly driven by common technologies, common vulnerabilities, or common service providers, unlike what would be observed in traditional insurance, is also supported by the literature (see, for example, \cite{bohme2006models} and \cite{boumezoued2023cyber}).  

Assumption \ref{hyp_independance} helps simplify the expression of the variance of $\frak{L}_s$, which becomes:

\begin{equation}
\label{eq:var}
\frak{V}_s(\bar{\mathbf{w}})=
\bar{\mathbf{w}}'\left(\begin{array}{ccc}\cdots & \cdots & \cdots \\
\cdots & \sigma_j \sigma_k r_{j,k} & \cdots \\
\cdots & \cdots & \cdots
\end{array}\right)\bar{\mathbf{w}}=\bar{\mathbf{w}}'\Sigma \bar{\mathbf{w}}.
\end{equation}

If systemic cloud failures were not an issue, then optimizing the portfolio would simply imply minimizing $\frak{V}_s(\bar{\mathbf{w}})$ subject to a constraint on the average loss $\frak{L}_s(\bar{\mathbf{w}})$. This framework corresponds to the classical portfolio theory framework of \cite{markowitz1991foundations}, where the assets would be the different cloud providers, and where the objective would be to determine the optimal proportion of the portfolio turnover to be, ideally, exposed to each of these providers.

This rough definition of an optimal portfolio in a context where diversification and mutualization are the main objectives is clearly not adapted to protecting this portfolio against systemic events. We therefore need to develop a specific modeling framework for these particular events. This will add constraints to the optimization problem mentioned above.

\section{Systemic risk measure, diversification and optimal insurance portfolio}
\label{sec:optim}

In Section \ref{sec:cloud}, we defined a way to measure the profitability and risk associated with a portfolio, based on the distribution $\bar{\mathbf{w}}$ of the exposed turnover across the different cloud providers. The purpose of the present section is to determine an optimization problem that allows us to identify the best configuration in terms of $\bar{\mathbf{w}}$ if one wants to achieve sufficient diversification in both standard and stressed regimes. Before defining this optimization problem in Section \ref{sec_optim_div}, we define in Section \ref{sec:systemic} the systemic risk measure in the stressed regime, as was done in Section \ref{sec:profit} for the standard regime.

\subsection{Systemic risk measure in the stressed case}
\label{sec:systemic}

At the portfolio level, in the case of a systemic interruption of the $j$-th cloud provider, the insurance company suffers the aggregated loss $\mathbf{L}^{(j)}=\sum_{i=1}^n w_{i,j}\tau_i L^{(j)}_i$,
where $L^{(j)}_i$ has been defined in Section \ref{sec:loss}, and depends on the same interruption duration $T^{(j)}$, which applies to every policyholder $i$. In this section, our aim is to measure how resilient the portfolio would be if such an event were to occur.

The distribution of $\mathbf{L}^{(j)}$ may be difficult to obtain analytically. A possibility is to use Monte Carlo simulations. However, this would lead to difficulties in understanding the impact of portfolio diversification: indeed, the distribution of the loss would depend on too many parameters, namely the proportions $w_{i,j}$ for each policyholder. On the other hand, our objective of analyzing how diversified the portfolio is requires a more macroscopic vision. Indeed, if we want to act to improve the diversification of the portfolio, it will be difficult to act at an individual level, that is, by changing $w_{i,j}$ for all $i$, or by recruiting new policyholders with a specific value of $w_{i,j}$. More detailed justification for this macroscopic vision is given in Section \ref{sec:compare}.

Therefore, we would ideally like a risk measure that only depends on $\bar{w}_j$, say $\mu(\bar{w}_j)$, since $\bar{w}_j$ represents the part of the turnover exposed to the failure of cloud provider $j$ in the portfolio. From an underwriting perspective, we believe that it would be easier to act on this allocation by trying to restructure the portfolio during new business acquisition. To build such a risk measure, a solution is to approximate the distribution of losses using asymptotic theory. This requires a sufficiently large number of policyholders $n$, which we assume to be the case.  

Let 
$$s_w^2=\sum_{i=1}^n \left(\frac{w_{i,j}\tau_i}{\bar{w}_j}\right)^2$$

We define:

\begin{eqnarray*}
\frak{m}_t &=& E\left[L_i^{(j)}|T^{(j)}=t\right], \\
\sigma^2_t &=& Var (L_i^{(j)}|T^{(j)}=t),
\end{eqnarray*} 

For sufficiently large $n$, we can approximate the distribution of $\mathbf{L}^{(j)}/\bar{w}_j$ by that of a random variable $\mathcal{L}^{(j)}$ such that:

\begin{equation}\mathcal{S}^{(j)}(l)=\mathbb{P}\left(\mathcal{L}^{(j)}\geq l\right)=\int_0^{\infty}\bar{\Phi}\left(\frac{l-\frak{m}_t}{s_w \sigma_t}\right)d\mathbb{P}_j(t)=\frak{f}(s_w,l), \label{eq:SJ}
\end{equation}
where $\mathbb{P}_j$ denotes the distribution of $T^{(j)}$. The proof of this approximation can be found in Section \ref{sec:proof}.

The interesting feature of this approximation is that the distribution of $\mathbf{L}^{(j)}$ is essentially reduced to its dependence on $\bar{w}_j$ and $s_w$. To measure how much a portfolio is exposed to the risk of massive losses generated by a cloud provider outage, it is natural to use some characteristics of the distribution of $\mathbf{L}^{(j)}$.

Based on this approximation, we then discuss three possible choices of risk measures:

\begin{enumerate}
\item Expectation of $\mathbf{L}^{(j)}$: this quantity is simply $\bar{w}_j E\left[\frak{m}_T\right]=\bar{w}_j \int_0^{\infty} \frak{m}_t d\mathbb{P}_j(t)$, which is already of the form $\mu(\bar{w}_j)$. Using this risk measure to identify the weakness of a portfolio in the context of a systemic cloud interruption has the disadvantage of focusing on the average loss scenario, which might be optimistic. The next two approaches are adapted to more pessimistic scenarios.  

\item Quantile of $\mathbf{L}^{(j)}$ (or Value-at-Risk): let $q_{\alpha_j}$ denote the $\alpha_j$-level upper quantile of $\mathbf{L}^{(j)}$, defined as the solution of
$\mathbb{P}(\mathbf{L}^{(j)}\geq q_{\alpha_j})=1-\alpha_j,$
with $\alpha_j\in (0,1)$, and essentially, $1-\alpha_j$ close to zero. Using $q_{\alpha_j}$ helps account for more pessimistic scenarios compared to the central version offered by the expectation. If we use approximation \ref{eq:SJ}, we get
$$q_{\alpha_j}\approx \bar{w}_j \mathcal{S}^{(j)-1}(1-\alpha_j).$$
This expression is not yet in the proper form, since, from \ref{eq:SJ}, we have $\bar{w}_j \mathcal{S}^{(j)-1}(1-\alpha_j)=\bar{w}_j\frak{f}^{-1}(s_w,1-\alpha_j)$. The dependence on $s_w$ does not fit with our need to have a representation of the risk measure only in terms of $\bar{w}_j$. Indeed, $s_w$ can be seen as a ``micro-level'' measure, since it measures the variation of weights among policyholders in the portfolio. Nevertheless, it can be easily verified that $s_w\rightarrow \frak{f}^{-1}(s_w,1-\alpha_j)$ is a non-decreasing function. Hence, if we consider that $s_w\leq s$, then we can use
$$\mu(\bar{w}_j)=\bar{w}_j\frak{f}^{-1}(s,1-\alpha_j),$$
as an upper bound of the previous quantity that does not depend on $s_w$. The constraint $s_w\leq s$ can be added to the set of constraints of the optimization problem (see Section \ref{sec_optim_div}), or relaxed. We choose the latter option in the numerical application and justify this choice by assuming that $s$ is sufficiently large.

\item CTE of $\mathbf{L}^{(j)}$: among risk measures, the Conditional Tail Expectation (CTE) is sometimes preferred to the quantile, since it provides information on the magnitude of extreme losses when they are larger than the quantile. The CTE is defined as
$$CTE(\alpha_j)=E\left[\mathbf{L}^{(j)}|\mathbf{L}^{(j)}\geq q_{\alpha_j}\right].$$
If we use approximation \ref{eq:SJ}, this CTE can be considered as proportional to $\bar{w}_j$, since
$$CTE(\alpha_j)\approx q_{\alpha_j}+\frac{\bar{w}_j}{1-\alpha_j}\int^{\infty}_{\mathcal{S}^{(j)-1}(1-\alpha_j)}\mathcal{S}^{(j)}(t)dt=\bar{w}_j\frak{g}(s_w,\alpha_j).$$
Again, the function $s_w\rightarrow \frak{g}(s_w,\alpha_j)$ is non-decreasing, and similarly to the Value-at-Risk, we can introduce the measure
$\mu(\bar{w}_j)=\bar{w}_j\frak{g}(s,\alpha_j).$
\end{enumerate}

We will therefore consider an aggregate risk measure of the form: 

\begin{equation}
\label{eq:gen}
\sum_{j=1}^d \mu_j(\bar{w}_j)=\mathbf{m}'\bar{\mathbf{w}},
\end{equation}

to describe the exposure of the portfolio to stress regimes and accumulation scenarios in which all or almost all policyholders are hit by a cloud interruption.

\begin{remark}
\label{rem:distribution}
In this section and in the rest of the study, we use a Gaussian approximation of the losses. This approximation is legitimate if the number of policyholders is high enough, and if the distribution of the loss is not heavy-tailed in the extreme value theory sense of the term (otherwise, high quantiles may be better approximated by a Generalized Pareto distribution, see \cite{mikosch1998}). We implicitly assume that the individual losses are moderate, and that the solvency of the portfolio is at risk more because of the number of victims than because of the value of a single claim. Relying on a Generalized Pareto distribution to approximate $\mathbf{L}^{(j)}$ is an alternative, but would not provide risk measures that are linear with respect to $\bar{\mathbf{w}}$ as in Equation \ref{eq:gen}. Moreover, let us note that the probabilities $(\alpha_j)_{j=1,\cdots,d}$ do not necessarily need to be as close to 1 as the $0.995$ value used in the Solvency II regulation for the Value-at-Risk: we are studying stress scenarios, whose probabilities of occurrence are not quantified in advance. The values $\alpha_j$ are essentially used to choose the level of conservativeness of the risk measure.
\end{remark}

\begin{remark}
In the stressed regime, we consider, for each cloud provider, the consequences of a failure of its solution that affects all its users. This is a very conservative approach that also embodies the less severe situation where a few users are not affected by the interruption. Adopting a more conservative position in our work aims to provide an optimal portfolio structure that would be resilient in both moderate and severe accumulation scenarios. 
\end{remark}

\subsection{Optimization problem}
\label{sec_optim_div}

In Sections \ref{sec:profit} and \ref{sec:systemic}, we defined two ways to measure risk at the portfolio level. The first one corresponds to the standard regime, where variance, as a quadratic function of $\bar{\mathbf{w}}$, is a way to measure uncertainty and the level of mutualization. On the other hand, in the case of a systemic cloud failure, the risk measures considered in Section \ref{sec:systemic} can be approximated by $\mathbf{m}' \bar{\mathbf{w}}$, for some vector $\mathbf{m}$. We propose to combine these two risk measures in a single criterion given by:
$$\frak{C}_{\lambda}(\bar{\mathbf{w}})=\frac{1}{2}\bar{\mathbf{w}}'\Sigma \bar{\mathbf{w}}+\lambda \mathbf{m}' \bar{\mathbf{w}},$$
where $\lambda>0$ is a parameter to be chosen by the insurer to quantify the relative importance of the risk measure in the standard case, materialized by $\bar{\mathbf{w}}'\Sigma \bar{\mathbf{w}}$, and the risk measure in the systemic case. We propose, in Section \ref{sec_lambda}, a way to calibrate this parameter $\lambda$.

For a given total exposure $W$ and an expected average loss equal to $\rho$, finding the optimal portfolio is then a quadratic optimization problem under constraints, that is:

\begin{align*}
\min_{\bar{\mathbf{w}}} \; & \frac{1}{2}\bar{\mathbf{w}}'\Sigma \bar{\mathbf{w}}+\lambda \mathbf{m}' \bar{\mathbf{w}}, \\
\text{s.t.} \;  
& \bm{\pi}'\bar{\mathbf{w}}=\rho,\\
& \bar{\mathbf{w}}\geq 0, \\
& \sum_{j=1}^d \bar{w}_j = W.
\end{align*}

Since $\Sigma$ is positive definite, the problem is a convex optimization problem and has a unique solution if the polyhedron defined by the constraints is nonempty. The literature provides many standard methods to solve this classical optimization problem (see, for example, \cite{floudas1995quadratic}).

\subsection{Practical choice of $\lambda$}
\label{sec_lambda}

The optimization criterion proposed in Section \ref{sec_optim_div} is designed to be simple: we want both the variance and the risk measure in the case of a catastrophe to be small, so we use a linear combination of the two objectives. However, the practical meaning of the parameter $\lambda$ is relatively unclear. This parameter is used to describe the importance of one part of the criterion with respect to the other, but these two parts are not of the same nature: the quadratic part is a variance, while the second part, related to the systemic risk measure, corresponds to an average, a quantile, or a conditional expectation.

The aim of this section is to provide guidelines that will help choose a realistic value of $\lambda$ corresponding to the needs of an insurer in an operational setting.

To do so, let us first consider the standard regime. If the size of the portfolio is large enough and if the policyholders are independent, we can approximate the distribution of the random total loss of the portfolio by a Gaussian random variable, that is, $\frak{L}_s\sim \mathcal{N}(\frak{L}_s(\bar{\mathbf{w}}),\frak{V}_s(\bar{\mathbf{w}}))$.
This Gaussian approximation is valid if, in addition to the large size of the portfolio, we focus on insured financial losses rather than raw financial losses. Indeed, the distribution of insured losses in a large portfolio can often be considered light-tailed, since insurance policies often include compensation limits that eliminate extreme losses for the insurer. In the particular case of cyber insurance, part of these extreme losses may also be transferred to reinsurance. These assumptions are even more reasonable for an insurer focusing on small and medium-sized companies. For heavy-tailed losses, other kinds of distributions could be used; see, for example, \cite{mikosch1998} for detailed results on approximations in collective risk models. However, the same reference also argues that the Gaussian approximation may remain legitimate even in the case of unlimited policies, provided that the loss quantiles considered are not very high. 

If we consider an insurer planning to allocate a maximal amount of capital $\frak{c} \in \mathds{R}^*_+$ to face the risk, the probability that this reserve is insufficient can be approximated by

\begin{equation}
\bar{\Phi}\left(\frac{\frak{c}-\frak{L}_s(\bar{\mathbf{w}})}{\frak{V}_s(\bar{\mathbf{w}})^{1/2}}\right), \label{eq:varisk}
\end{equation} 
where $\bar{\Phi}$ is the survival function of a $\mathcal{N}(0,1)$ random variable. Under the constraint that the expected loss is $\rho$, which is a constraint of the optimization problem, the probability of exceeding the reserve in (\ref{eq:varisk}) will be less than a given value $\eta$ provided that

\begin{equation}
\label{cond1}
\frac{1}{2}\frak{V}_s(\bar{\mathbf{w}})=\frac{1}{2}\bar{\mathbf{w}}'\Sigma \bar{\mathbf{w}}\leq \frac{(\frak{c}-\rho)^2}{2\bar{\Phi}^{-1}(\eta)^2}.
\end{equation}

On the other hand, if we consider the systemic case discussed in Section \ref{sec:systemic}, we see that the three risk measures presented could correspond in practice to an amount of reserve or capital: the average loss in the case of accumulation, a high loss quantile, or the average value of the loss beyond this quantile. We then consider that the insurer does not want these values to exceed $\frak{c}$, where $\frak{c}$ is the same limit value as the one used in the first situation, where we were only considering the standard regime. This means that the optimal portfolio is expected to satisfy the condition $\mathbf{m}'\bar{\mathbf{w}}\leq d\frak{c}$, since we have $d$ cloud providers and $\mathbf{m}'\bar{\mathbf{w}}$ corresponds to the sum of capital amounts associated with failure scenarios for each of the providers.

An optimal portfolio for the program defined in Section \ref{sec_optim_div} is associated with a low variance and a low value of the systemic risk measure. If we expect the optimal portfolio to have a variance close to the bound in condition \ref{cond1}, and a value of the systemic risk measure also close to the limit $d\frak{c}$, and if we give equal importance to both objectives, then, in $\frak{C}_{\lambda}$, the quadratic part and the linear part should have the same importance at the optimum. We therefore expect to have

\begin{equation}
\lambda d\frak{c}=\frac{(\frak{c}-\rho)^2}{2\bar{\Phi}^{-1}(\eta)^2}. 
\label{eq:lambda}
\end{equation}

This is the methodology we propose to set a reasonable value for $\lambda$ in practice. However, in an operational setting, any other rule that makes sense from a business perspective could be used. Expert judgment could also intervene to decide the level of importance to give to each risk measure.

\subsection{Comparison with other diversification measures}
\label{sec:compare}

We made the choice to study the standard regime using a quadratic function, and to study the stressed regime separately. This leads to a modification of the classical criterion used in modern portfolio theory. The introduction of this additional factor is motivated by the need to protect the portfolio in case of systemic episodes. The same motivation has led to other strategies in the portfolio optimization literature, such as strategies based on optimizing the CTE or Conditional Value-at-Risk (see \cite{angelelli2008comparison}, \cite{ray2017particle} and \cite{cesarone2011portfolio}). In these approaches, the authors do not distinguish between two regimes, but directly model the tail of the distribution of the loss variables. To use such approaches, one could, for example, approximate the tail of the loss of the portfolio using extreme value theory (see \cite{mainik2015portfolio}).

However, this would have many operational drawbacks that are directly linked to our insurance context and to the lack of data on systemic cyber events: due to limited statistical experience with extreme claims, historical data may not be sufficient to correctly estimate the tail of the distribution. Moreover, this distinction between two regimes corresponds to a distinction between catastrophes, which escape the framework of statistical modeling and whose probability is impossible to evaluate accurately, since it refers to events that have not yet been encountered or that are yet to come, and a situation where classical rules of insurance apply.

In this optimization problem, we focus on finding $\bar{\mathbf{w}}$. Another optimization approach could require determining the optimal value of $w_{i,j}$ for all $i$ and $j$ (see \cite{gunjan2023brief}). Considering this would considerably increase the dimension and complexity of the problem. Also, the proposed approach is aligned with the broader systemic-risk literature. Although our setting is not a financial contagion model in the strict sense, as in the models studied by \cite{caccioli2018network} and \cite{jackson2021systemic}, the mechanism is analogous: a common cloud dependency acts as an overlapping exposure across policyholders. This analogy supports our choice to control the aggregate exposure to each provider rather than focusing on individual policyholder-level risk.

Finally, this choice to solve for the aggregated indicator $\bar{\mathbf{w}}$ in the optimization problem is also motivated by the fact that we are studying an insurance portfolio in which policyholders should not be targeted or rejected in a discriminatory manner. Considering an optimization with respect to $w_{i,j}$ could lead to the targeting and rejection during underwriting of policyholders whose activity is largely covered by a service provider that does not improve the diversification of the portfolio. Optimizing the aggregated indicator $\bar{\mathbf{w}}$ is therefore preferable, since it is more efficient for risk management and for the transparent and non-discriminatory improvement of diversification. Indeed, the solution obtained would help monitor the general evolution of the distribution of cloud providers within the portfolio without targeting individual policyholders.

\section{Numerical illustration}
\label{sec:empirical}

In this section, we illustrate, through a practical example, the modeling of the impact of cloud failures on a realistic synthetic cloud insurance portfolio. We also study how solving the optimization problem of Section \ref{sec_optim_div} can lead to actionable underwriting guidelines for a cloud interruption insurance product. We start by describing, in Section \ref{sec:distr}, the distributions chosen to model cloud outages. In Section \ref{sec:proba}, we present the data, the interruption probabilities and costs, and the covariance structure used in our numerical analysis. Section \ref{sec:resultats} discusses the optimal portfolio, the targeted underwriting rules, and their evolution with respect to the parameters of the cloud interruption model.

\subsection{Distribution of the loss}
\label{sec:distr}

The calibration of cloud-outage scenarios remains difficult because public data on provider failures are limited, heterogeneous, and often incomplete. In addition to data scarcity, technology changes quickly, and historical observations may not be representative of future systemic scenarios. For this reason, several cyber risk studies combine limited empirical data with enhanced modeling, stress scenarios, and expert assumptions. The numerical illustration of this study should therefore be interpreted as a transparent modeling and stress-testing exercise designed to be reproduced on an industry insurance portfolio to guide underwriting.

To build the loss of Equation \ref{loss} in our synthetic portfolio, we need to specify five variables:

\begin{enumerate}
\item the cloud interruption time $T_i^{(j)}$. We consider a Weibull distribution, that is, $\mathbb{P}(T_i^{(j)}\geq t)=\exp(-(\gamma t)^{a})$. We consider three values of $a$ ($0.5$, $1$, and $1.5$) and set the value of $\gamma$ to obtain the same expected interruption time of 2 days in the main case, or 5 days for comparison, for all values of $a$. The goal of this choice is to compare situations where the average interruption time is the same, but where the hazard rate of the distribution is different: decreasing for $a=0.5$, constant for $a=1$, which corresponds to the particular case of an exponential distribution, and increasing for $a=1.5$.

\item the time before the reaction of the policyholder's information system, that is, the start of a backup plan or a recovery procedure, $U_i$. As explained in Section \ref{sec:timeline}, there is a probability that the policyholder does not have a backup plan or that this backup plan does not start before the resolution of the cloud interruption. We set this probability to $\mathbb{P}(U_i=\infty)=0.2$, which is close to the average value given by \cite{lloyds}. Still following the findings of this report, we set $E[U_i|U_i<\infty]=2.1$ and assume that $U_i|U_i<\infty$ follows the same Weibull distribution as $T_i^{(j)}$, with $\gamma$ chosen such that $E[U_i|U_i<\infty]=2.1 \; \forall \; a$.

\item the distribution of $\frak{a}_{i,j}$, which reflects heterogeneity between policyholders and between service providers in terms of the cost of damages due to cloud interruption. We consider two cases: a deterministic base case where, for each service provider $j$, $\frak{a}_{i,j}$ has a constant value for all policyholders. These constant values are given in Table \ref{tab:critic}. We also consider a case where $\frak{a}_{i,j}$ is a random variable, as specified in Section \ref{sec:loss}. In this case, we choose a normal distribution for $\frak{a}_{i,j}$ with standard deviation in the set $\{0.1, 0.75, 1.4\}$ and averages for each service provider given in Table \ref{tab:critic}.

\item the efficiency of the backup plan $1-\frak{b}_i$. The distribution of $\frak{b}_i$ is also inspired by the \cite{lloyds} report and is taken as a uniform distribution on $[0.3, 0.5]$.

\item the time $V_i$ before the resumption of full capacity after the cloud outage is resolved. The distribution of $V_i$ is assumed to be identical to that of $T_i^{(j)}$.
\end{enumerate}

Table \ref{tab:random_vars} summarizes the distributions and parameters used to build the dataset of the study. In addition to the \cite{lloyds} report, the distributions and parameter choices were also inspired by official reports, including AMRAE's LUCY report (see \cite{amrae2024lucy}) and AON's Global 2025 Cyber Risk report (see \cite{aon2025global}). These choices were also made to ensure consistency with market trends.

\begin{table}[h!]
\centering
\begin{tabular}{|c|c|c|}
\hline
\textbf{Random Variable} & \textbf{Distribution} & \textbf{Parameters} \\
\hline
$T_i^{(j)}$ & \multirow{3}{*}{Weibull: $S(x; a, \gamma) = e^{-(\gamma x)^a}$} & \multirow{3}{6cm}{$a \in \{0.5, 1, 1.5\}$, $\gamma$ is such that mean durations are constant for all $a$} \\
\cline{1-1} 
$V_i$ & & \\
\cline{1-1}
$U_i \mid U_i < +\infty$ & &  \\
\hline
$\mathbb{P}(U_i = +\infty)$ & Bernoulli & $p = 0.2$ \\
\hline
$\frak{b}_i$ & Uniform on $[u_1, u_2]$ & $u_1 = 0.3$, $u_2 = 0.5$ \\
\hline
\multirow{2}{*}{$\frak{a}_{i,j}$} & Deterministic & See Table \ref{tab:critic} for values \\
\cline{2-3}
& Normal: $\mathcal{N}(\nu,\sigma^2)$ & $\sigma \in \{0.1, 0.75, 1.4\}$, see Table \ref{tab:critic} for $\nu$ \\
\hline
\end{tabular}
\caption{Summary of random variables, distributions, and parameters}
\label{tab:random_vars}
\end{table}

To simplify the analysis of our results, we consider the same distribution for each cloud provider in each setting. In practice, this may not be the case: some cloud providers may operate critical functions of the information systems, while less important tasks may be left to others. This is where a careful analysis of the way these solutions are used among policyholders could provide valuable additional information. Some market analyses, such as \cite{Flexera}, could be used since they provide statistics on the types of workloads performed by the different cloud solutions, although these statistics may be biased compared to the specific population of the considered portfolio.

\subsection{Descriptive statistics and probability of interruption}
\label{sec:proba}

\subsubsection{Descriptive statistics}
\label{sec:desc_stats}
In our numerical illustration, we assume that there are $d = 5$ cloud service providers, named 1 to 5, and $n = 5\,000$ policyholders. The small number of cloud providers on the market (5) is inspired by the report \cite{Flexera}, which shows that 5 companies (namely, in alphabetical order, AWS, Azure, Google Cloud, IBM Cloud, and Oracle Cloud) shared more than 95\% of the market in 2022. In this numerical application, the policyholders we focus on are small or medium-sized enterprises with relatively small turnovers and who are frequently affected by light to moderately severe cloud interruptions. This justifies the choices of the values of daily average losses (see Table \ref{tab:critic}) and probabilities of interruption (see Table \ref{tab:prob_fail}). As explained in Section \ref{sec:distr}, we consider that the tasks performed by the different cloud providers do not have the same level of criticality. The mean daily losses of each service provider given in Table \ref{tab:critic} reflect this difference to a certain extent.

\begin{table}[!h]
    \centering
    \begin{tabular}{|c | c|}
    \hline
    Cloud provider $j$ & $E[\frak{a}_{i,j}]$ in $10^3$ euros \\
    \hline
        1 & 2.7 \\
        2 & 3.1 \\
        3 & 2.4 \\
        4 & 2.6 \\
        5 & 2.3 \\
        \hline
    \end{tabular}
    \caption{Mean daily losses in $10^3$ euros for the cloud providers considered in the study. Recall that in the deterministic case, the $\frak{a}_{i,j}$ are equal to the values of this table for all policyholders $i$. In the stochastic case, these values are the means of the distribution used to generate different values of $\frak{a}_{i,j}$ for policyholders.}
    \label{tab:critic}
\end{table}

Summary statistics of the final dataset used in the next sections are presented in Table \ref{tab:desc}. We observe that for all values of $a$, the presence of a backup plan reduces the average loss of policyholders. For instance, when $a=1$, which corresponds to an exponential distribution for durations, the average loss for policyholders with a backup plan is around 5\,300 euros, compared to about 6\,200 euros for policyholders with no backup plan. Regarding the evolution of losses with the value of $a$, we observe that although increasing the value of $a$ reduces the heaviness of the tail of the Weibull distribution of the interruption duration, there is an increase in the expected financial loss (5\,770 euros for $a=0.5$, 5\,860 euros for $a=1$, and 6\,180 euros for $a=1.5$). To explain this, recall that the mean interruption duration is kept fixed across the different Weibull shapes. Changing the shape parameter therefore redistributes probability mass across short, medium, and long interruption durations. For small shape parameters, the distribution combines many very short interruptions with a few very long ones. Since very short interruptions generate little loss and very long interruptions may be partly mitigated by the activation of a backup plan, the resulting average loss is lower. Conversely, larger shape parameters produce lighter tails but concentrate more probability mass around the mean duration\footnote{The densities of interruption durations are given in Figure \ref{fig:loss_densities} in the appendix.}, increasing the frequency of interruptions long enough to generate substantial business interruption before or around backup activation. Hence, the expected loss increases even though the interruption-duration tail becomes lighter.

\begin{table}[h!]
    \centering
    \begin{tabular}{|c|c|c|c|c|}
        \hline
       \multicolumn{5}{|c|}{\textbf{$a = 0.5$}}\\
       \hline
       Variable  & Mean & Minimum & Maximum & Standard deviation  \\
       \hline
       $Y$ (in $10^3$ euros)  & 5.77 & 0.15 & 43.23 & 4.54 \\
       $T$ (in days) & 1.92 & 0.01 & 13.76 & 2.03 \\
       \hline
       $Y,\; \text{Backup}$ & 5.39 & 0.25 & 26.28 & 3.94 \\
       $T,\; \text{Backup}$ & 2.85 & 0.11 & 13.76 & 2.38   \\
       \hline
       $Y,\; \text{No backup}$ & 6.02 & 0.15 & 43.23 & 4.89\\
       $T,\; \text{No backup}$ & 1.30 & 0.01 & 10.10 & 1.47 \\
        \hline
        \multicolumn{5}{|c|}{\textbf{$a = 1.0$}}\\
       \hline
       Variable  & Mean & Minimum & Maximum & Standard deviation  \\
       \hline
       $Y$ (in $10^3$ euros)  & 5.86 & 0.12 & 49.41 & 4.62 \\
       $T$ (in days) & 1.91 & 0.01 & 15.15 & 1.99 \\
       \hline
       $Y,\; \text{Backup}$ & 5.31 & 0.27 & 23.75 & 3.55 \\
       $T,\; \text{Backup}$ & 2.76 & 0.03 & 14.50 & 2.19   \\
       \hline
       $Y,\; \text{No backup}$ & 6.22 & 0.12 & 49.41 & 5.15\\
       $T,\; \text{No backup}$ & 1.38 & 0.01 & 15.15 & 1.64 \\
        \hline
        \multicolumn{5}{|c|}{\textbf{$a = 1.5$}}\\
       \hline
       Variable  & Mean & Minimum & Maximum & Standard deviation  \\
       \hline
       $Y$ (in $10^3$ euros)  & 6.18 & 0.85 & 25.32 & 3.11 \\
       $T$ (in days) & 1.94 & 0.02 & 7.71 & 1.36 \\
       \hline
       $Y,\; \text{Backup}$ & 5.84 & 0.85 & 18.44 & 2.46 \\
       $T,\; \text{Backup}$ & 2.72 & 0.18 & 7.71 & 1.37   \\
       \hline
       $Y,\; \text{No backup}$ & 6.41 & 0.88 & 25.32 & 3.45\\
       $T,\; \text{No backup}$ & 1.44 & 0.02 & 6.44 & 1.09 \\
        \hline
    \end{tabular}
    \caption{Summary statistics of the study data. The statistics are presented for $a=0.5$, $a=1$, and $a=1.5$, which correspond respectively to decreasing, constant, and increasing hazard rate Weibull distributions for interruption durations.}
    \label{tab:desc}
\end{table}

\begin{figure}[!h]
    \centering
    \begin{subfigure}[b]{0.5\textwidth}
        \centering
        \includegraphics[width=\linewidth]{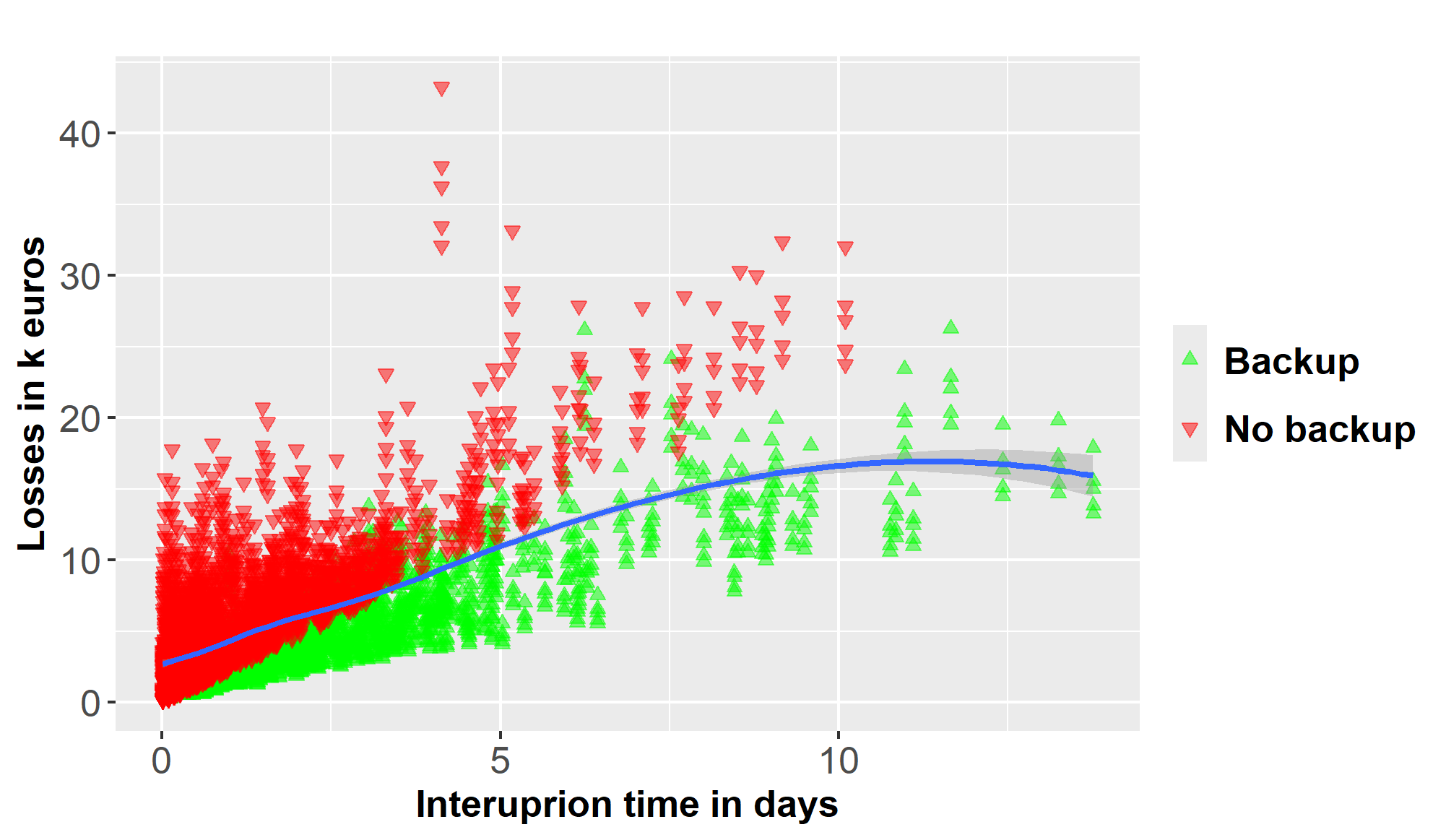}
        \caption{$a = 0.5$.}
    \end{subfigure}%
    \begin{subfigure}[b]{0.5\textwidth}
        \centering
        \includegraphics[width=\linewidth]{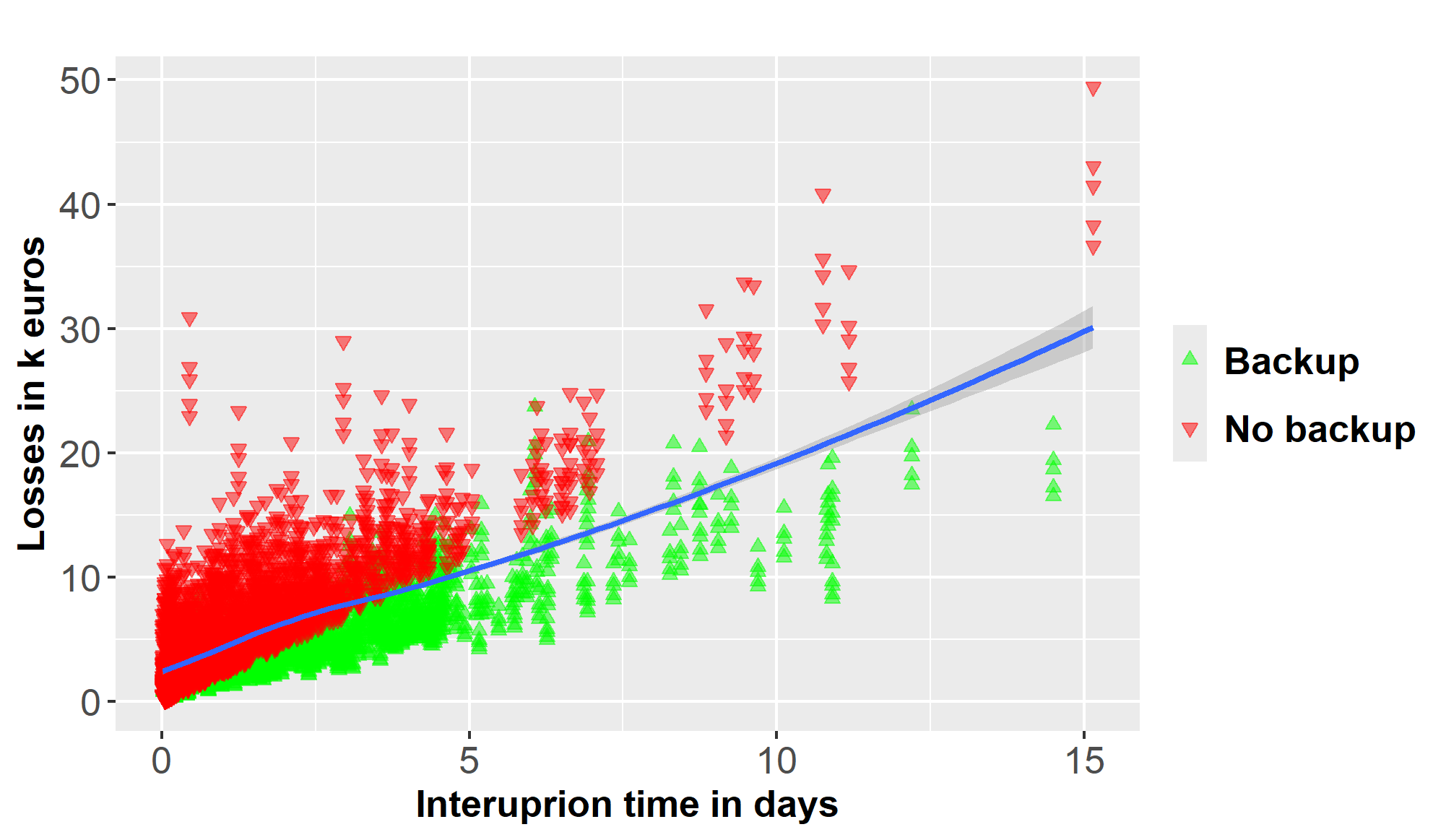}
        \caption{$a = 1.0$.}
    \end{subfigure}\\[20pt]
    \centering
    \begin{subfigure}[b]{0.5\textwidth}
        \centering
        \includegraphics[width=\linewidth]{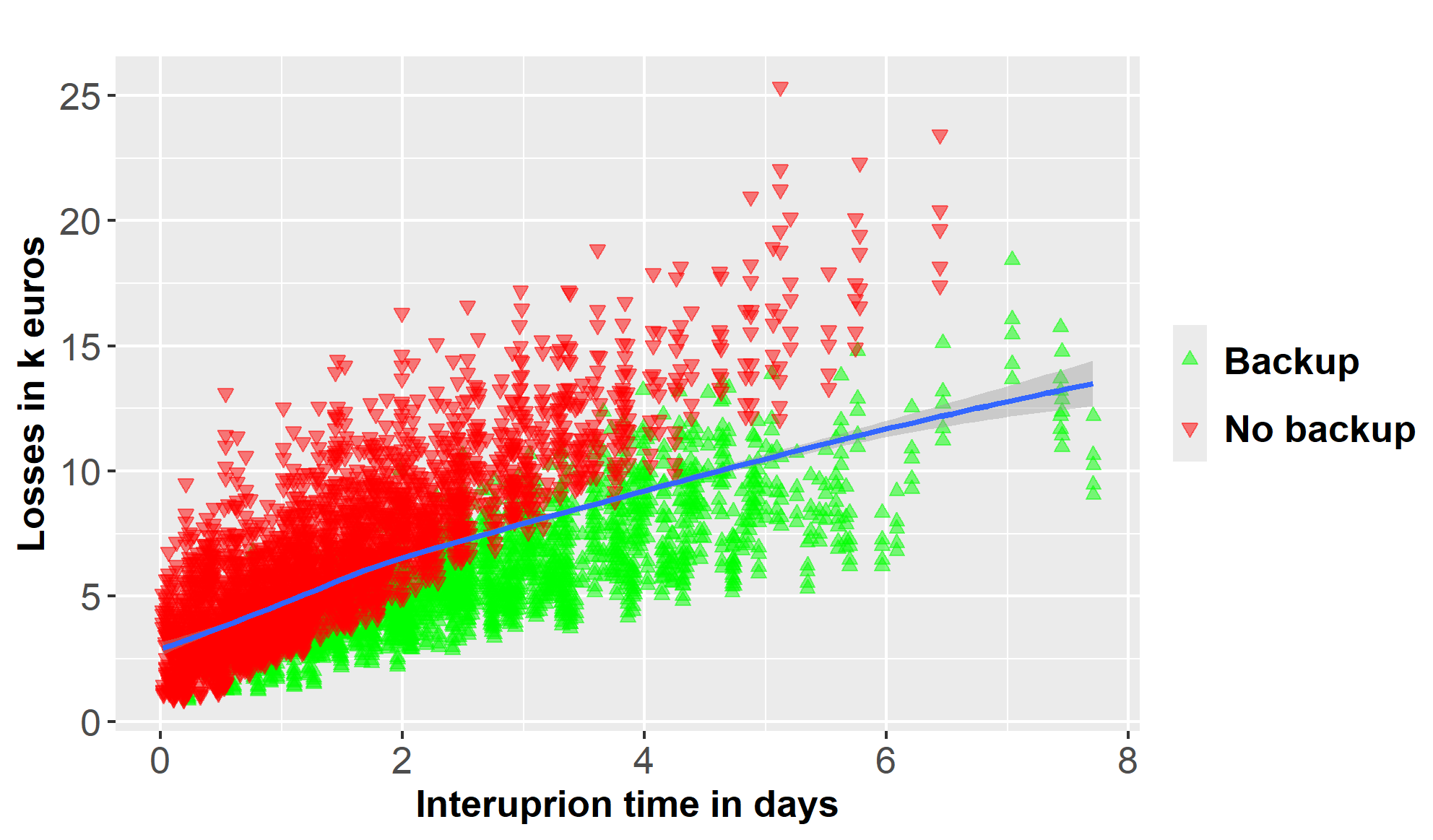}
        \caption{$a = 1.5$.}
    \end{subfigure}%
    \caption{Overview of the relationship between financial losses due to cloud interruptions and interruption durations. There seems to be a positive correlation between financial losses and interruption duration. The presence of a backup plan seems to reduce the severity of losses.}
    \label{fig:loss_vs_int}
\end{figure}

Figure \ref{fig:loss_vs_int} shows plots of financial losses as a function of interruption durations for various values of $a$. As expected, there appears to be a clear positive correlation between the duration of business interruption and the severity of the event. The relationship between the duration of business interruption and the resulting economic loss is a key factor in assessing the consequences of cyber incidents, as highlighted in several studies, including \cite{tam2023quantifying}. Naturally, the impact of a given duration of service interruption depends heavily on the sector of activity and the size of the affected company. In this simplified example, we consider a portfolio composed of policyholders with homogeneous profiles, that is, operating in the same sector and of similar size.

\subsubsection{Probability of interruption}
\label{sec:prob_interup}

Recall that in this study, we consider two regimes: a standard regime where the probability of undergoing a cloud interruption is strictly less than 1, and a catastrophe regime in which all or almost all policyholders are affected by an interruption. The annual probabilities of interruption for each cloud service provider in the standard regime are given in Table \ref{tab:prob_fail}. These probabilities provide a second distinction between service providers, the first being their different average daily interruption costs, whose aim is to quantify their level of risk and reliability. Based on the analysis made in the \cite{lloyds} report, these probabilities are chosen so that our study includes moderately risky cloud service providers (providers 1 and 2), a risky cloud service provider (provider 4), and very risky cloud service providers (providers 3 and 5). Note that our study addresses catastrophe from a frequency perspective and not from a severity perspective (see Remark \ref{rem:distribution}). Therefore, we assume a catastrophe event if all or almost all policyholders in the portfolio are affected, even if the severity of the losses is low.

\begin{table}[!h]
    \centering
    \begin{tabular}{|c|c|c|c|c|c|}
    \hline
        Cloud provider $j$ & 1 & 2 & 3 & 4 & 5  \\
        \hline
        $p_j$ & 0.36 & 0.22 & 0.70 & 0.49 & 0.81 \\
        \hline
    \end{tabular}
    \caption{Probabilities of interruption for the 5 cloud service providers of the study. Providers 1 and 2 are considered moderately risky, provider 4 is considered risky, and providers 3 and 5 are considered very risky.}
    \label{tab:prob_fail}
\end{table}

The dependence between cloud providers is represented by the covariance matrix $\Sigma$. Its entries depend on the probability of interruption, the cost of interruption for the various cloud providers, and a term describing the association between these providers. Indeed, if $\Delta_{j,k}=E[\delta_{i,j}\delta_{i,k}]$ and $l_j=E[L_i^{(j)}]$, the coefficient $\frak{s}_{j,k}$ (line $j$, column $k$) of $\Sigma$ is:

$$\frak{s}_{j,k}=l_jl_k(\Delta_{j,k}-p_jp_k).$$ 

The matrix $\Delta$, whose coefficients are given in Table \ref{tab:rhojk}, is introduced as a way to account for dependence between incident indicators associated with different cloud providers. This dependence is motivated by the fact that cyber risks are rarely independent across policyholders or across technological providers (see \cite{bohme2006models}). Common software components, shared vulnerabilities, common authentication systems, network dependencies, third-party services, and operational interconnections could create simultaneous or correlated failures. This is consistent with the cyber-insurance literature, which identifies correlation and accumulation as central features of cyber risk. It is also consistent with regulatory and actuarial analyses of cloud risk, which emphasize that reliance on a limited number of cloud service providers can create dependencies and simultaneous business interruption across firms (see \cite{iais2020cyber}). The covariance matrix $\Sigma$, obtained from the coefficients of the matrix $\Delta$ in Table \ref{tab:rhojk}, was verified to be positive definite.

\begin{table}[!h]
    \centering
    \begin{tabular}{|c|c|c|c|c|c|}
    \hline
       \diagbox{j}{k}  & 1 & 2 & 3 & 4 & 5 \\
       \hline
        1 &  & 0.16 & 0.27 & 0.23 & 0.31 \\
        2 &  &  & 0.20 & 0.14 & 0.19 \\
        3 &  &  &  & 0.36 & 0.58 \\
        4 &  &  &  &  & 0.43 \\
        5 &  &  &  &  &  \\
        \hline
    \end{tabular}
    \caption{Matrix $\Delta$, where the coefficient in line $j$ and column $k$ is $\Delta_{j,k}$. Since the matrix is symmetric, we only show the upper part for readability. We also do not show the diagonal, since $\Delta_{j,j}=p_j$ can be deduced from Table \ref{tab:prob_fail}.}
    \label{tab:rhojk}
\end{table}

Although the choice of the coefficients of $\Delta$ is inspired by the \cite{lloyds} report, the matrix should not be interpreted as an empirically estimated matrix. Rather, it is a plausible dependence matrix designed to represent positive or weak association between provider-related incidents in a context where historical cloud-outage data are scarce. Also, recall that the purpose of this illustration is not to estimate the true joint failure probabilities of specific providers. For this type of estimation, see \cite{eling2018copula}, \cite{wang2021fast}, \cite{bohme2006models}, and \cite{herath2007cyber}. The objective is instead to study how the proposed optimization framework leads to enhanced mutualization on a plausible cloud interruption insurance portfolio, including in the presence of accumulation. The choice of positive or weak correlation between interruptions of service providers is realistic, since according to \cite{bohme2006models}, cyber incidents are either positively correlated across firms due to common risk factors or not correlated at all. Negative correlation is rare and difficult to justify in practice.

\subsection{Optimal portfolio}
\label{sec:resultats}

The aim of this work is to improve the level of diversification in a cloud insurance portfolio. This is done by identifying the optimal configuration of the portfolio in terms of the proportions of policyholders' turnover exposed to each cloud provider on the market. To obtain a portfolio structure that guarantees an acceptable level of diversification and mutualization, the optimization problem introduced in Section \ref{sec_optim_div} is solved numerically. In this section, we study the results of this numerical optimization on the dataset described in Section \ref{sec:desc_stats}. The optimization problem combines a criterion controlling the volatility of standard-regime losses and a criterion controlling the severity of losses affecting all or a substantial part of the portfolio. The resulting optimal structure is therefore intended to ensure diversification both in the standard regime and in a stressed regime where a cloud interruption affects a large share of the portfolio.

For the rest of the analysis in this section, a baseline value for $\lambda$ is set using an average capital amount of $\frak{c}=25\,000$ euros (see Section \ref{sec_lambda}). This choice is aligned with the fact that our illustration focuses on small and medium-sized companies whose activities are frequently affected by minor cloud interruptions. However, in Section \ref{sec:imp_lambda}, we study the impact of variations in $\lambda$ on the optimal portfolio. Most of the results in this section are reported in terms of the expected loss (or profitability) constraint $\rho$ defined in Section \ref{sec:profit}. For the quantile (VaR) and Conditional Tail Expectation risk measures, we consider $\alpha_j=\alpha$, with a baseline value of $0.95$ for all $j \in \{1,\cdots,5\}$, and we also study the impact of variations in $\alpha$ on our final results in Section \ref{sec:imp_alpha}. As discussed in Section \ref{sec:distr}, we explore in detail the case where the average service unavailability is 2 days and present a brief overview of the case where this average duration is 5 days in Section \ref{sec:5_day_int}.

\subsubsection{General optimization solution}

\begin{figure}[!h]
    \centering
    \begin{subfigure}[b]{0.33\textwidth}
        \centering
        \includegraphics[width=\linewidth]{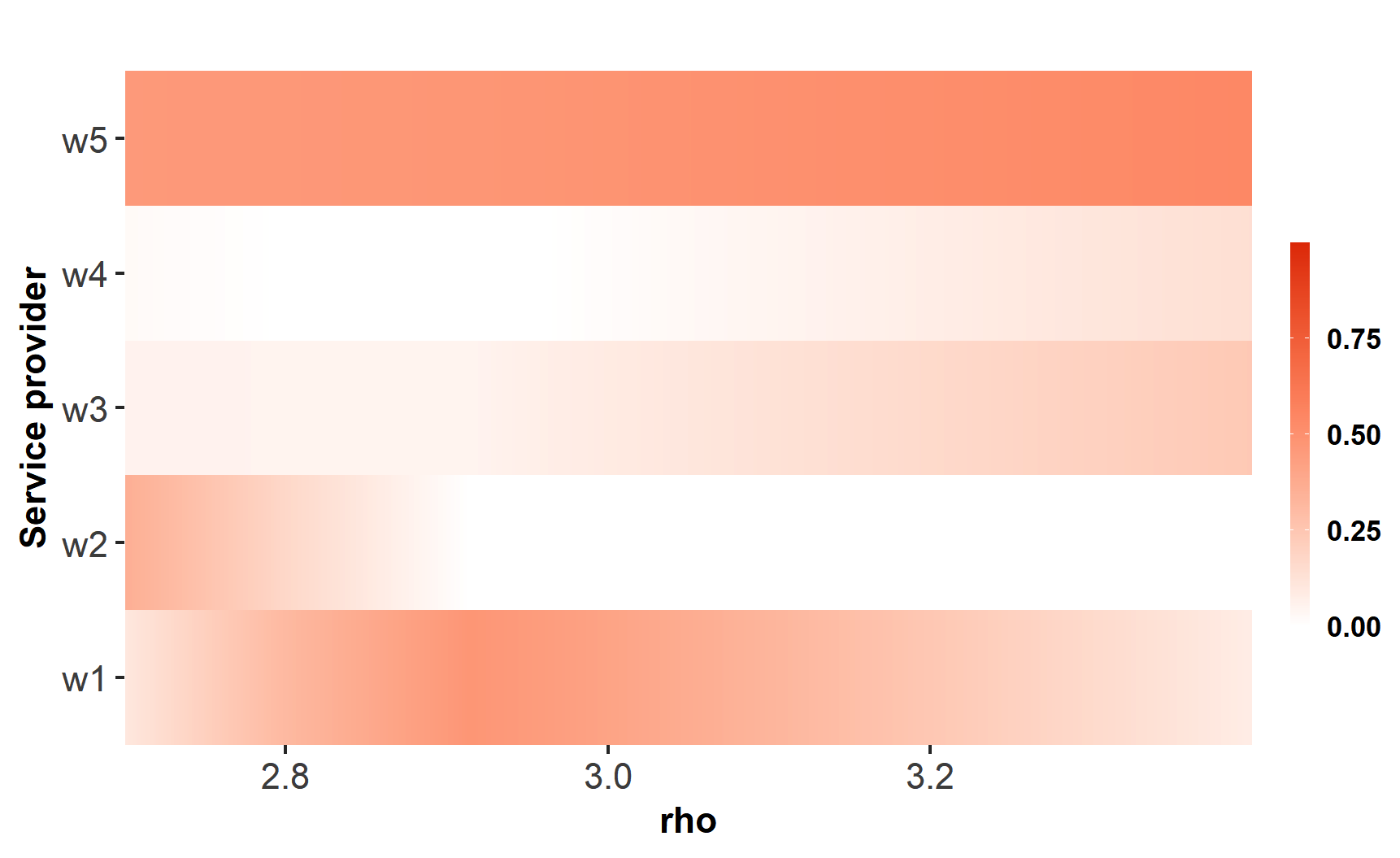}
        \caption{$a=0.5$, mean.}
    \end{subfigure}%
    \begin{subfigure}[b]{0.33\textwidth}
        \centering
        \includegraphics[width=\linewidth]{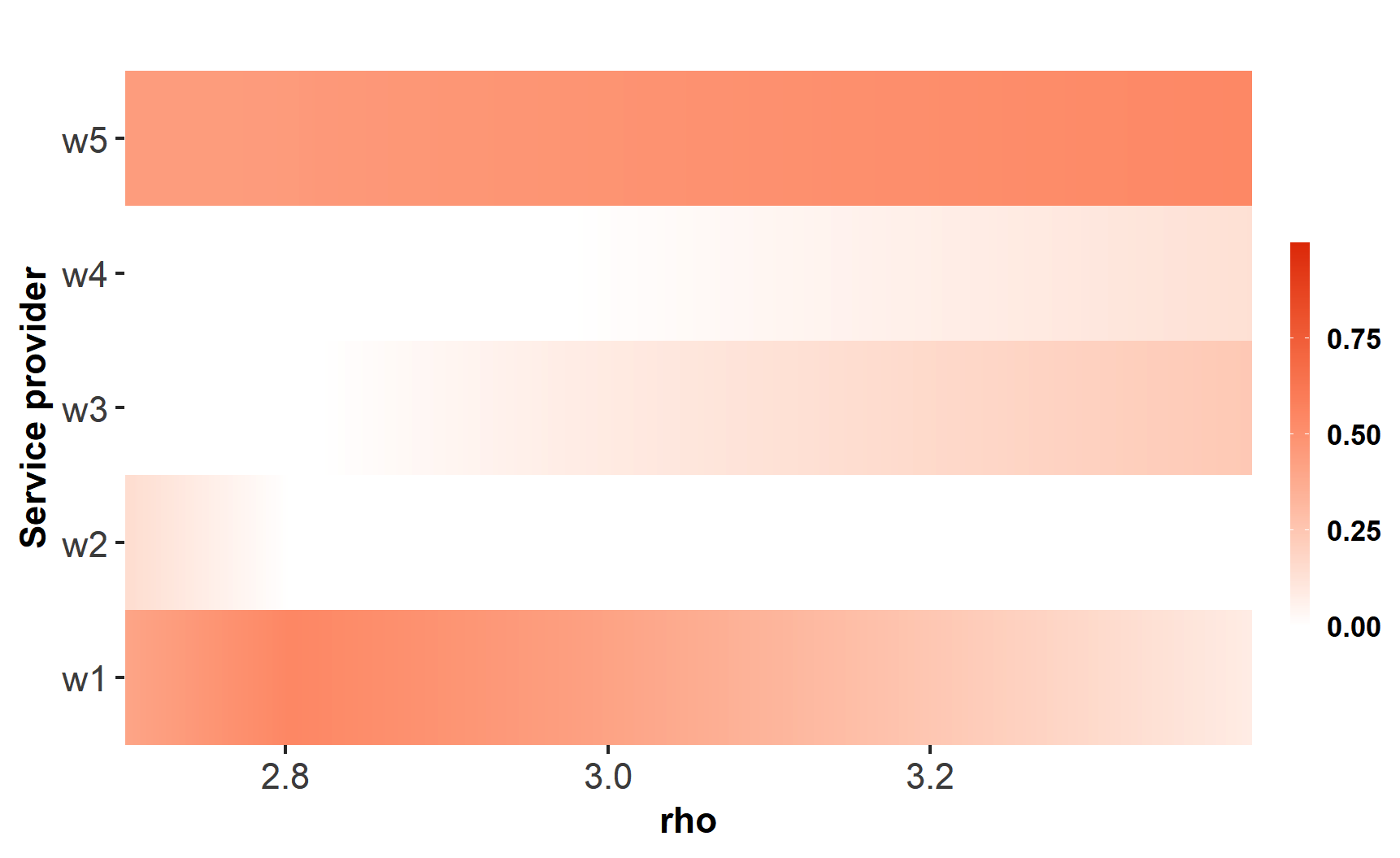}
        \caption{$a=0.5$, VaR.}
    \end{subfigure}
    \begin{subfigure}[b]{0.33\textwidth}
        \centering
        \includegraphics[width=\linewidth]{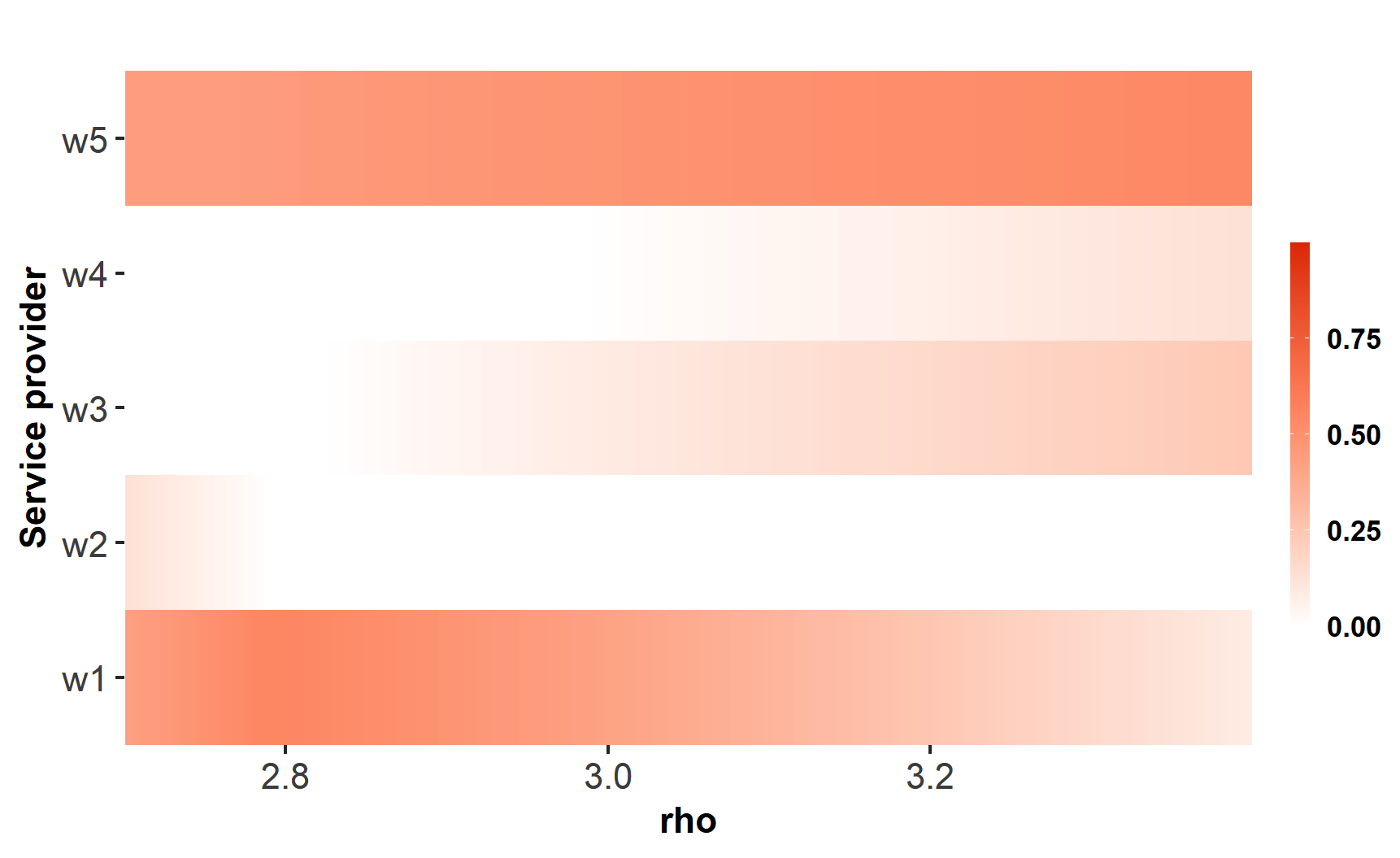}
        \caption{$a=0.5$, CTE.}
    \end{subfigure}
    \\[20pt]
    \begin{subfigure}[b]{0.33\textwidth}
        \centering
        \includegraphics[width=\linewidth]{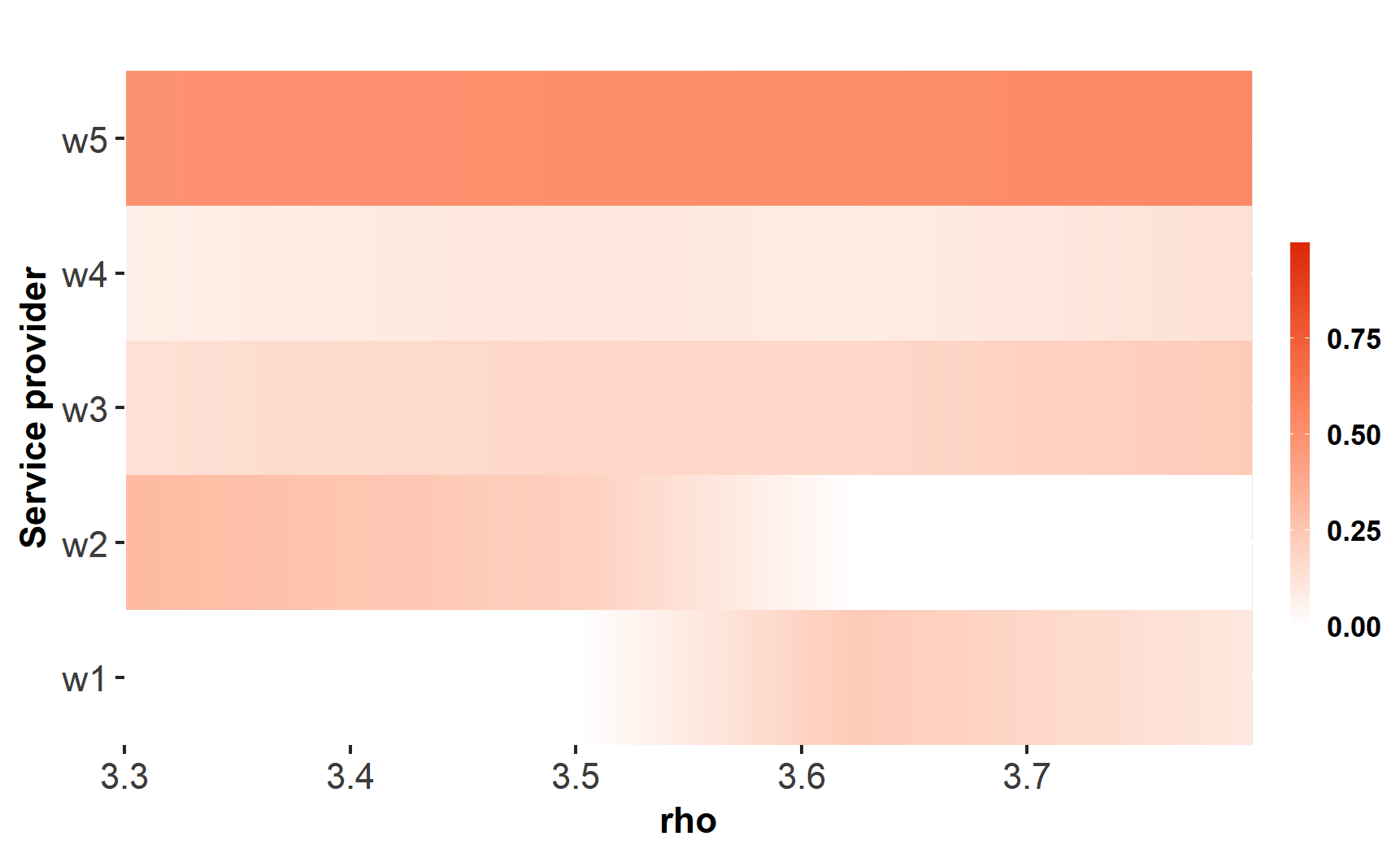}
        \caption{$a=1$, mean.}
    \end{subfigure}%
    \begin{subfigure}[b]{0.33\textwidth}
        \centering
        \includegraphics[width=\linewidth]{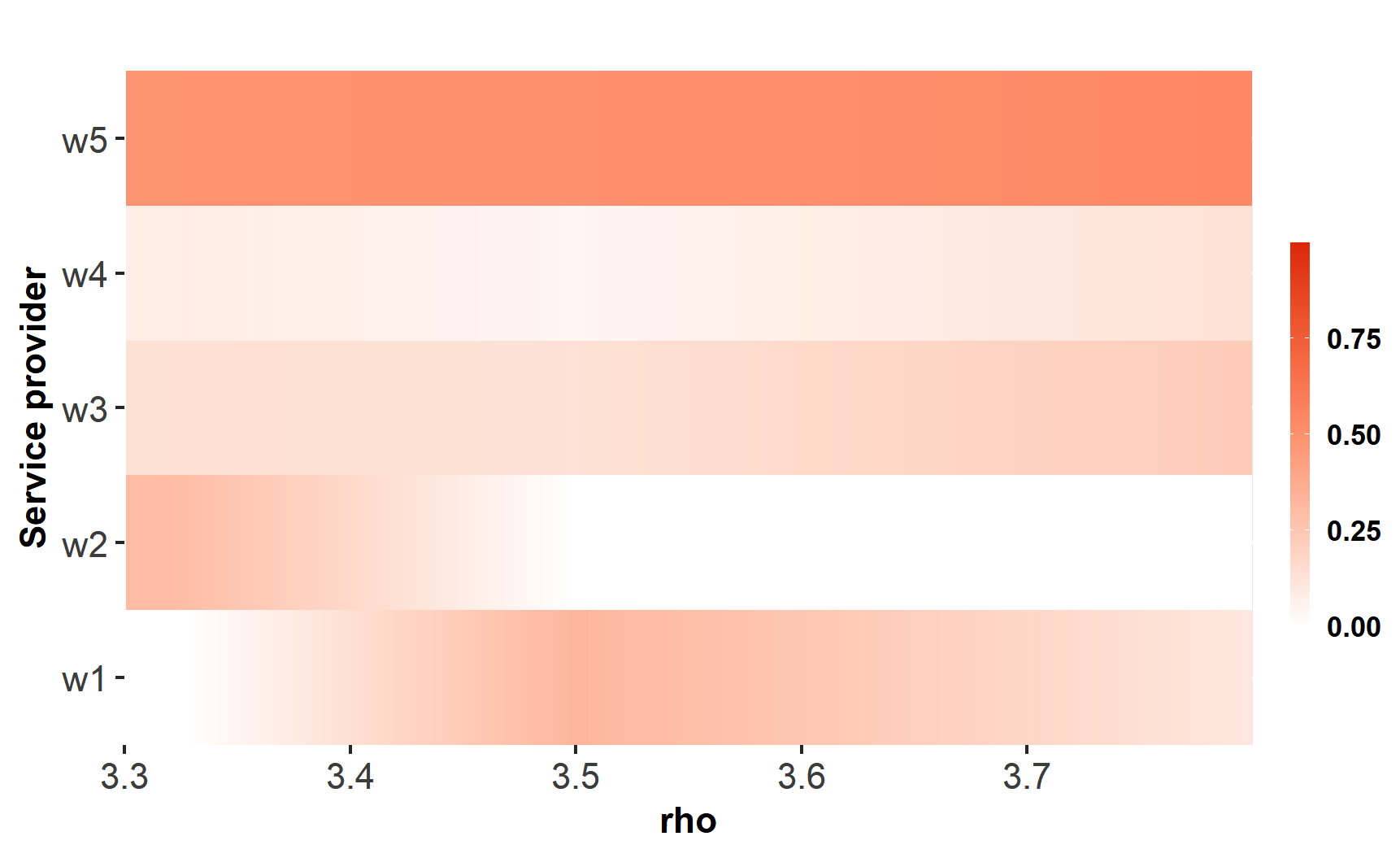}
        \caption{$a=1$, VaR.}
    \end{subfigure}
    \begin{subfigure}[b]{0.33\textwidth}
        \centering
        \includegraphics[width=\linewidth]{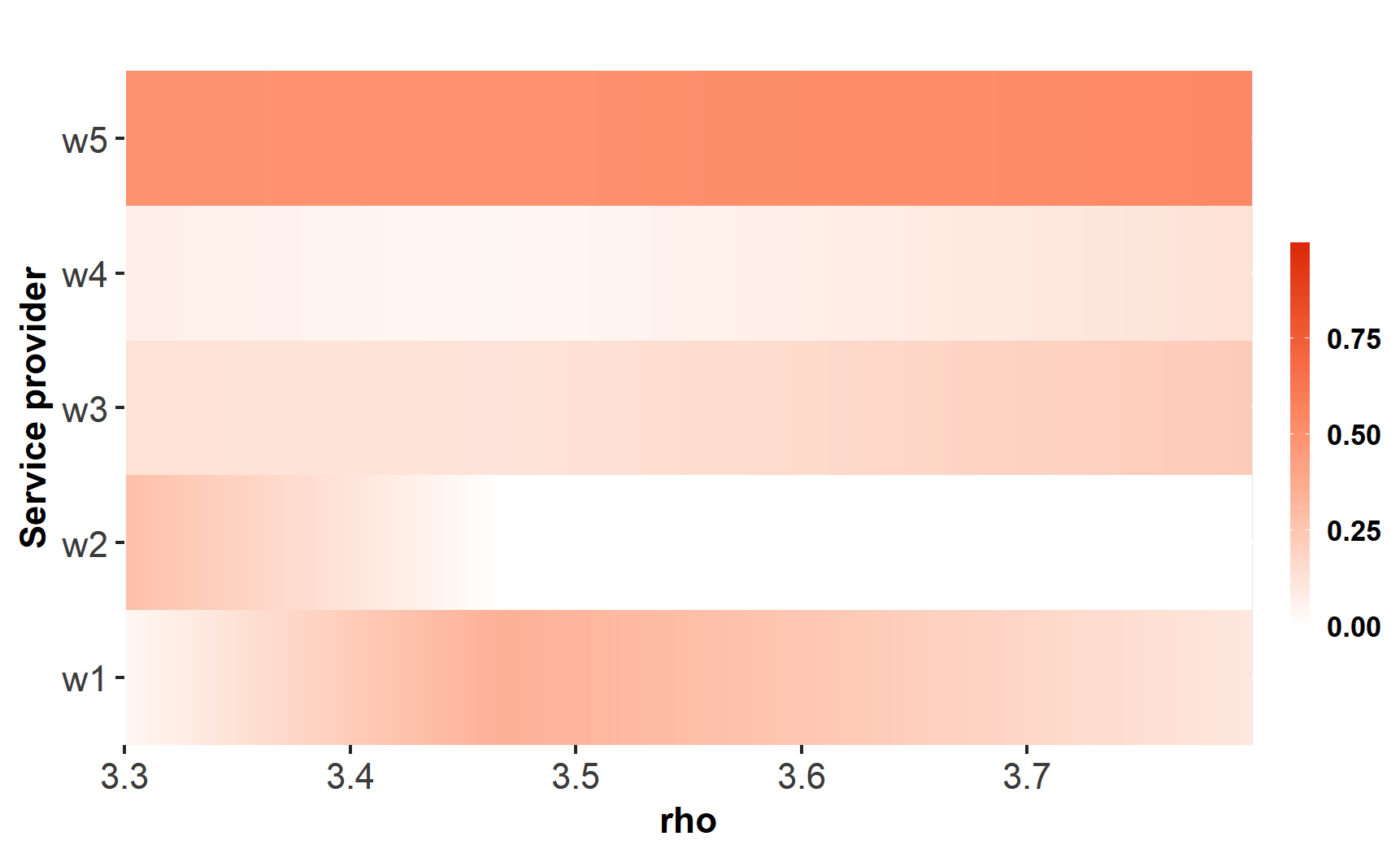}
        \caption{$a=1$, CTE.}
    \end{subfigure}
    \\[20pt]
    \begin{subfigure}[b]{0.33\textwidth}
        \centering
        \includegraphics[width=\linewidth]{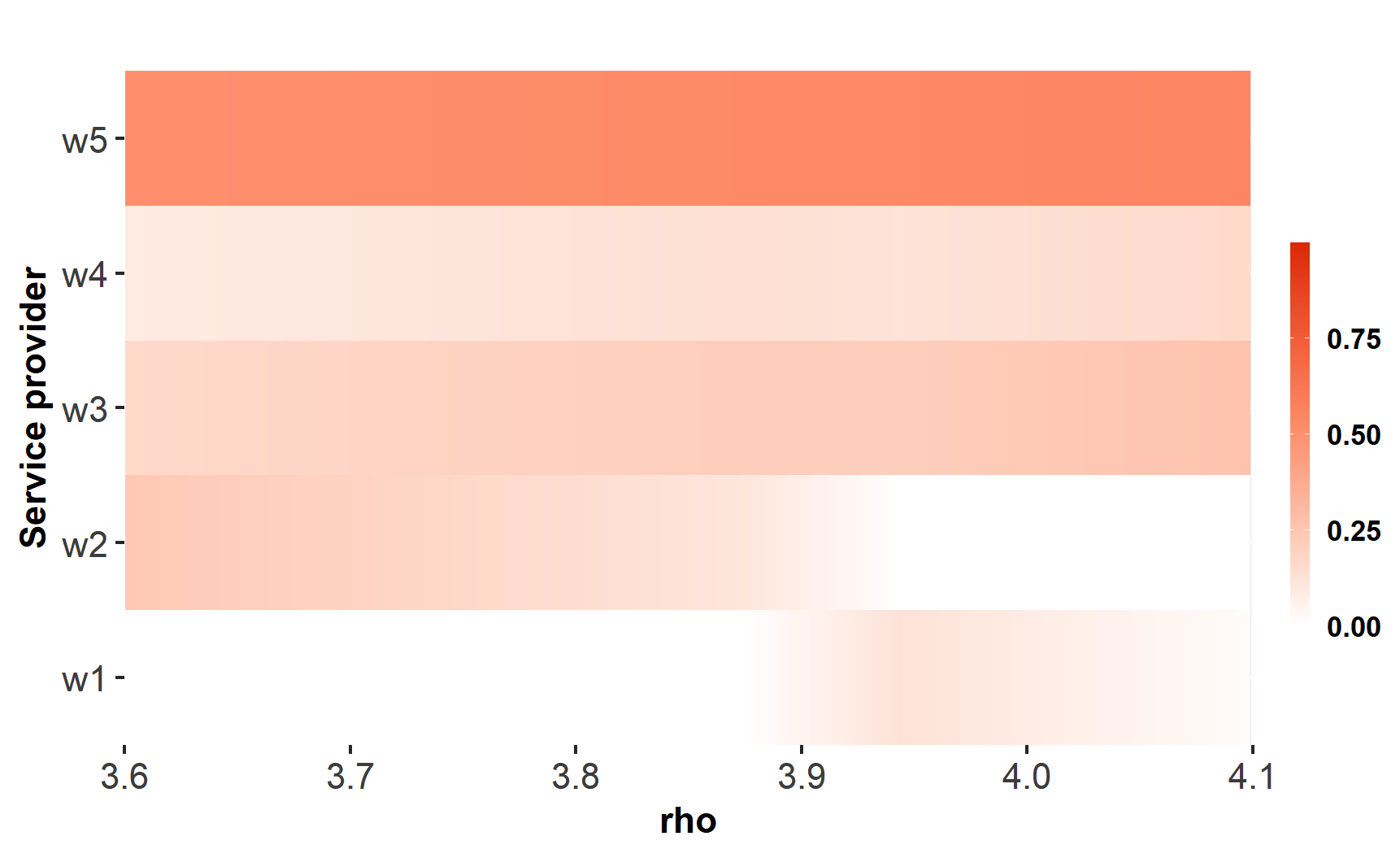}
        \caption{$a=1.5$, mean.}
    \end{subfigure}%
    \begin{subfigure}[b]{0.33\textwidth}
        \centering
        \includegraphics[width=\linewidth]{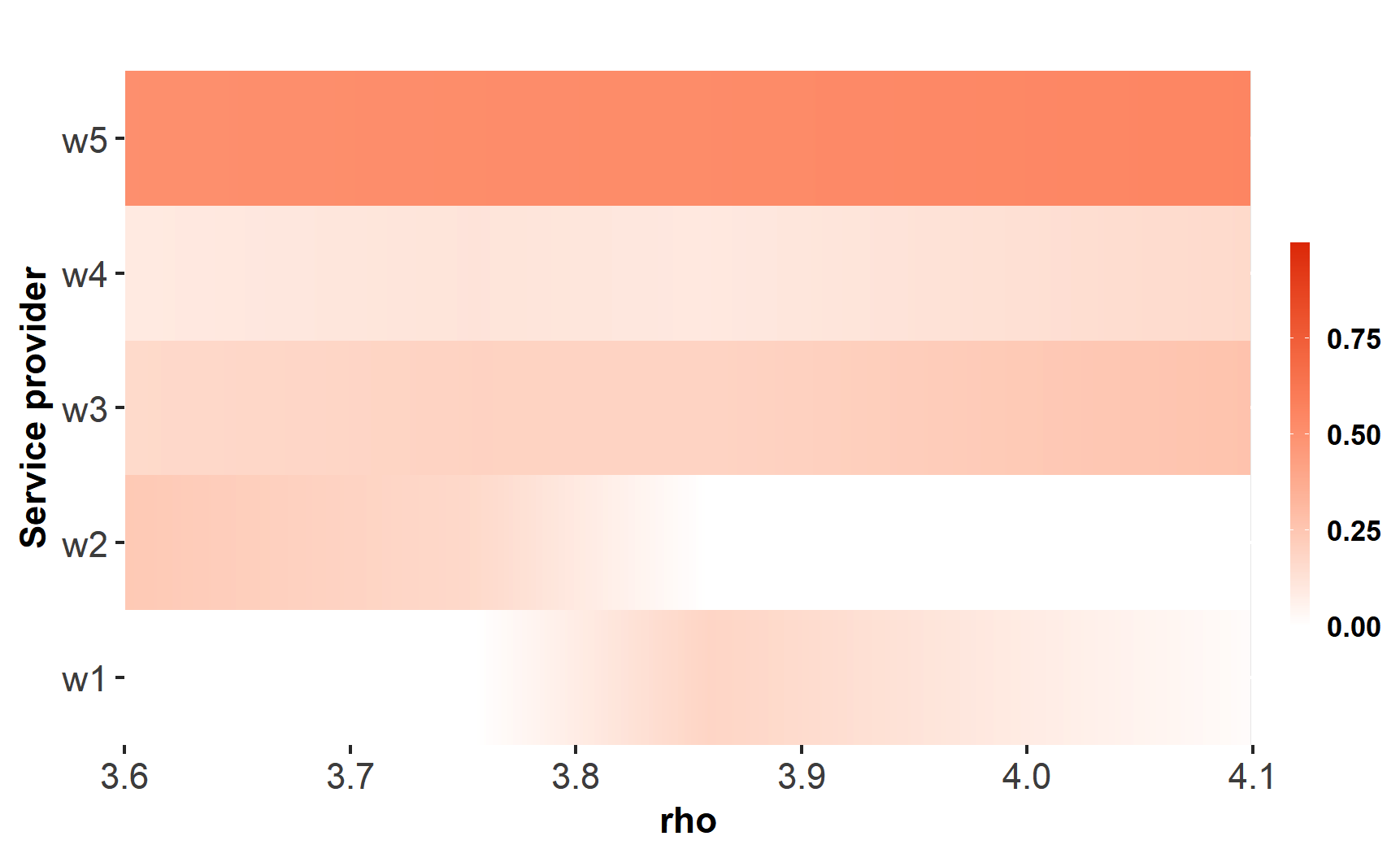}
        \caption{$a=1.5$, VaR.}
    \end{subfigure}
    \begin{subfigure}[b]{0.33\textwidth}
        \centering
        \includegraphics[width=\linewidth]{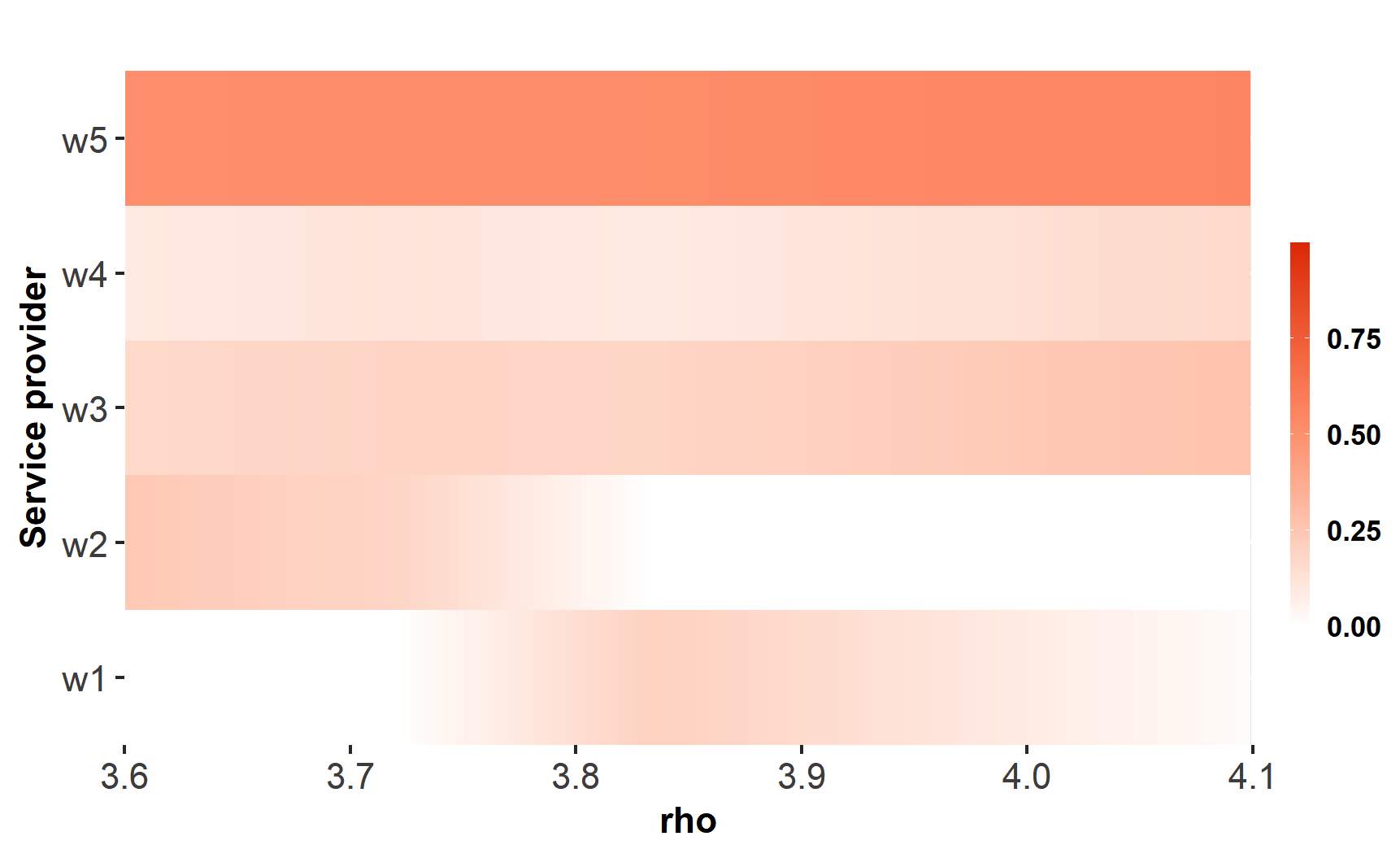}
        \caption{$a=1.5$, CTE.}
    \end{subfigure}
    \caption{Evolution of the weights of the optimal portfolio as a function of $\rho$. The average duration of the interruption is set to 2 days. The left column displays the results using the mean as the risk measure, the middle column shows the results for VaR, and the right column shows the results for Conditional Tail Expectation.}
    \label{fig_rho}
\end{figure}

The first numerical experiment studies the structure of the optimal portfolio when the expected loss, or equivalently expected profitability, constraint $\rho$ varies. The results are reported in Figure \ref{fig_rho}, for the three risk measures considered in the stressed regime: the mean loss, the Value-at-Risk, and the Conditional Tail Expectation. The analysis is conducted under the baseline assumption that the average cloud interruption duration is equal to two days and that the average daily cost of interruption $\frak{a}_{i,j}$ is deterministic. Three Weibull shape parameters are considered for the interruption duration: $a=0.5$, $a=1$, and $a=1.5$.

A first important observation is that the optimization does not lead to a uniform allocation across cloud providers. This is expected because the providers are not assumed to be equivalent. They differ in terms of probability of failure, daily loss intensity, covariance with other providers, and contribution to the stressed-regime risk measure. Therefore, the optimal allocation reflects a trade-off between the expected profitability of the standard regime and the need to limit exposure to severe cloud-interruption scenarios, where a large share of the portfolio is affected at once. In other words, the optimal portfolio is not the one that simply spreads exposure equally across all providers, but the one that balances profitability, volatility, and systemic vulnerability.

Provider $\mathbf{5}$ receives the largest weight in most configurations. This result is consistent with the construction of the numerical example. Provider $\mathbf{5}$ has the highest failure probability and therefore contributes strongly to the expected loss in the standard regime. Since the premium is assumed to be proportional to the expected loss, this also mechanically makes provider $\mathbf{5}$ attractive from a profitability perspective. The optimization therefore tends to allocate a large share of the portfolio to this provider, as long as the variance penalty and the stressed-regime penalty do not fully offset this profitability effect. This illustrates a central message of this work: a provider can be attractive in the standard insurance regime while simultaneously creating a concentration issue in the stressed regime. The role of the optimization criterion is precisely to control this tension.

The effect of $\rho$ is also informative. Increasing $\rho$ corresponds to imposing a higher expected loss, and therefore a higher expected profitability under the pricing assumption retained in our setting. As $\rho$ increases, the optimizer is progressively forced to move toward providers that generate higher expected returns. This explains why some providers become less represented, or even disappear from the optimal allocation, when the profitability constraint becomes more demanding. In particular, in the exponential case $a=1$, the weight associated with provider $\mathbf{2}$ vanishes beyond a certain value of $\rho$. This means that, under the chosen assumptions, provider $\mathbf{2}$ is no longer efficient once the insurer requires a sufficiently high profitability level: it does not contribute enough to the expected return relative to its contribution to variance and systemic risk.

The comparison across the three risk measures shows that the general allocation pattern remains broadly stable, but the exact weights vary with the definition of the stressed-regime objective. This is expected. The mean stressed loss measures the average severity of a cloud failure, while VaR and CTE focus on adverse outcomes. Consequently, the quantile and CTE criteria tend to penalize providers that contribute more strongly to extreme stressed losses. The CTE is particularly conservative because it accounts not only for the threshold exceeded in the tail, but also for the average severity beyond this threshold. This is consistent with the portfolio-risk literature, where CTE, or CVaR, is often preferred to VaR when the objective is to control the magnitude of tail losses rather than only the probability of exceeding a given threshold.

The Weibull shape parameter has a visible impact on the optimal portfolio. As discussed in Section \ref{sec:desc_stats}, when $a=0.5$, the interruption duration has a heavier tail but also a larger mass close to zero. When $a=1$, the distribution corresponds to the exponential case. When $a=1.5$, the tail becomes lighter and the duration is more concentrated around its mean. Because the loss function includes backup activation and recovery effects, the expected financial loss is not only driven by the tail of the interruption duration. This explains why changing the shape of the interruption duration modifies the optimal allocation even when the average interruption duration is kept fixed.

Overall, Figure \ref{fig_rho} confirms that the proposed optimization framework effectively produces a portfolio structure that reflects both standard-regime diversification and stressed-regime resilience. The allocation is sensitive to profitability requirements, the selected tail-risk measure, and the assumed distribution of outage durations. This sensitivity confirms that the proposed framework can be used as a stress-testing tool: by changing the assumptions on interruption duration and risk measurement, the insurer can identify which providers become sources of concentration and which allocations remain robust across scenarios.

\subsubsection{Impact of the confidence level $\alpha$ for VaR and CTE risk measures}
\label{sec:imp_alpha}

\begin{figure}[!h]
    \centering
    \begin{subfigure}[b]{0.5\textwidth}
        \centering
        \includegraphics[width=\linewidth]{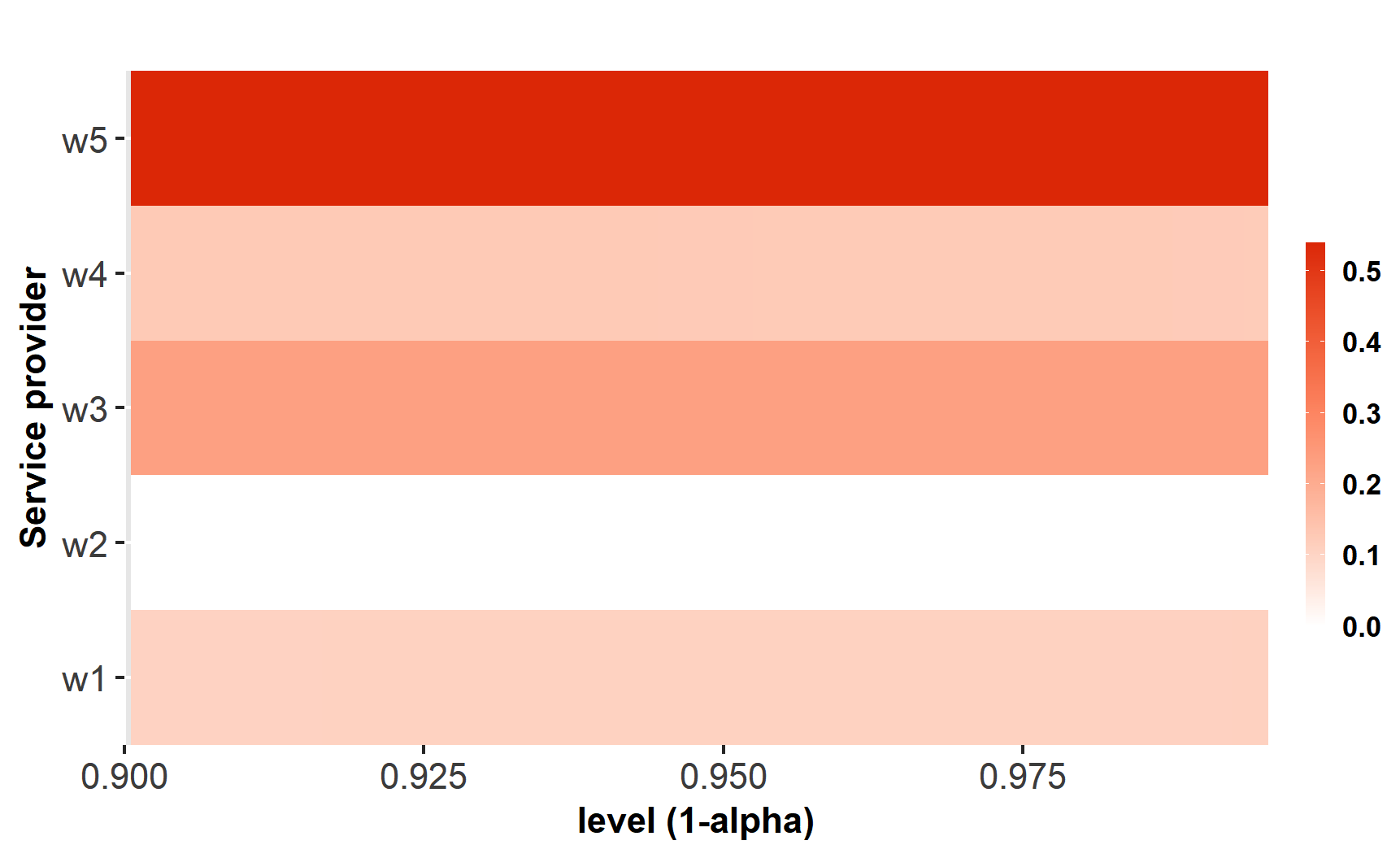}
        \caption{$a=0.5$, VaR.}
    \end{subfigure}%
    \begin{subfigure}[b]{0.5\textwidth}
        \centering
        \includegraphics[width=\linewidth]{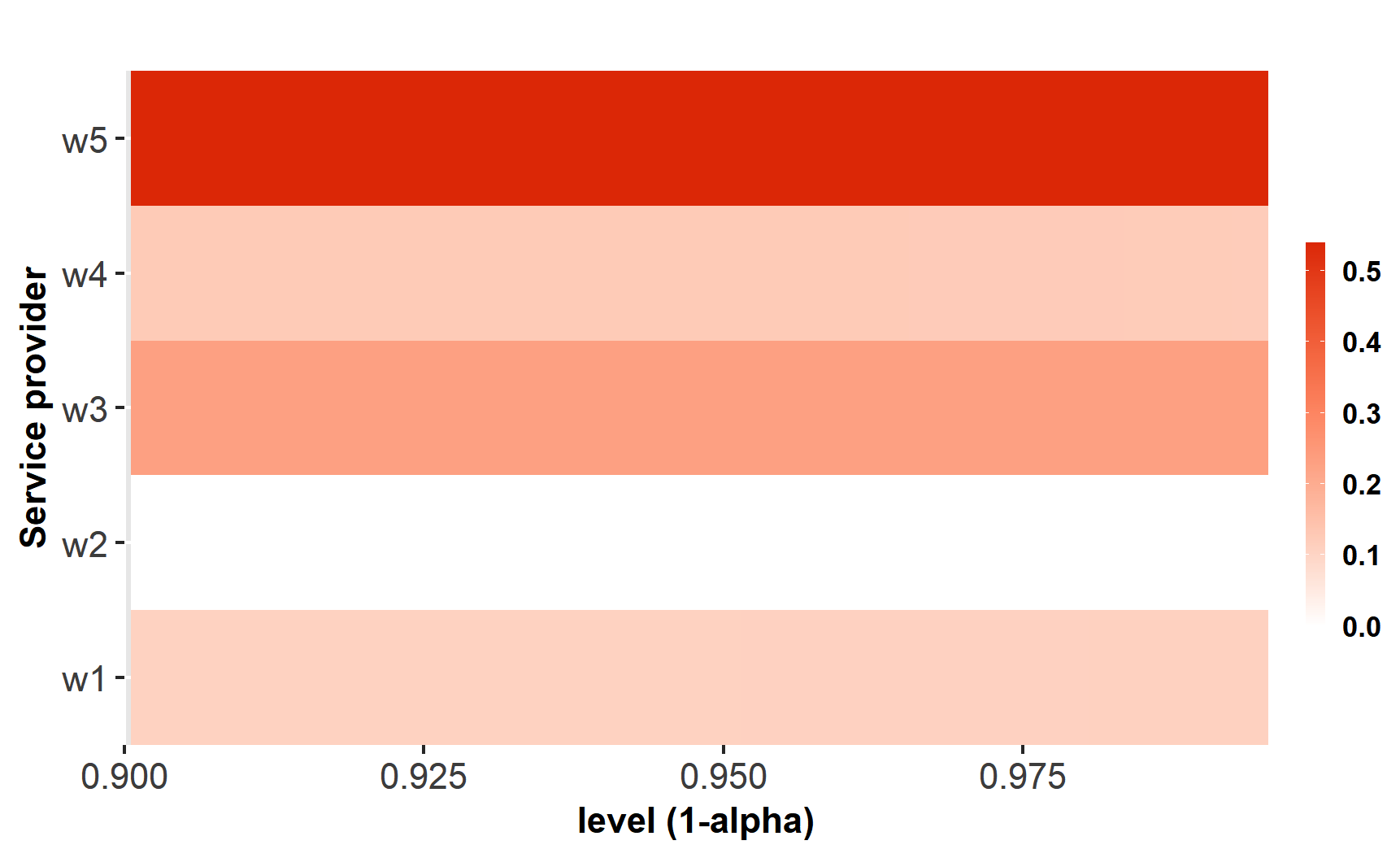}
        \caption{$a=0.5$, CTE.}
    \end{subfigure}\\[20pt]
    \begin{subfigure}[b]{0.5\textwidth}
        \centering
        \includegraphics[width=\linewidth]{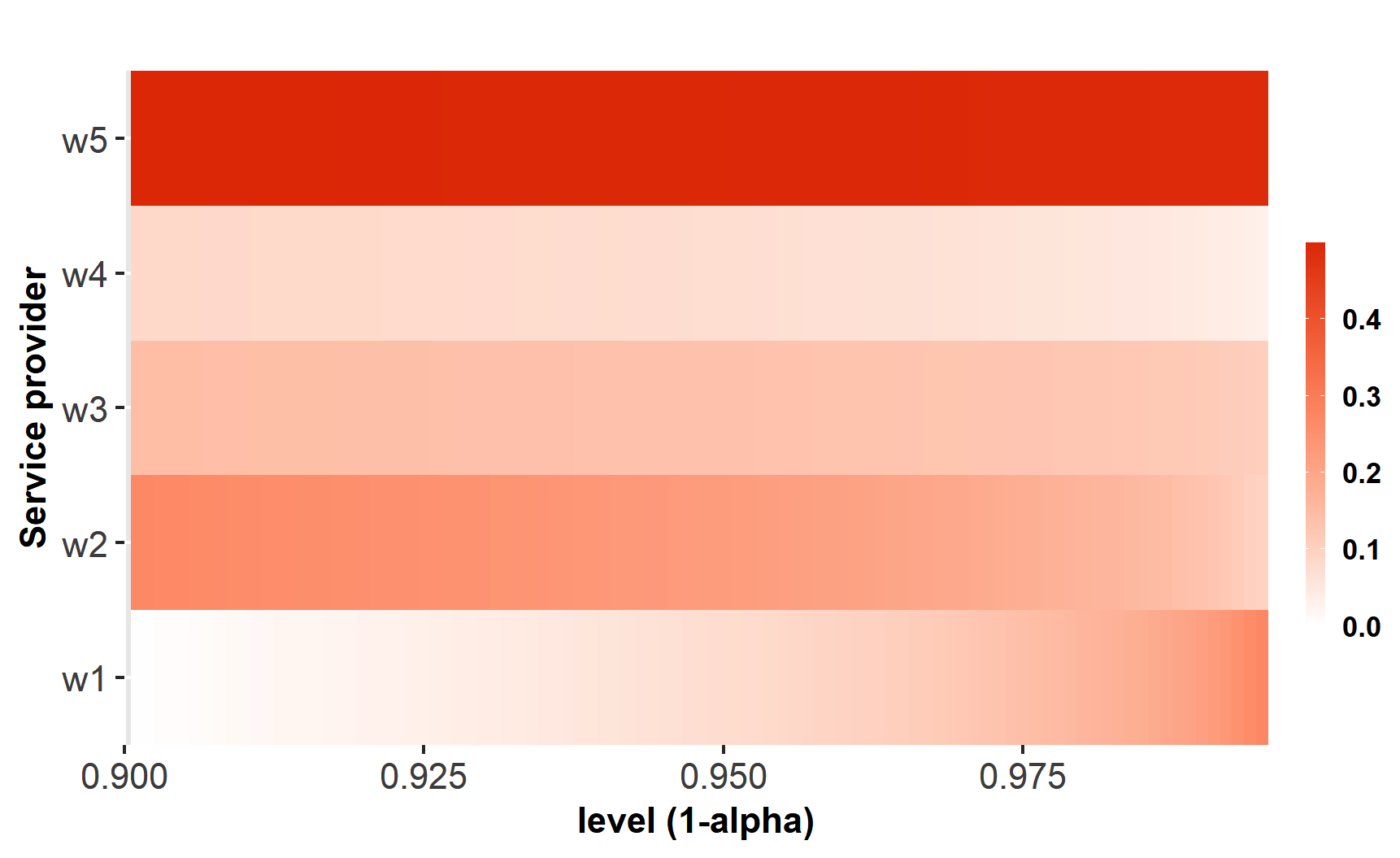}
        \caption{$a=1$, VaR.}
    \end{subfigure}%
    \begin{subfigure}[b]{0.5\textwidth}
        \centering
        \includegraphics[width=\linewidth]{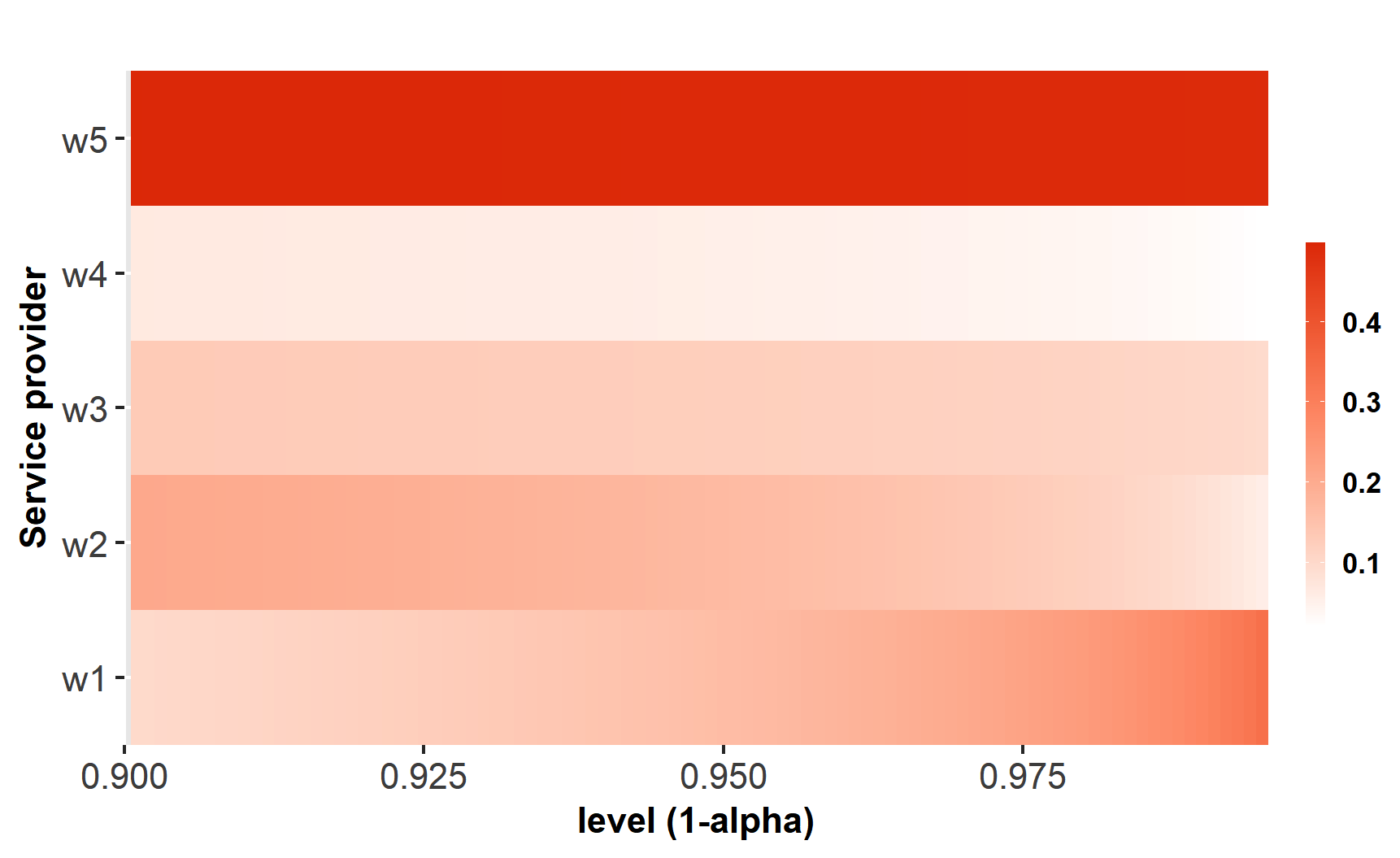}
        \caption{$a=1$, CTE.}
    \end{subfigure}\\[20pt]
    \begin{subfigure}[b]{0.5\textwidth}
        \centering
        \includegraphics[width=\linewidth]{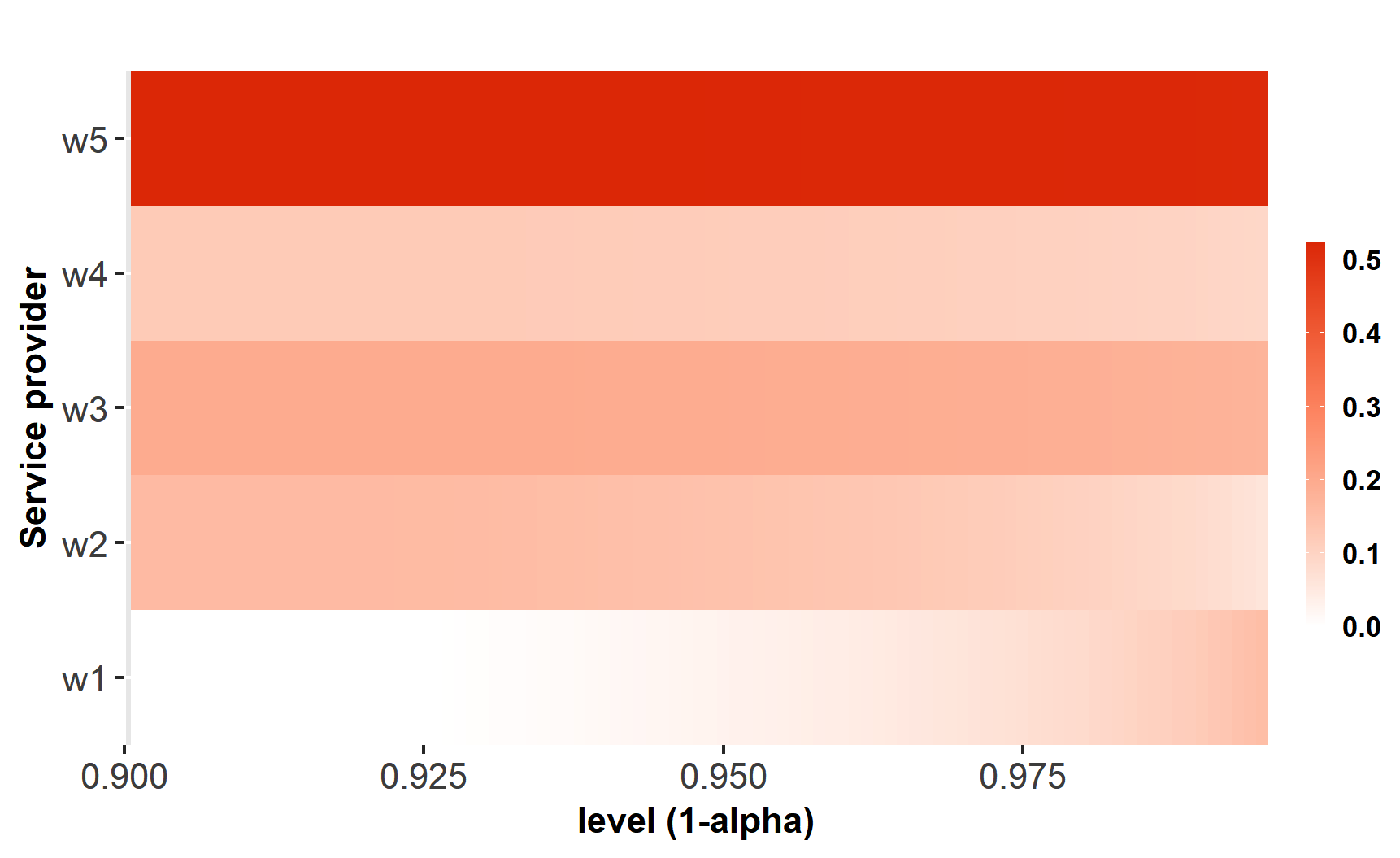}
        \caption{$a=1.5$, VaR.}
    \end{subfigure}%
    \begin{subfigure}[b]{0.5\textwidth}
        \centering
        \includegraphics[width=\linewidth]{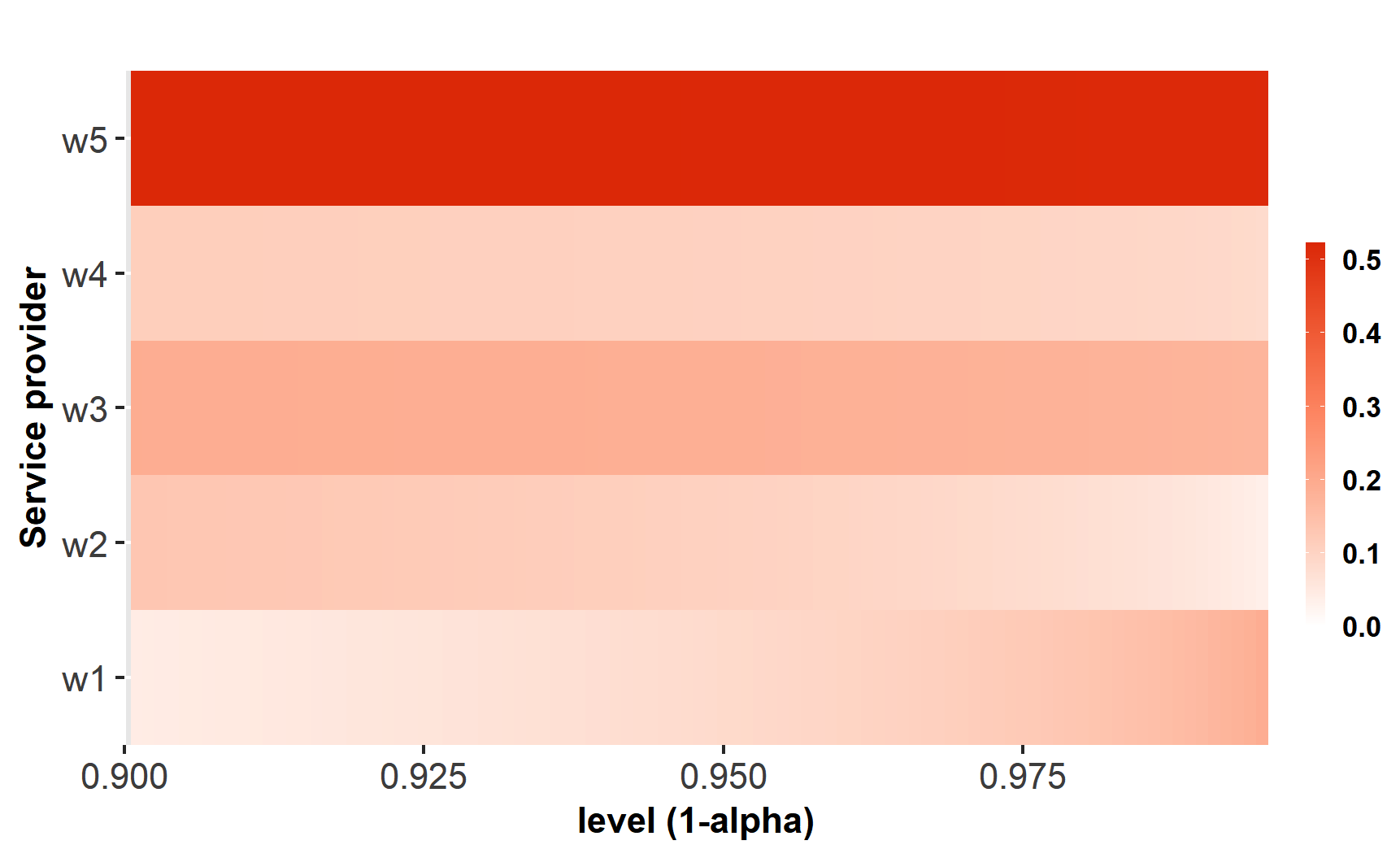}
        \caption{$a=1.5$, CTE.}
    \end{subfigure}
    \caption{Evolution of the weights of the optimal portfolio as a function of $\alpha$. The average duration of the interruption is set to 2 days. The left column displays the results using Value-at-Risk as the risk measure, while the right column shows the results for Conditional Tail Expectation.}
    \label{fig_alpha}
\end{figure}

Figure \ref{fig_alpha} studies the effect of the confidence level $\alpha$ on the optimal allocation when the stressed-regime risk measure is either the Value-at-Risk or the Conditional Tail Expectation. This analysis is important because $\alpha$ controls the severity of the stressed scenario considered by the insurer. A higher confidence level corresponds to a more conservative view of cloud-outage losses.

The results show that the optimal allocation is relatively stable with respect to $\alpha$ for a fixed expected profitability $\rho$, although some weights change gradually when the confidence level increases. This stability indicates that, under the baseline assumptions, the ranking of providers in terms of risk-return contribution is not fully reversed when the insurer moves from a moderately adverse scenario to a more severe one. This is an important operational result: it suggests that the underwriting guidelines derived from the optimization are not excessively sensitive to small changes in the selected confidence level.

However, the effect of $\alpha$ is not negligible. When $\alpha$ increases, the optimizer gives more importance to the most adverse realizations of the stressed loss distribution. Providers whose outage losses have a more severe tail, or whose contribution to portfolio-level concentration is larger, become less attractive. Conversely, providers that offer a better trade-off between expected profitability and tail severity receive higher weights. This effect is more pronounced for CTE than for VaR, since CTE captures the expected loss beyond the quantile, while VaR only identifies the loss threshold associated with the chosen confidence level.

The comparison between the Weibull cases again highlights the importance of the duration distribution. For $a=0.5$, the interruption duration has a heavier tail, so increasing $\alpha$ gives more importance to rare but very long interruptions. However, because the loss function incorporates backup activation, very long interruptions do not necessarily translate linearly into losses: after the backup plan is activated, the loss rate is reduced. For $a=1$ and $a=1.5$, the distribution places relatively more mass around moderate-to-long durations, which produces different allocation patterns. This illustrates that the sensitivity to $\alpha$ depends not only on the tail of $T^{(j)}$, but also on the interaction between the outage duration, the backup mechanism, and the recovery period.

From a risk-management perspective, these results show that the confidence level should not be interpreted as a purely technical parameter. It reflects the insurer's degree of conservatism with respect to systemic cloud failures. A low level of conservatism may lead to an allocation that is efficient under average stressed conditions but insufficiently protective in extreme cases. A high level of conservatism leads to stronger protection against tail events, but at the cost of reducing the allocation to providers that are profitable in the standard regime. The choice of $\alpha$ should therefore be linked to the insurer's risk appetite, solvency position, and underwriting strategy.

\subsubsection{Impact of the volatility of $\frak{a}_{i,j}$ in the stochastic case}

\begin{figure}[!h]
    \centering
    \begin{subfigure}[b]{0.5\textwidth}
        \centering
        \includegraphics[width=\linewidth]{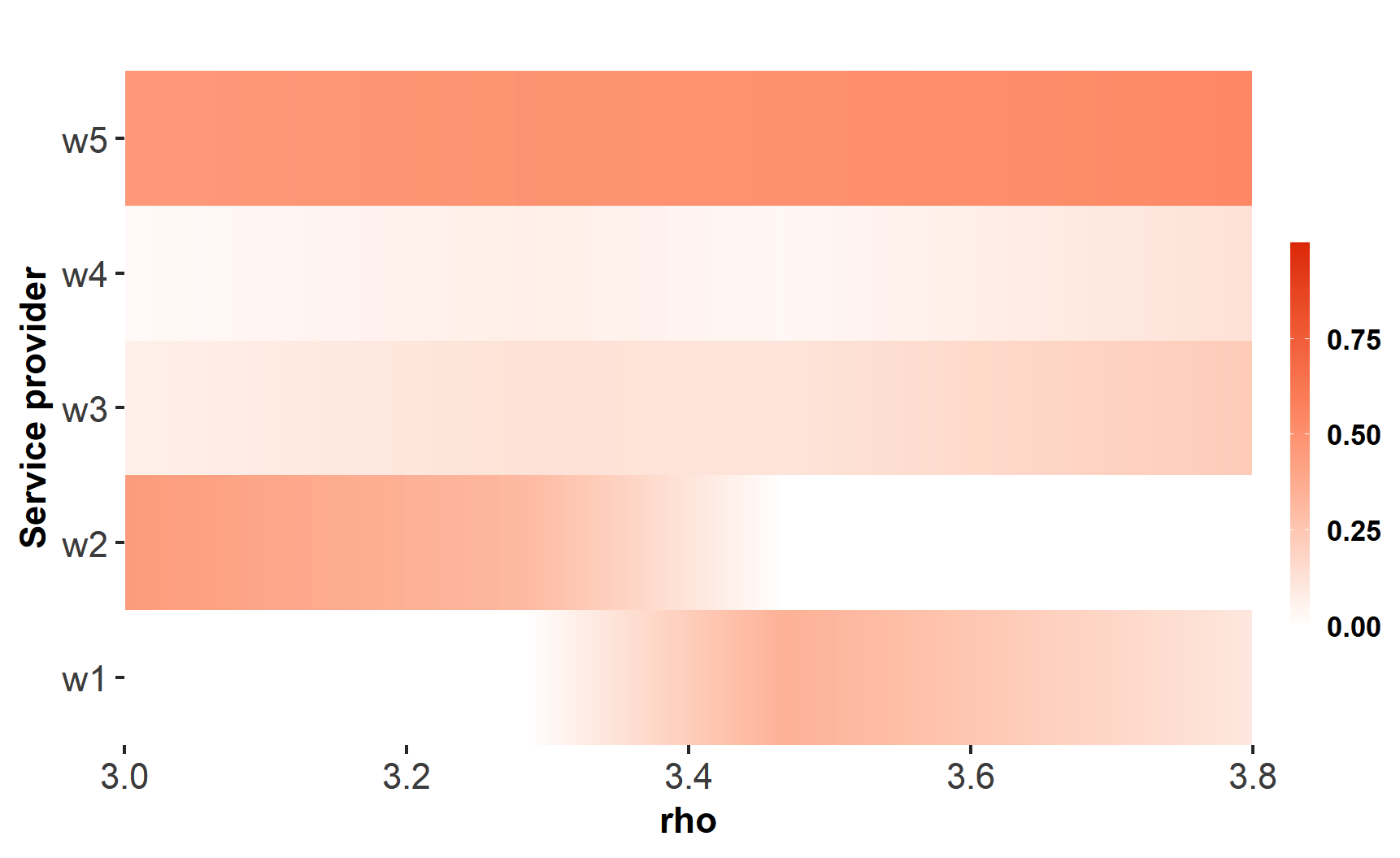}
        \caption{Standard deviation $0.1$.}
    \end{subfigure}%
    \begin{subfigure}[b]{0.5\textwidth}
        \centering
        \includegraphics[width=\linewidth]{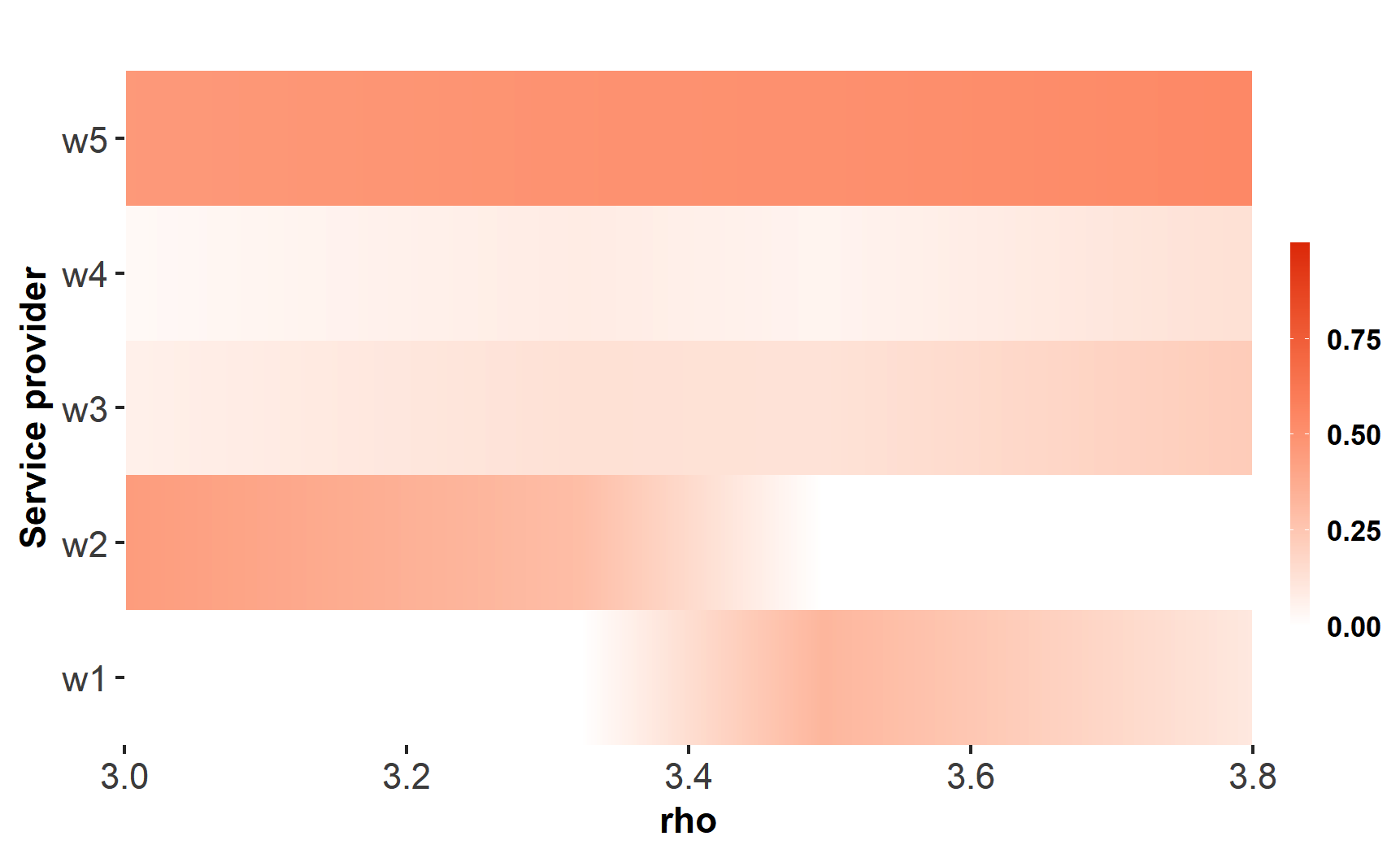}
        \caption{Standard deviation $0.75$.}
    \end{subfigure}\\[20pt]
    \centering
    \begin{subfigure}[b]{0.5\textwidth}
        \centering
        \includegraphics[width=\linewidth]{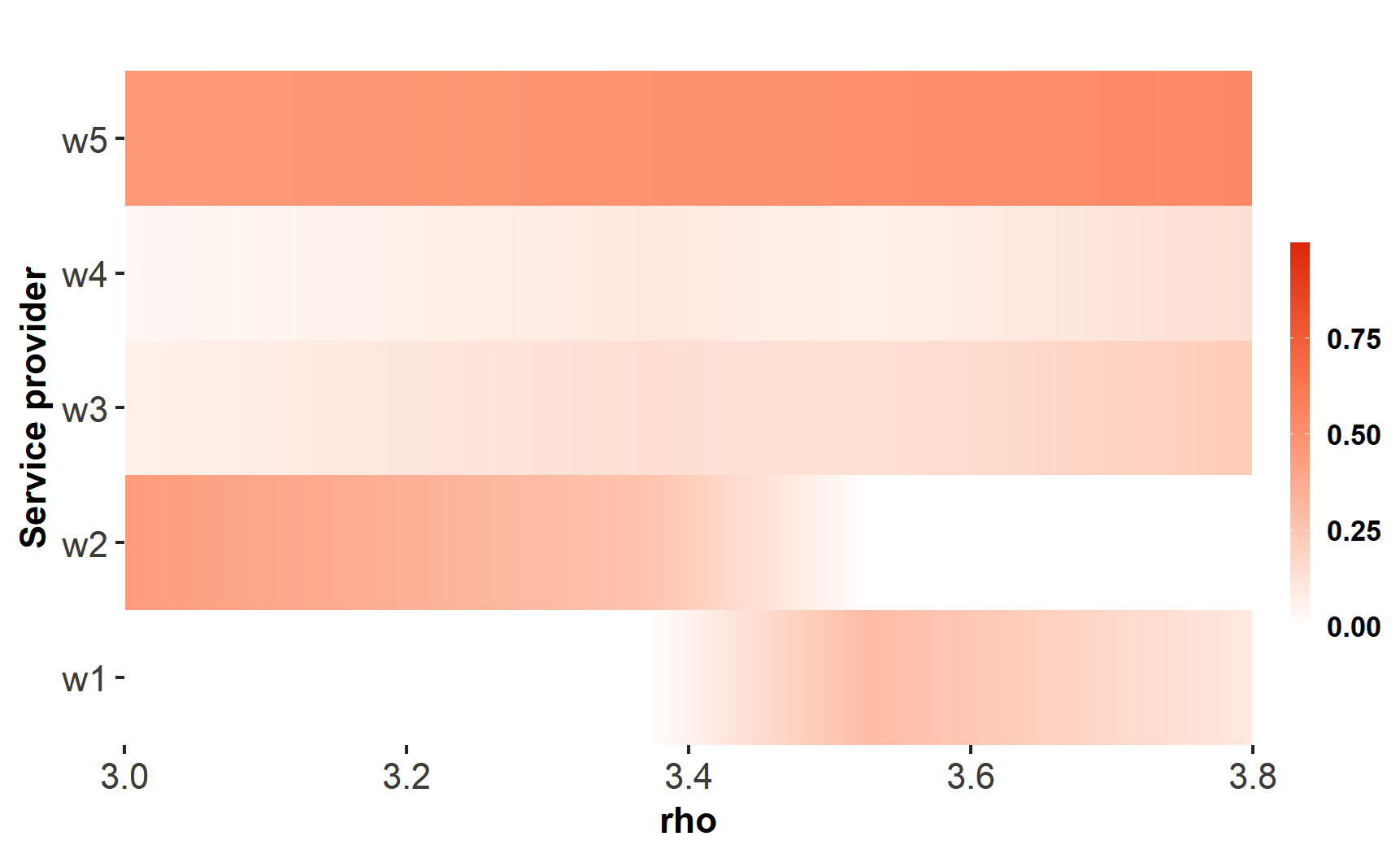}
        \caption{Standard deviation $1.4$.}
    \end{subfigure}%
    \caption{Evolution of the weights of the optimal portfolio as a function of $\rho$ in the case where $a=1$ (exponential distribution), $\alpha=0.95$, the risk measure is CTE, and $\frak{a}_{i,j}$ is random. The distribution of $\frak{a}_{i,j}$ is Gaussian, with mean defined as in Table \ref{tab:critic}, and different values of the standard deviation are considered.}
    \label{fig_random}
\end{figure}

The previous analyses rely mainly on a deterministic daily loss intensity $\mathfrak{a}_{i,j}$ for each provider. Figure \ref{fig_random} relaxes this assumption by allowing $\mathfrak{a}_{i,j}$ to be stochastic, with a provider-specific mean and different levels of standard deviation. This extension is important because, in practice, two policyholders exposed to the same cloud provider could experience very different financial consequences from the same outage. Their vulnerability depends on their business model, operational dependence on the cloud, ability to switch to alternative procedures, sector of activity, digital maturity, and crisis-management capabilities.

The results show that increasing the volatility of $\mathfrak{a}_{i,j}$ affects the composition of the optimal portfolio. The general allocation pattern remains comparable to the deterministic case, but some weights become more sensitive to the profitability constraint $\rho$. This result is consistent with the objective of the optimization problem. The quadratic component of the criterion penalizes volatility in the standard regime. Therefore, when the variability of daily loss intensity increases, the covariance matrix $\Sigma$ changes and the optimizer reduces the exposure to providers that contribute more strongly to portfolio volatility. This is particularly the case for providers $\mathbf{1}$ and $\mathbf{2}$, which, according to Table \ref{tab:critic}, have the highest average daily losses and, according to the formula for the variance in Section \ref{sec:prob_interup}, make a significant contribution to portfolio volatility.

The stochastic specification of $\mathfrak{a}_{i,j}$ also has an important interpretation for underwriting. If the insurer has little information about the actual dependence of each policyholder on a given provider, the volatility of $\mathfrak{a}_{i,j}$ could be interpreted as a measure of underwriting uncertainty. A low volatility corresponds to a homogeneous portfolio where policyholders suffer relatively similar daily losses from a cloud outage. A high volatility corresponds to a heterogeneous portfolio where some policyholders may suffer much larger losses than others. In this second case, diversification across providers alone may not be sufficient: the insurer should also diversify across sectors, business models, and levels of cloud criticality.

\subsubsection{Impact of the penalty parameter $\lambda$}
\label{sec:imp_lambda}
\begin{figure}[!h]
    \centering
    \begin{subfigure}[b]{0.5\textwidth}
        \centering
        \includegraphics[width=\linewidth]{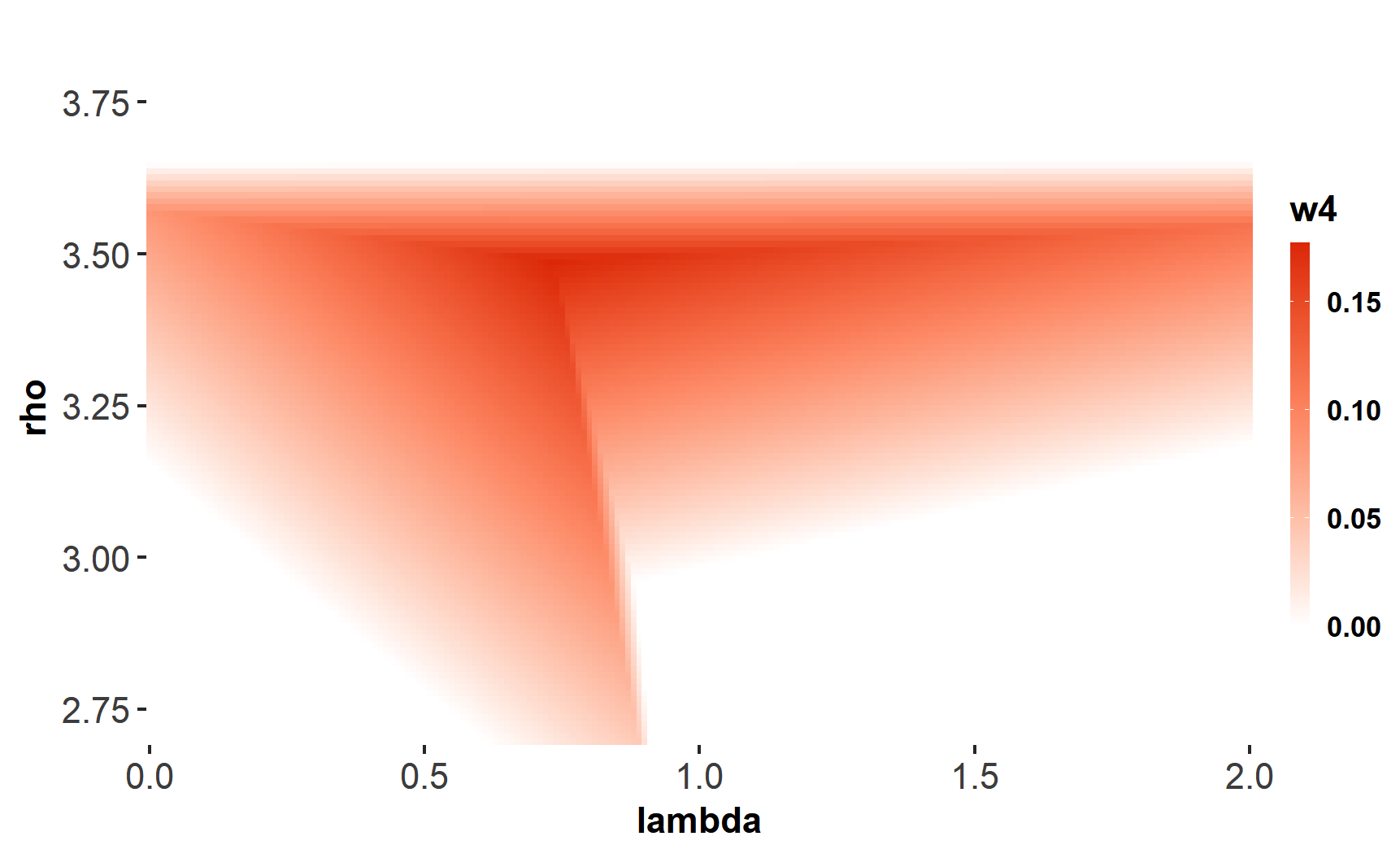}
        \caption{$a=0.5$.}
    \end{subfigure}%
    \begin{subfigure}[b]{0.5\textwidth}
        \centering
        \includegraphics[width=\linewidth]{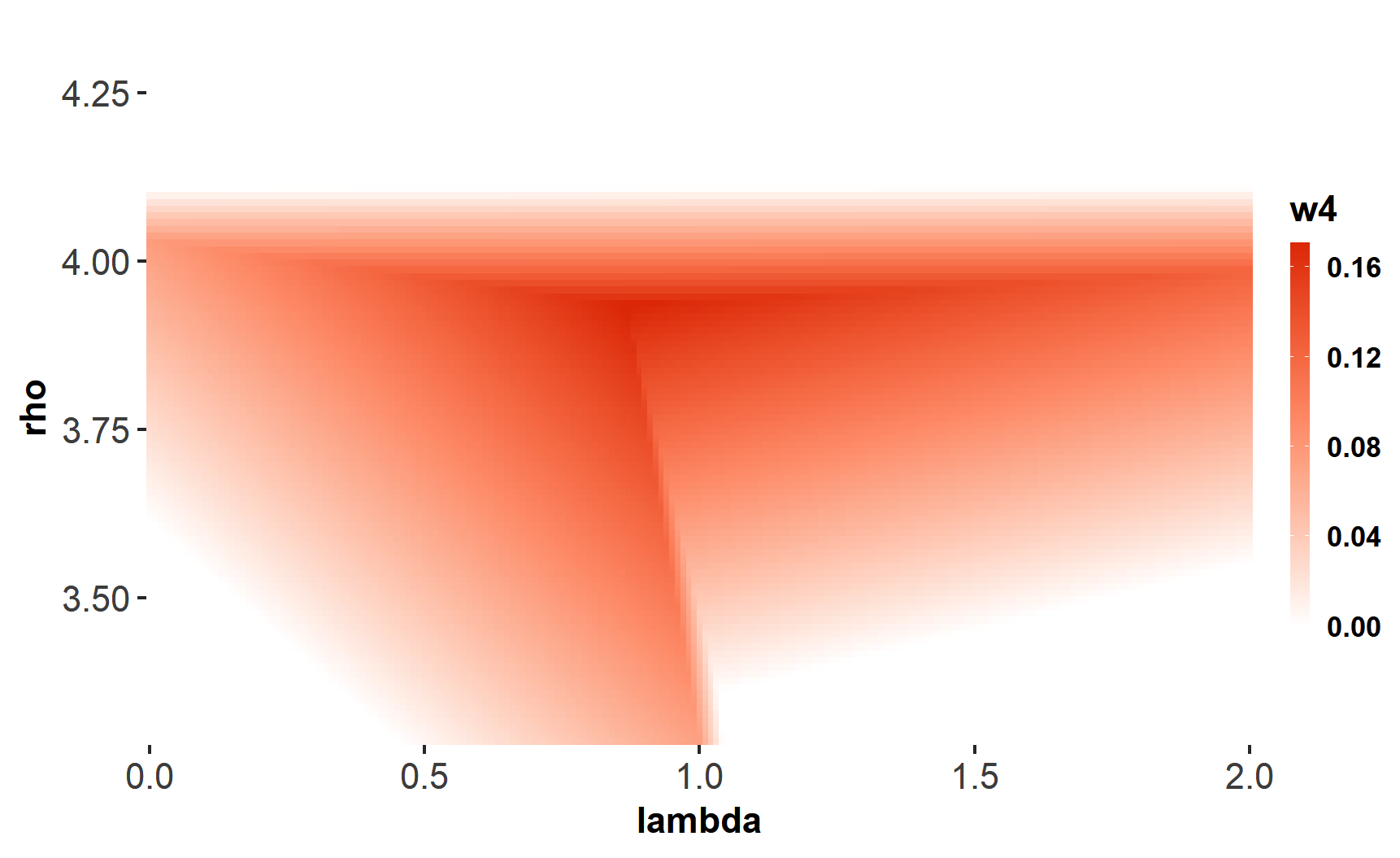}
        \caption{$a=1$.}
    \end{subfigure}\\[20pt]
    \centering
    \begin{subfigure}[b]{0.5\textwidth}
        \centering
        \includegraphics[width=\linewidth]{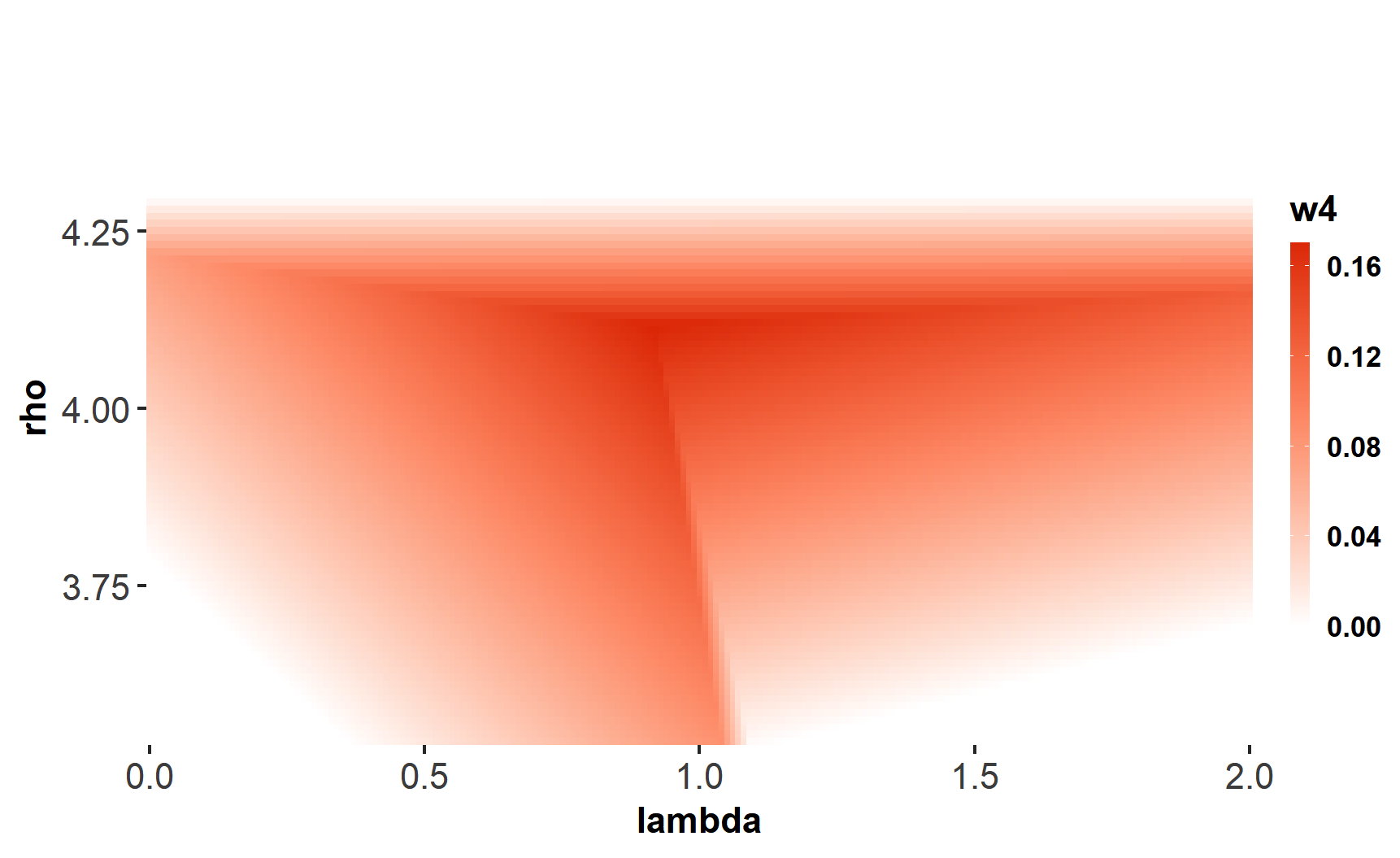}
        \caption{$a=1.5$.}
    \end{subfigure}%
    \caption{Evolution of the weight $w_4$ using CTE as the risk measure with $\alpha=0.95$.}
    \label{fig_bof}
\end{figure}

Figure \ref{fig_bof} analyzes the role of the penalty parameter $\lambda$. This parameter controls the relative importance given to the stressed-regime risk measure in the optimization criterion. When $\lambda$ is small, the optimization is mainly driven by the variance of standard-regime losses in the standard regime. When $\lambda$ increases, the optimizer gives more weight to the potential severity of cloud-outage scenarios affecting a large part of the portfolio.

The results show that $\lambda$ has a direct influence on the optimal allocation. A higher value of $\lambda$ discourages excessive concentration on providers that generate large stressed-regime losses, even if they are attractive from the point of view of standard-regime profitability. Conversely, a lower value of $\lambda$ allows the optimizer to remain closer to a classical mean-variance allocation, where diversification is mainly understood through volatility reduction.

The calibration rule proposed earlier in the paper provides a practical way to choose $\lambda$ by linking it to the capital amount the insurer is willing to allocate to these risks. Nevertheless, the sensitivity analysis in Figure \ref{fig_bof} is useful because it shows how the portfolio would change if the insurer adopted a more or less conservative view of systemic cloud interruption risk.

From an underwriting perspective, the interpretation is straightforward. If the insurer is mainly concerned with short-term profitability and ordinary claims volatility, it should select a relatively low value of $\lambda$. The resulting allocation may be efficient in the standard regime, but it can leave the portfolio more exposed to a cloud-outage accumulation scenario. If the insurer is more concerned with solvency resilience under systemic cyber events, it should select a higher value of $\lambda$. The resulting allocation may sacrifice some standard-regime profitability, but it should provide stronger protection against concentration risk.

\subsubsection{An overview of a 5-day average interruption duration}
\label{sec:5_day_int}

\begin{figure}[!h]
    \centering
    \begin{subfigure}[b]{0.5\textwidth}
        \centering
        \includegraphics[width=\linewidth]{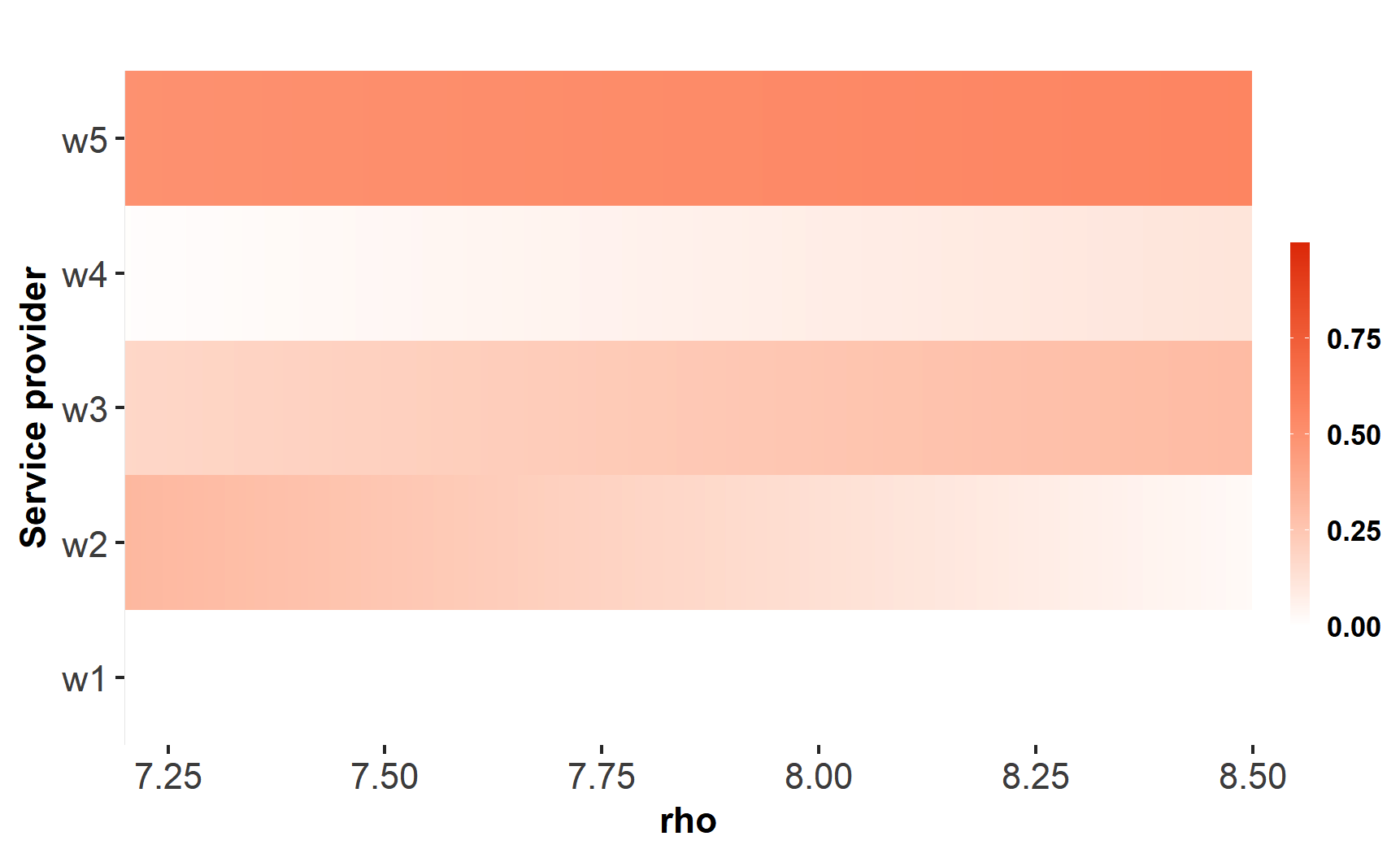}
        \caption{Mean.}
    \end{subfigure}%
    \begin{subfigure}[b]{0.5\textwidth}
        \centering
        \includegraphics[width=\linewidth]{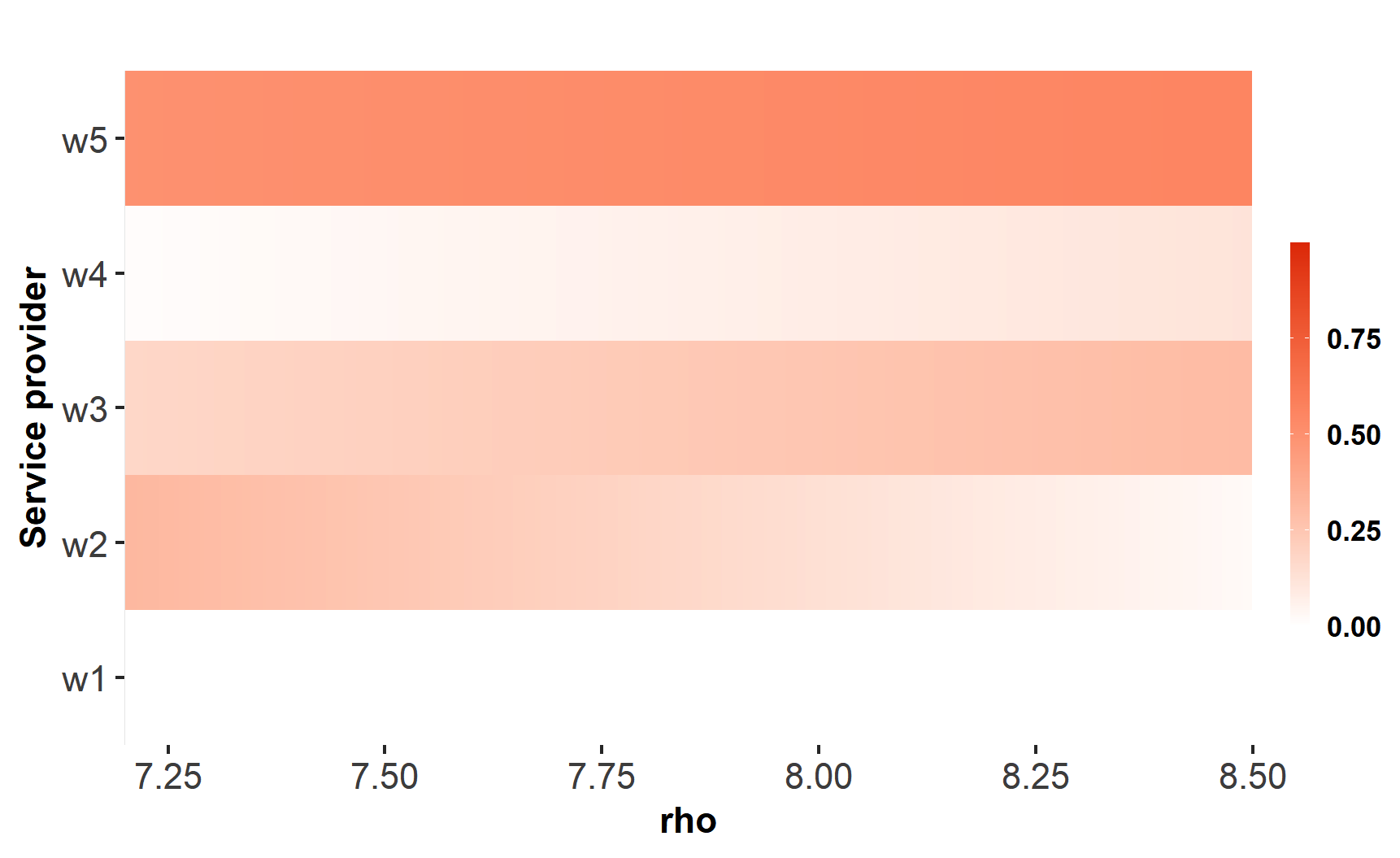}
        \caption{VaR.}
    \end{subfigure}\\[20pt]
    \centering
    \begin{subfigure}[b]{0.5\textwidth}
        \centering
        \includegraphics[width=\linewidth]{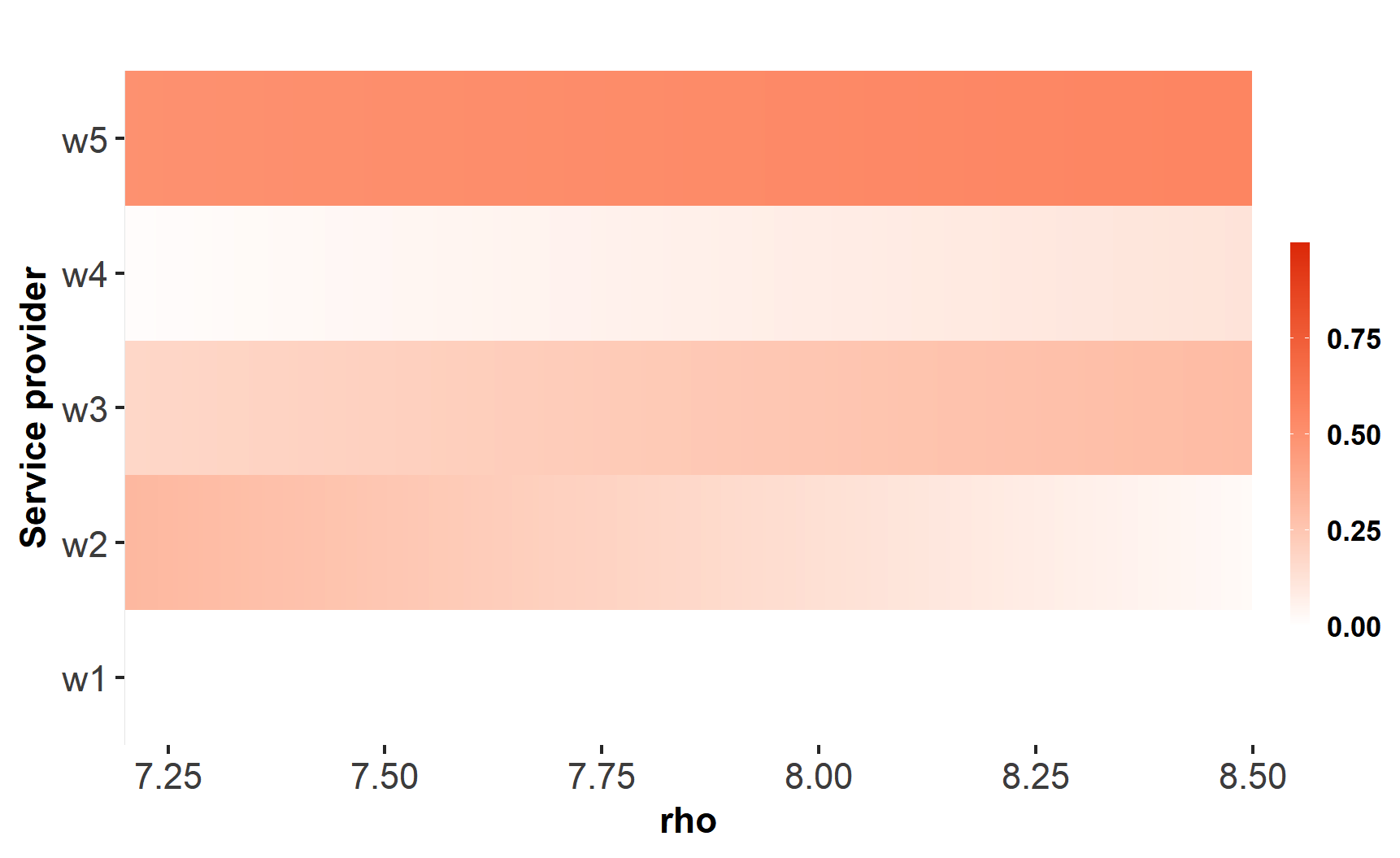}
        \caption{CTE.}
    \end{subfigure}%
    \caption{Evolution of the weights of the optimal portfolio as a function of $\rho$. The average duration of the interruption is set to 5 days. Only the results for the exponential distribution ($a=1$) are reported for the different risk measures.}
    \label{fig_ext}
\end{figure}

\begin{figure}[!h]
    \centering
    \begin{subfigure}[b]{0.5\textwidth}
        \centering
        \includegraphics[width=\linewidth]{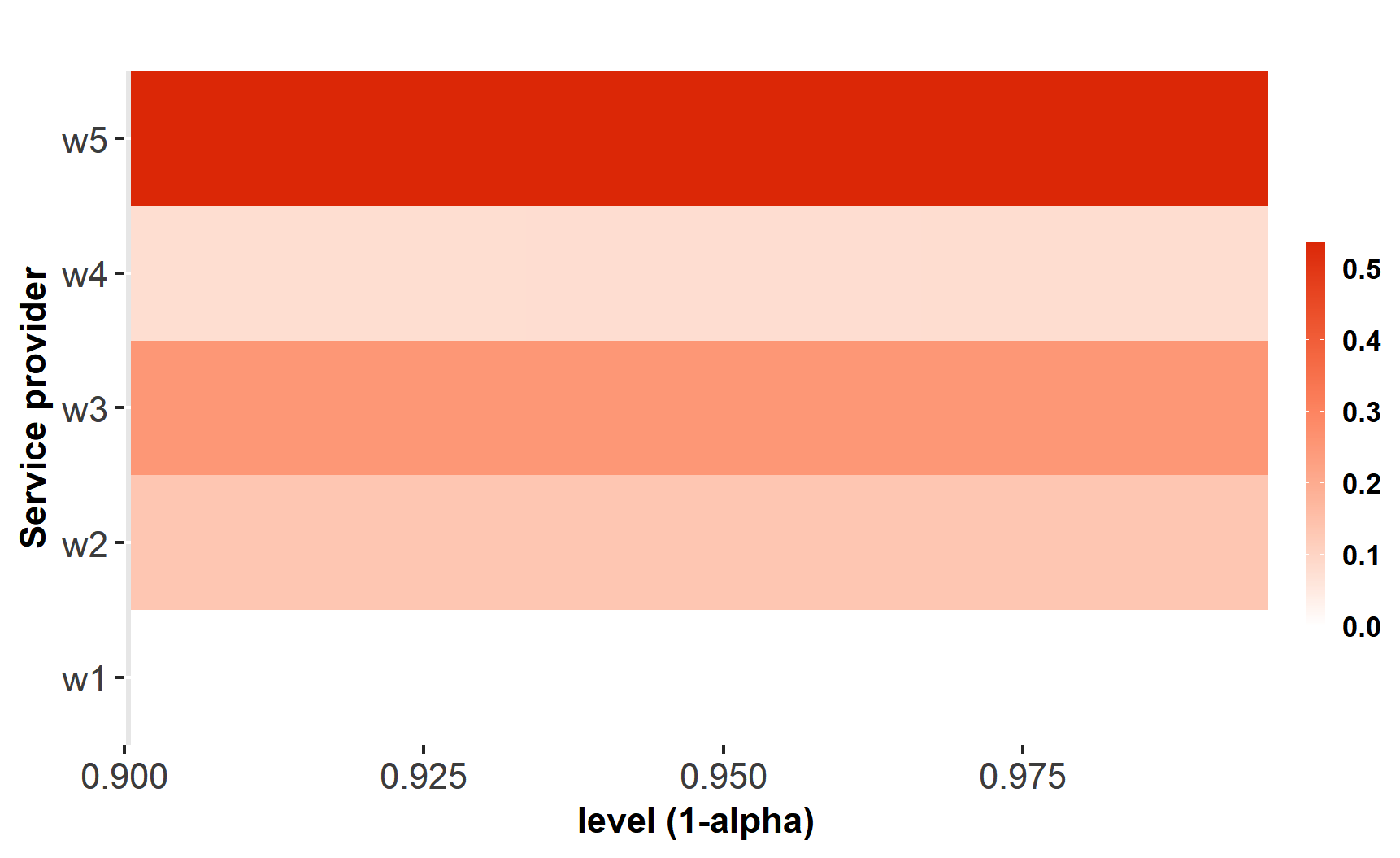}
        \caption{VaR.}
    \end{subfigure}%
    \begin{subfigure}[b]{0.5\textwidth}
        \centering
        \includegraphics[width=\linewidth]{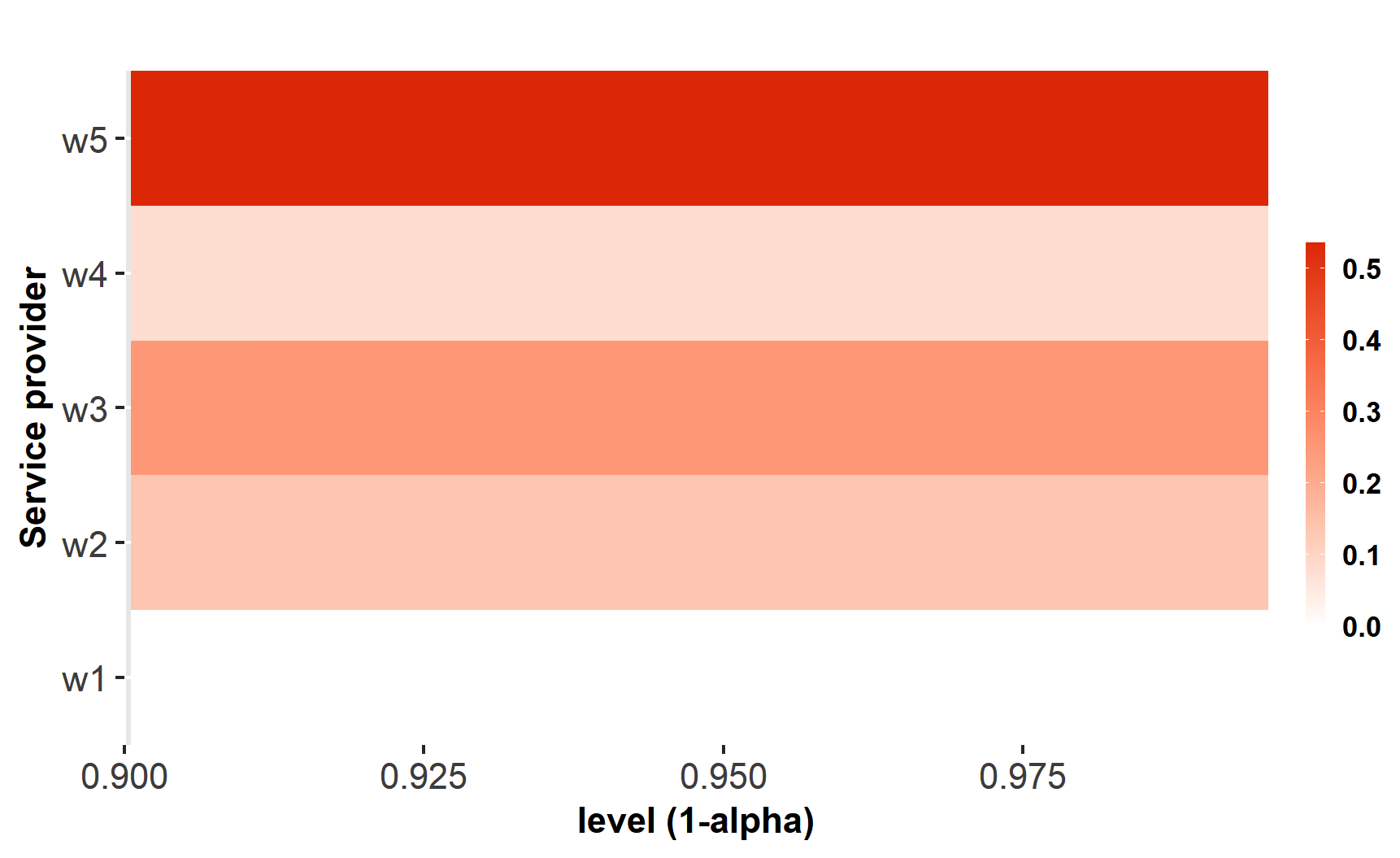}
        \caption{CTE.}
    \end{subfigure}
    \caption{Evolution of the weights of the optimal portfolio as a function of $\alpha$. The average duration of the interruption is set to 5 days. Only the results for the exponential distribution ($a=1$) are reported for the different risk measures.}
    \label{fig_ext2}
\end{figure}

Figures \ref{fig_ext} and \ref{fig_ext2} extend the baseline analysis by considering an average cloud-interruption duration of five days instead of two days. This scenario is more severe and corresponds to a longer disruption of the cloud service. It is therefore useful for assessing whether the qualitative conclusions obtained under the baseline scenario remain valid when the interruption becomes more persistent.

The first observation is that increasing the average interruption duration modifies the structure of the optimal portfolio. This is expected because the loss function is directly linked to the duration of business interruption. A longer average outage increases the probability that the interruption lasts beyond the backup activation time and also increases the contribution of the recovery phase. As a consequence, the stressed-regime risk associated with each provider becomes larger, and the optimizer adjusts the allocation to control the resulting accumulation risk.

The comparison with the two-day scenario shows that some providers become less attractive when the average interruption duration increases. The reason is that a longer outage magnifies differences in provider criticality, loss intensity, and dependence structure. This is the case for provider $\mathbf{1}$, which has the second-highest daily cost of interruption and a relatively high probability of interruption. More generally, providers with higher daily loss intensity or stronger contributions to the covariance matrix become more costly from a diversification perspective. The optimizer therefore reallocates exposure toward providers that offer a better balance between expected profitability and stressed-regime resilience under longer outages.  

Overall, the five-day scenario confirms the main conclusion of the numerical illustration. The optimal portfolio structure is not fixed once and for all. It depends on the insurer's profitability target, its confidence level for tail-risk measurement, its assumptions on policyholder heterogeneity, the penalty assigned to systemic risk, and the assumed duration of cloud interruption. The proposed optimization framework provides a coherent way to translate these assumptions into aggregate exposure targets by provider. It can therefore support underwriting guidelines aimed at preserving mutualization in ordinary conditions while limiting concentration in systemic cloud-outage scenarios.

\begin{remark}
Due to fairness and regulatory considerations limiting discriminatory underwriting decisions, as mentioned in Section \ref{sec:compare}, the insurer might be required to choose a target portfolio structure in which all service providers in the market are represented. This imposes a constraint on the expected profitability that the insurer can target. In the setting of Figure \ref{fig_rho}, for example, under the VaR risk measure with $a=1$ (Panel e), this expected profitability target should be around 3\,400 euros to ensure that all five service providers in the market are represented in the portfolio.
\end{remark}

\section{Conclusion}
\label{sec:conclusion}

Cloud interruption is one of the most concerning scenarios for the collective risk management of cyber risk. It can destabilize an insurance portfolio and is identified by EIOPA as one of the situations that should be modeled in cyber risk stress-testing frameworks. Measuring the consequences of such an episode first requires understanding the impact of such a disruption on affected policyholders. We propose a framework, inspired by \cite{lloyds}, that studies the different stages of business interruption and identifies the main quantities that need to be modeled to assess this impact.

A first strength of the proposed approach is that it explicitly distinguishes two regimes. The first corresponds to the standard insurance regime, where losses remain sufficiently isolated for mutualization to operate. The second corresponds to a stressed regime, where a common cloud-service failure affects a large share of the portfolio. This distinction is important because a portfolio that appears diversified under ordinary loss conditions may still be highly vulnerable to an accumulation scenario generated by a common technological dependency. The model therefore provides a way to analyze both the profitability and volatility of standard-regime losses and the exposure of the portfolio to systemic cloud-outage events.

A second strength of the framework is its operational simplicity. Instead of optimizing the exposure of each policyholder to each provider separately, the proposed approach works with the aggregate exposure of the portfolio to each cloud provider. This aggregation makes the model easier to interpret and more directly usable for underwriting. In practice, an insurer cannot freely redesign its portfolio at the individual policyholder level. However, it can monitor the distribution of aggregate exposure by provider and use this information to define underwriting guidelines, limits, or alerts when exposure to a given provider becomes too concentrated.

The numerical illustration confirms the relevance of this approach. The optimal portfolio is sensitive to the expected profitability constraint, the selected risk measure in the stressed regime, the confidence level used for VaR or CTE, the volatility of the daily interruption cost, the penalty parameter associated with systemic risk, and the average duration of the cloud interruption. These results show that the optimal portfolio structure cannot be defined independently of the insurer's risk appetite and stress-testing assumptions. They also show that the framework can be used as a sensitivity-analysis tool to identify which providers or assumptions drive concentration risk.

The model nevertheless has several limitations. First, the numerical application is based on the \cite{lloyds} report and on realistic assumptions without a complete empirical calibration. This is partly due to the scarcity of reliable historical data on systemic cloud outages and cyber accumulation events. The probabilities of failure, dependence structure, interruption durations, and daily loss intensities should therefore be interpreted as scenario assumptions rather than as estimates of the true risk profile of specific providers. Second, the model assumes that policyholders can be summarized through aggregate exposure, turnover, backup activation time, recovery time, and daily loss intensity. Although this structure is useful for portfolio-level analysis, it necessarily simplifies the diversity of firms' information systems and operational dependencies.

Another limitation concerns the modeling of heterogeneity. We stress that the consequences of a cloud interruption may be very different for two policyholders operating in different industrial sectors. Depending on the nature of the activity, time may not have the same value: some firms may postpone operations with limited damage, whereas others may face immediate and severe business interruption. Therefore, a specific analysis should be developed to study the link between policyholder characteristics and the cost of one day of business interruption. This is not the main objective of the present paper, but the model can be naturally adapted to include covariates such as sector of activity, digital maturity, dependence on online sales, existence of alternative providers, and quality of business-continuity planning.

Several extensions could be considered in future research. A first direction would be to improve the empirical calibration of the model by using insurer portfolio data, cloud-dependency questionnaires, incident databases, or external information on cloud-service outages. This would make it possible to estimate more accurately the distribution of interruption durations, the effectiveness of backup plans, and the dependence structure between providers. A second direction would be to introduce a more detailed modeling of policyholder heterogeneity, for example by allowing the parameters of the loss function to depend on observable characteristics of the insured firm. This would help identify which sectors or types of policyholders contribute most to cloud accumulation risk. Finally, future work could connect the proposed optimization criterion with solvency capital requirements, reinsurance design, or cyber accumulation limits, in order to translate the model outputs into practical risk-management and capital-management decisions.

Overall, the proposed framework should be viewed as a transparent and flexible stress-testing tool for cloud-outage accumulation risk. Its objective is not to predict the exact probability or cost of a future cloud catastrophe, but to help insurers understand how concentration in cloud-service dependencies may threaten mutualization and how underwriting guidelines can be designed to improve portfolio resilience.

\section{Appendix}

\subsection{Proof of Equation \ref{eq:SJ}}
\label{sec:proof}

In case of a systemic event impacting the cloud provider $j$, $T_i^{(j)}$ is the same for all $i,$ that $T_i^{(j)}=T^{(j)}.$ On the event $\{T^{(j)}=t\},$ the portfolio loss $\mathbf{L}^{(j)}$ can be rewritten as
$$\frac{\mathbf{L}^{(j)}}{\bar{w}_j}=\sum_{i=1}^n \frac{w_{i,j}\tau_i}{\bar{w}_j}\varphi_t(\mathbf{Z}^{(j)}_i),$$
where $\mathbf{Z}_i^{(j)}=(\frak{a}_{i,j},\frak{b}_{i},U_i,V_i).$ $(\mathbf{Z}^{(j)}_i)_{1\leq i \leq n}$ are i.i.d. and independent of $T^{(j)}$ by assumption. On the other hand,
$E[\varphi_t(\mathbf{Z}^{(j)}_i)]=\frak{m}_t,$ and $Var(\mathbf{L}^{(j)}/\bar{w}_j)=s^2_w\sigma^2_t.$

From an asymptotic point of view, if
\begin{equation}
\sum_{i=1}^n \left(\frac{w_{i,j}\tau_i}{\bar{w}_j}\right)^4<\infty, \label{cond4}
\end{equation}
the weighted Central Limit Theorem applies (see \cite{WEBER20061482}), and
$$\mathbb{P}\left(\frac{\mathbf{L}^{(j)}}{\bar{w}_j}\geq l|T^{(j)}=t\right)\sim_{n\rightarrow \infty} \bar{\Phi}\left(\frac{l-\frak{m}_t}{s_w\sigma_t}\right).$$

Equation \ref{eq:SJ} then comes from the fact that
$$\mathbb{P}\left(\frac{\mathbf{L}^{(j)}}{\bar{w}_j}\geq l\right)=\int_0^{\infty}\mathbb{P}\left(\frac{\mathbf{L}^{(j)}}{\bar{w}_j}\geq l|T^{(j)}=t\right)d\mathbb{P}_j(t).$$

Let us note that condition \ref{cond4} could be weakened, see Theorem 1 in \cite{WEBER20061482}.

\subsection{Densities of financial losses and interruption durations}

\begin{figure}[!h]
    \centering
    \begin{subfigure}[b]{0.5\textwidth}
        \centering
        \includegraphics[width=\linewidth]{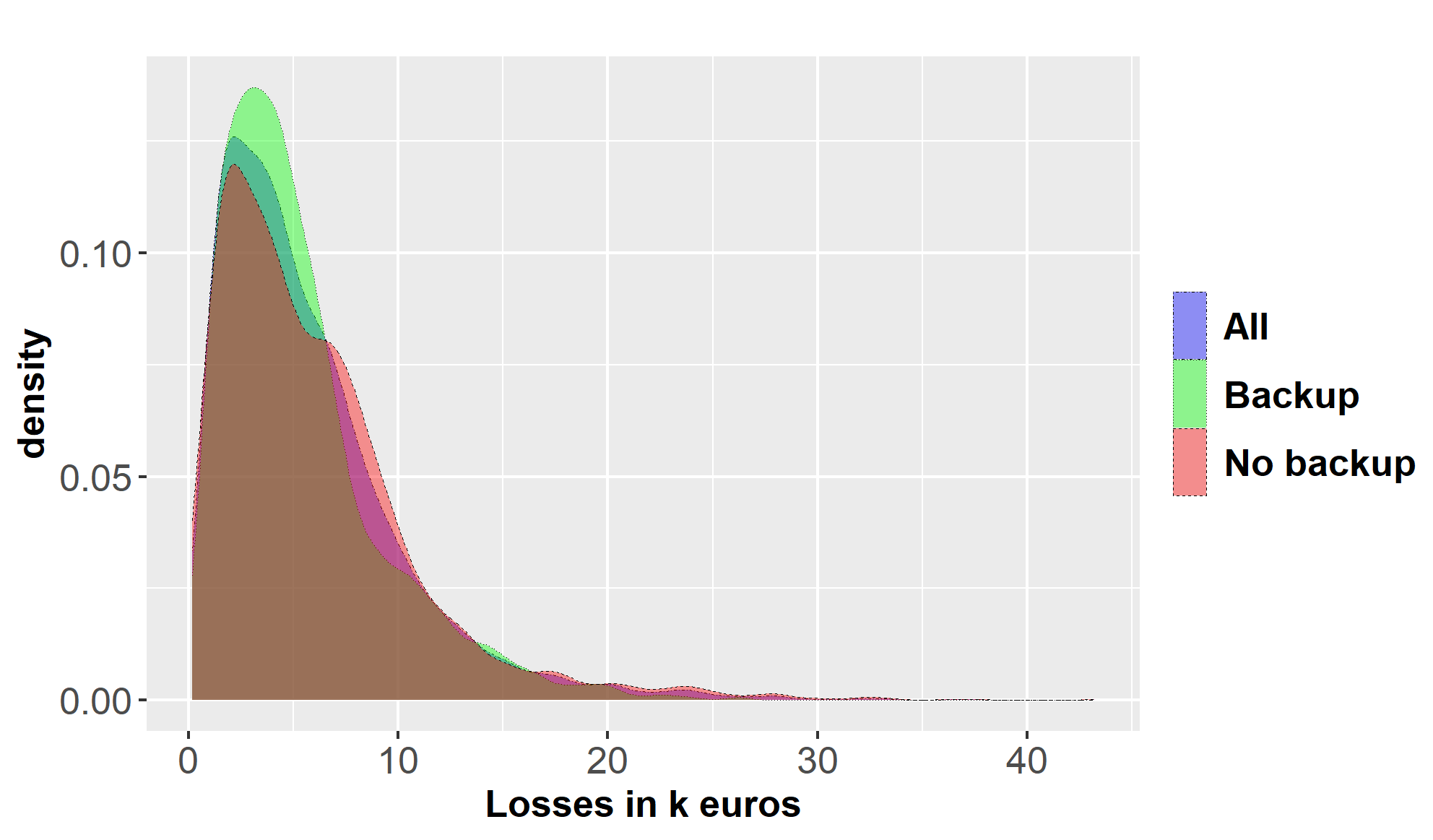}
        \caption{$a=0.5$, Financial loss.}
    \end{subfigure}%
    \begin{subfigure}[b]{0.5\textwidth}
        \centering
        \includegraphics[width=\linewidth]{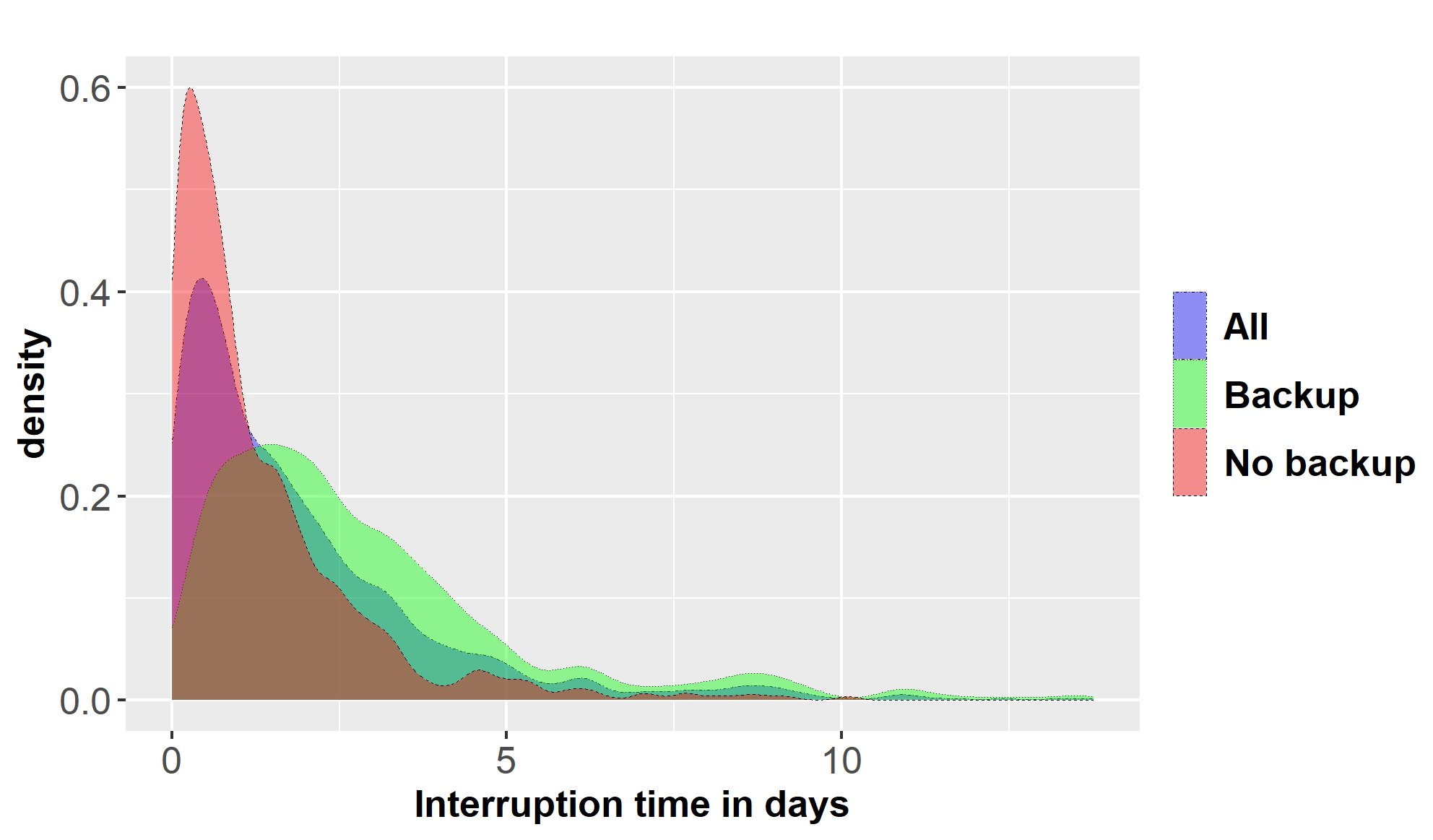}
        \caption{$a=0.5$, Interruption duration.}
    \end{subfigure}\\[20pt]
    \begin{subfigure}[b]{0.5\textwidth}
        \centering
        \includegraphics[width=\linewidth]{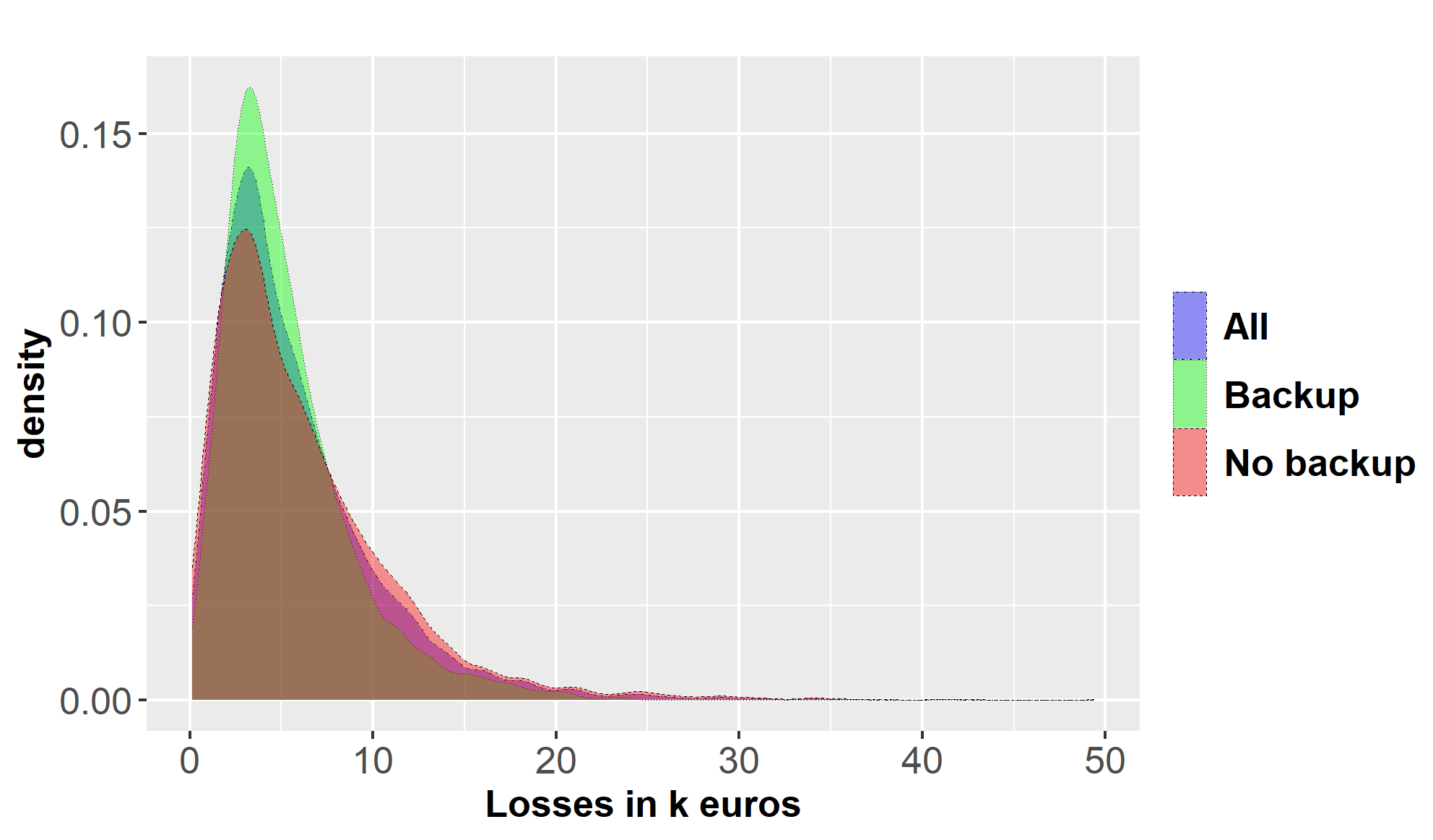}
        \caption{$a=1.0$, Financial loss.}
    \end{subfigure}%
    \begin{subfigure}[b]{0.5\textwidth}
        \centering
        \includegraphics[width=\linewidth]{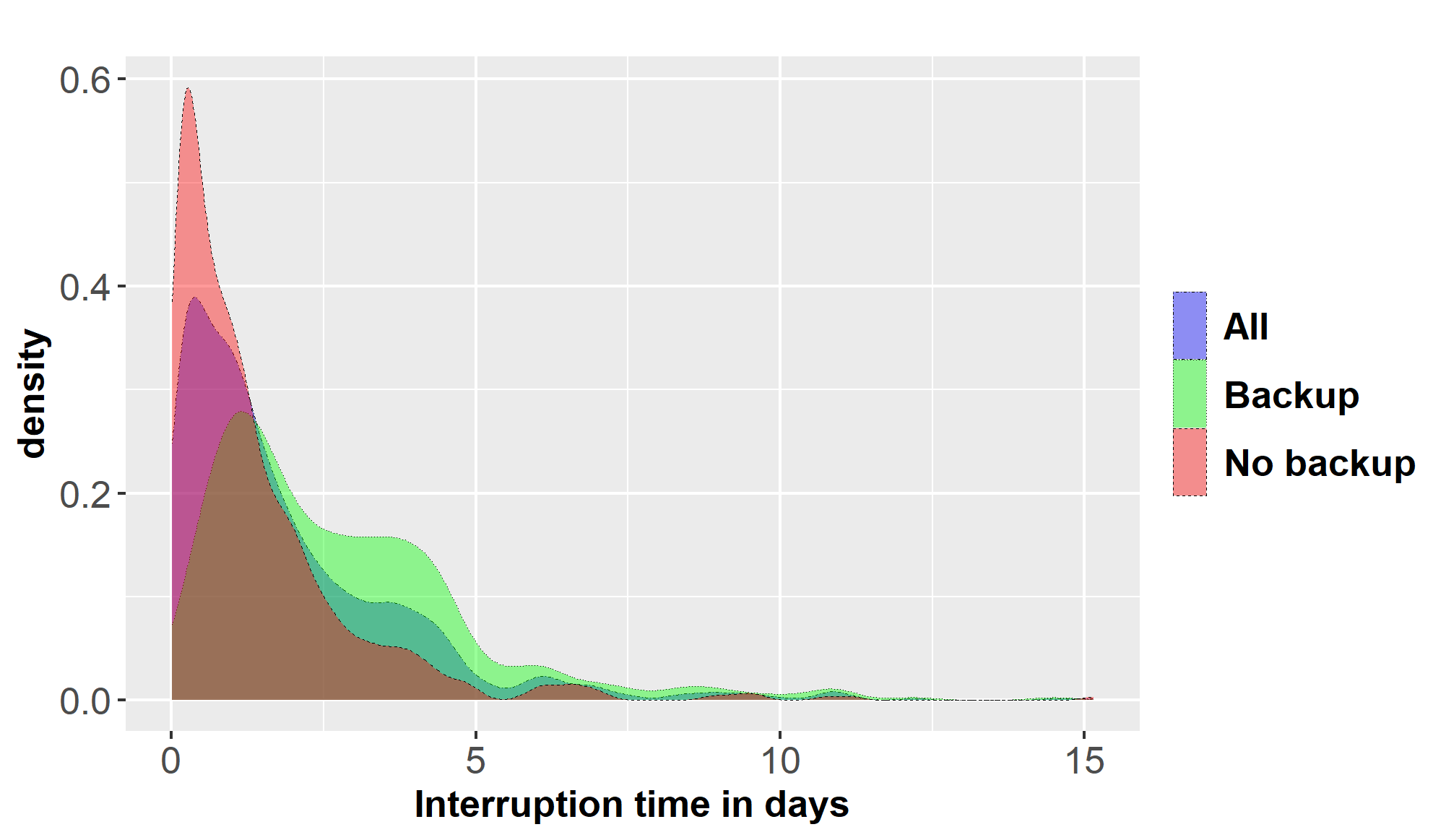}
        \caption{$a=1.0$, Interruption duration.}
    \end{subfigure}\\[20pt]
    \begin{subfigure}[b]{0.5\textwidth}
        \centering
        \includegraphics[width=\linewidth]{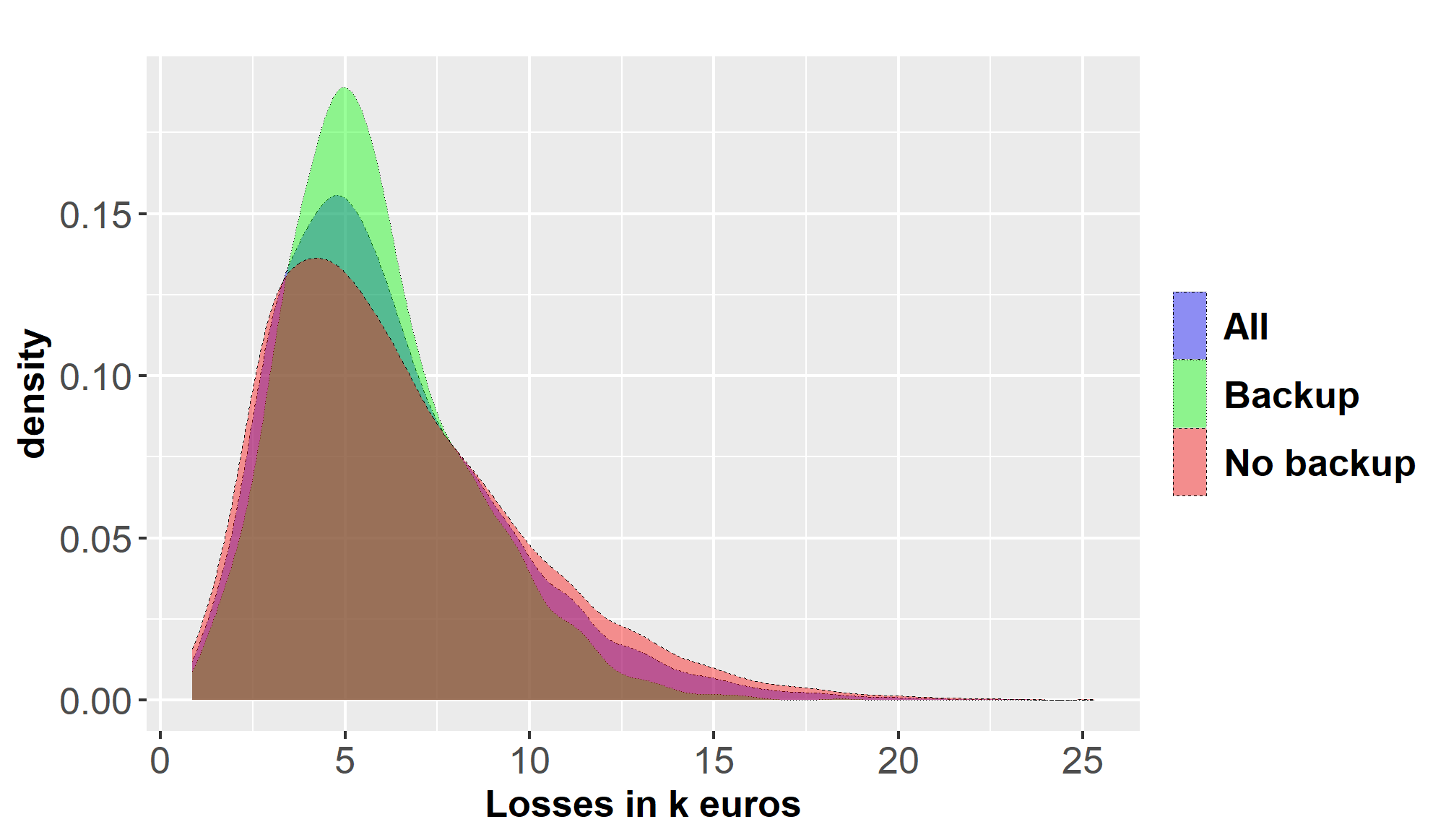}
        \caption{$a=1.5$, Financial loss.}
    \end{subfigure}%
    \begin{subfigure}[b]{0.5\textwidth}
        \centering
        \includegraphics[width=\linewidth]{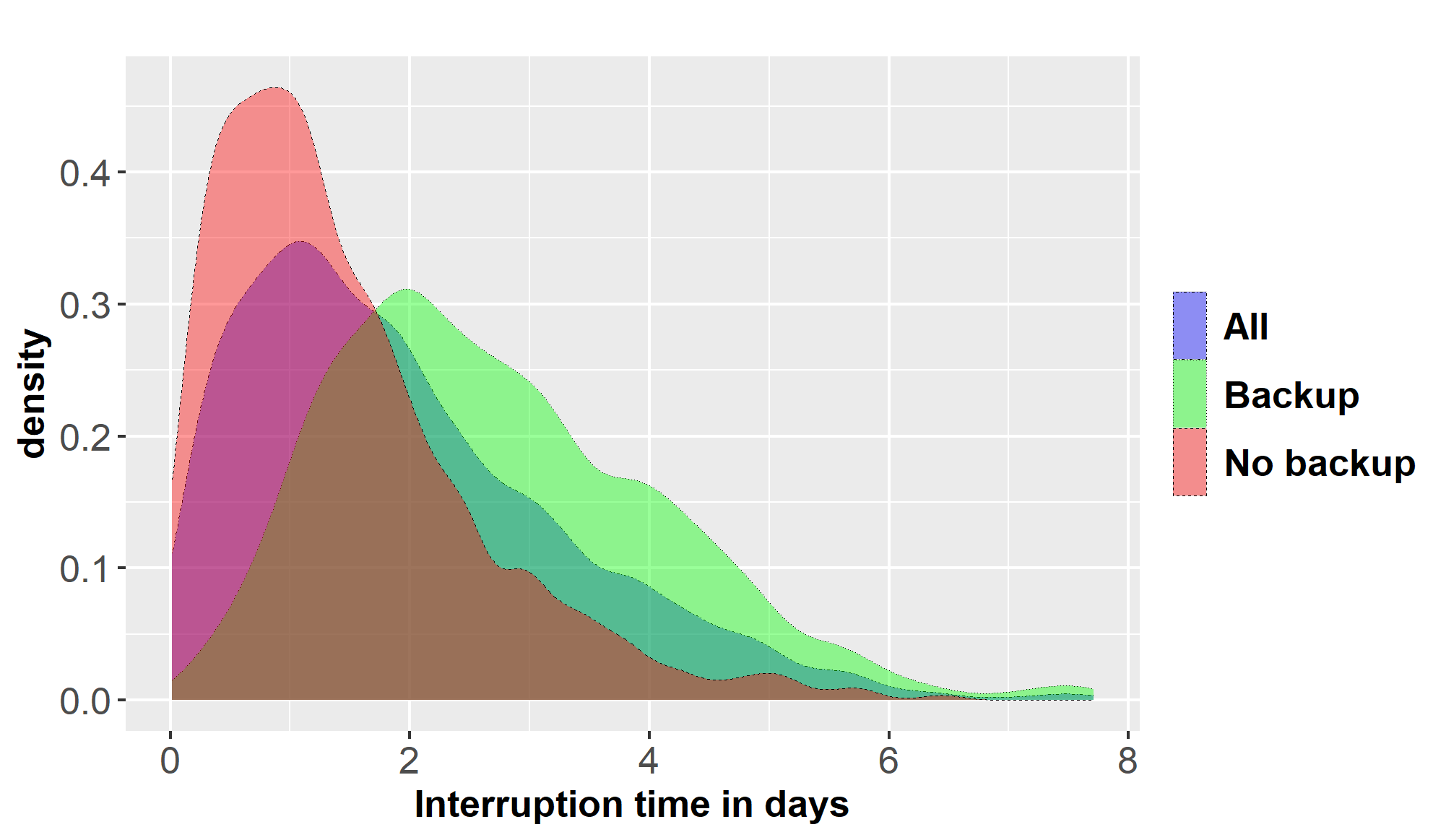}
        \caption{$a=1.5$, Interruption duration.}
    \end{subfigure}
    \caption{Densities of financial losses due to cloud interruptions (Panels a, c, and e) and interruption durations (Panels b, d, and f). The densities are presented for $a=0.5$, $a=1$, and $a=1.5$, which correspond respectively to Weibull distributions with decreasing, constant, and increasing hazard rates for interruption durations.}
    \label{fig:loss_densities}
\end{figure}

\textbf{Funding information:} {\textit{Olivier Lopez received funding from the Excellence chair CARE under the aegis of Fondation du Risque, in partnership with GENES, and with the support of Allianz, and from the project CyFi (Cyber Financialization) between Fondation du Risque, Citalid and Bpi France.}}

\bibliographystyle{abbrvnat}

\end{document}